\documentclass[a4paper,twocolumn,11pt]{quantumarticle}
\pdfoutput=1
\usepackage[utf8]{inputenc}
\usepackage[english]{babel}
\usepackage[T1]{fontenc}
\usepackage{hyperref}
\usepackage{amsmath}
\usepackage{amssymb}
\usepackage{amsthm}
\usepackage{calc}
\usepackage{tikz-cd}
\usepackage{braket}
\usepackage{array}
\usepackage{multirow}
\usepackage{makecell}
\usepackage{mathtools}
\usepackage{comment}

\newtheorem{theorem}{Theorem}
\newtheorem*{theorem*}{Theorem}
\newtheorem{definition}{Definition}
\newtheorem*{definition*}{Definition}
\newcolumntype{C}[1]{>{\centering\arraybackslash}m{#1}}
\newcolumntype{L}[1]{>{\raggedright\arraybackslash}m{#1}}

\newcommand{\Opposite}[1]{{\raisebox{\depth}{\rotatebox[origin=c]{180}{$#1$}}}}

\newcommand{\HilSpace}{{\mathsf{H}}}
\newcommand{\StSpace}{{\mathsf{M}}}
\newcommand{\ObSpace}{\Opposite{\StSpace}}
\newcommand{\StOps}{\StSpace}
\newcommand{\ObOps}{\ObSpace}
\newcommand{\StDim}{{m}}
\newcommand{\ObDim}{{w}}
\newcommand{\identity}{{\imath}}
\newcommand{\braid}{{\tau}}
\newcommand{\unita}{{\eta}}
\newcommand{\counita}{\Opposite{\unita}}
\newcommand{\producta}{{\nabla}}
\newcommand{\coproducta}{{\Delta}}
\newcommand{\invol}{\mbox{\large$\star$}}
\newcommand{\coinvol}{\Opposite{\invol}}
\newcommand{\antipode}{{s}}

\DeclareMathOperator{\aut}{aut}
\DeclareMathOperator{\tr}{tr}

\DeclareMathOperator{\pdf}{pdf}
\DeclareMathOperator{\cdf}{cdf}

\DeclareMathOperator{\Hom}{Hom}

\newcommand{\homspace}[4][]{{\hom_{#1}^{#2}[#3,#4]}}
\newcommand{\Homspace}[4][]{{\Hom_{#1}^{#2}[#3,#4]}}

\graphicspath{{images}}

\mathtoolsset{showonlyrefs}

\begin{document}

\title{Information and Arbitrage: Applications of Quantum Groups in Mathematical Finance}

\author{Paul McCloud}
\affiliation{Department of Mathematics, University College London, UK}
\orcid{0000-0001-9531-5045}
\email{p.mccloud@ucl.ac.uk}
\maketitle

\begin{abstract}
The relationship between expectation and price is commonly established with two principles: {\em no-arbitrage}, which asserts that both maps are positive; and {\em equivalence}, which asserts that the maps share the same null events. Constructed from the Arrow-Debreu securities, classical and quantum models of economics are then distinguished by their respective use of classical and quantum logic, following the program of von Neumann.

In this essay, the operations and axioms of quantum groups are discovered in the minimal preconditions of stochastic and functional calculus, making this the natural domain for the axiomatic development of mathematical finance. Quantum economics emerges from the twin pillars of the Gelfand-Naimark-Segal construction, implementing the principle of no-arbitrage, and the Radon-Nikodym theorem, implementing the principle of equivalence.

Exploiting quantum group duality, a holographic principle that exchanges the roles of state and observable creates two distinct economic models from the same set of elementary valuations. Advocating on the grounds that this contains and extends classical economics, noncommutativity is presented as a modelling resource, with novel applications in the pricing of options and other derivative securities.
\end{abstract}

\section{Introduction}

The data associated with a system is presented in this essay as a pair of complementary $\ast$-algebras, comprising the states and observables that are combined in the investigation of the system. The state space $\StSpace$ and the observable space $\ObSpace$ are paired by a bilinear map:
\begin{equation}
\bullet:\StSpace\times\ObSpace\rightarrow\mathbb{C}
\end{equation}
that sends the state $\mathsf{z}\in\StSpace$ and the observable $\mathsf{a}\in\ObSpace$ to their valuation $\mathsf{z}\bullet\mathsf{a}\in\mathbb{C}$. With the emphasis on empirical determination, the states and observables are completed as locally convex topological spaces that include all the experiments uniquely identified by their valuations. Using the operations of the $\ast$-algebras, stochastic and functional calculus can then be developed on the category of empirical systems, hereafter referred to as the {\em empirical category}.

Contextualising this to mathematical finance, the state represents a market and the observable represents a security; the pairing is then the price of the security in the market. The two-dimensional nature of this data construct is evident in the following example market for interest rate swaps on three consecutive days:
\begin{center}
\begin{tabular}{r|ccc}
& {\sf mkt1} & {\sf mkt2} & {\sf mkt3} \\ \hline
&&&\\[-2.3ex]
{\sf sw1y} & $5.118\%$ & $5.075\%$ & $5.008\%$ \\
{\sf sw2y} & $4.395\%$ & $4.322\%$ & $4.248\%$ \\
{\sf sw5y} & $3.629\%$ & $3.563\%$ & $3.497\%$ \\
{\sf sw10y} & $3.449\%$ & $3.399\%$ & $3.351\%$ \\
{\sf sw20y} & $3.402\%$ & $3.373\%$ & $3.335\%$
\end{tabular}
\end{center}
The pairing is lookup in this elementary market, as with the sample valuation:
\begin{equation}
{\sf mkt3}\bullet{\sf sw2y}=4.248\%
\end{equation}
where the state is the market {\sf mkt3} on the third day and the observable is the swap rate {\sf sw2y} maturing in two years. The {\em a priori} structure comprises the basis markets and securities that are available to the market agent, and the {\em a posteriori} structure is the data that populates the table. Analysis then identifies patterns within the data, reducing its dimensionality and supporting algorithmic trading and risk management.

\begin{figure*}[!t]
\centering
\setlength{\tabcolsep}{0.0\linewidth}
\begin{tabular}{C{0.166\textwidth}C{0.166\textwidth}C{0.166\textwidth}|C{0.166\textwidth}C{0.166\textwidth}C{0.166\textwidth}}
Stasis & Accumulate & Reverse & Unit & Multiply & Conjugate \\ \hline
&&&&&\\[-1.5ex]
$[1]$ & $[xy]$ & $[x^{-1}]$ & & $[X]\hspace{0.35cm}\otimes\hspace{0.35cm}[Y]$ & $[X]$ \\
\includegraphics[width=0.13\textwidth]{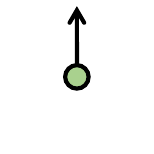} & \includegraphics[width=0.13\textwidth]{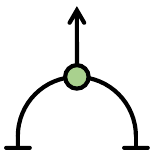} & \includegraphics[width=0.13\textwidth]{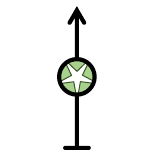} & \includegraphics[width=0.13\textwidth]{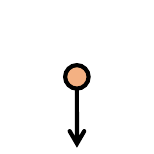} & \includegraphics[width=0.13\textwidth]{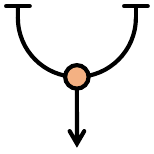} & \includegraphics[width=0.13\textwidth]{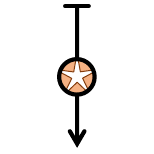} \\[-0.5ex]
& $[x]\hspace{0.4cm}\otimes\hspace{0.4cm}[y]$ & $[x]$ & $1$ & $[X\cap Y]$ & $[X]$
\end{tabular}
\caption{The operations of $\ast$-algebra are defined on the point measures and digital functions of a classical group, linearly and topologically completed as the convolution algebra of measures and the pointwise algebra of functions. The definition of the quantum group abstracts these operations, enabling noncommutativity of both products simultaneously. This extension from classical to quantum economics introduces Heisenberg uncertainty as a novel source of volatility, with applications in the pricing of options.}
\label{fig:classicalgroup}
\end{figure*}

Mathematical finance is here developed within the empirical category, adopting no more than the algebraic operations and relations necessary to assert the elementary principles of pricing. A feature-complete model of quantum economics is generated from the category of quantum groups, defined as dual $\ast$-algebras satisfying the Hopf axioms, with its subcategory of classical groups generating the submodel of classical economics.

Applications of the quantum extension exploit noncommutativity as a source of volatility, creating algorithms for option pricing that are efficiently implemented on classical hardware. The approach also supports a diagrammatic language that overlaps with the string diagrams of the ZX calculus, suggesting that further efficiencies could be extracted by implementing the algorithms on quantum hardware.

\begin{figure*}[!t]
\centering
\setlength{\tabcolsep}{0.0\linewidth}
\begin{tabular}{C{0.5\textwidth}|C{0.5\textwidth}}
Stochastic Integration & Stochastic Differentiation \\ \hline
\includegraphics[width=0.4\textwidth]{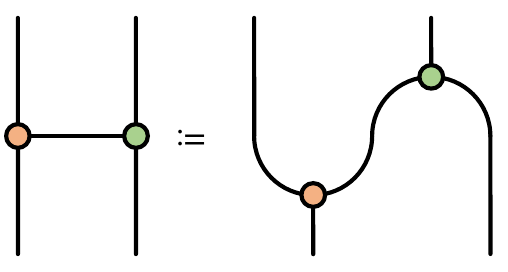} & \includegraphics[width=0.4\textwidth]{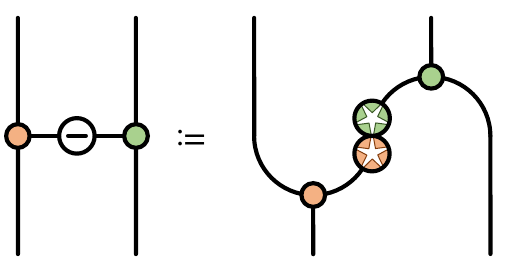} \\
$\mathsf{E}:[y_1]\otimes[y_2]\mapsto[y_1]\otimes[y_1y_2]$ & $\mathsf{D}:[x_1]\otimes[x_2]\mapsto[x_1]\otimes[x_1^{-1}x_2^{\vphantom{-1}}]$ \\[1ex]
\includegraphics[width=0.4\textwidth]{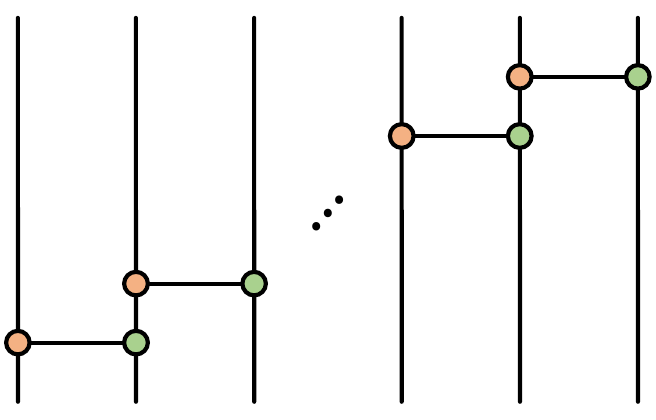} & \includegraphics[width=0.4\textwidth]{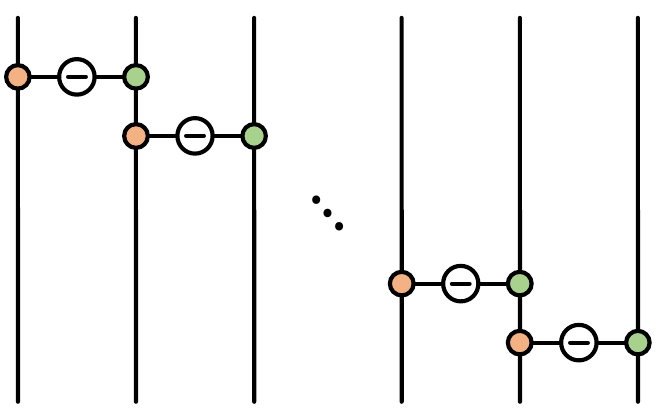} \\
$\mathsf{E}:\bigotimes_{i=1}^n[y_i]\mapsto\bigotimes_{i=1}^n[y_1\ldots y_i]$ & $\mathsf{D}:\bigotimes_{i=1}^n[x_i]\mapsto\bigotimes_{i=1}^n[x_{i-1}^{-1}x_i^{\vphantom{-1}}]$
\end{tabular}
\caption{The products of states and observables are combined to create reversible operations that perform stochastic integration and differentiation on discrete schedules. Implemented on quantum hardware, these are entanglement and disentanglement operations.}
\label{fig:integrationanddifferentiation}
\end{figure*}

Pursuing this perspective in the context of classical economics, the pairing of the Arrow-Debreu model is set membership:
\begin{equation}
[x]\bullet[X]:=(x\in X)
\end{equation}
assigning the numerical values $\top:=1$ and $\bot:=0$ to the expression on the right. Evaluated on the point measure $[x]$ supported on the event $x$, the digital function $[X]$ indicating the subset of events $X$ returns one when the subset contains the event and returns zero otherwise. Linear and topological completion on both sides of the pairing then creates measures and functions that are paired by the integral. Dynamics is incorporated with the additional assumption that events can be accumulated, thereby acting as a group on the economy. Classical economics is thus founded on its group of events.

The operations of the $\ast$-algebra of states are the unit, product and involution of discrete measures, acting upward in the string diagrams on the left of figure \ref{fig:classicalgroup} with the first two operations extended linearly and the last operation extended antilinearly. The unit models stasis, the product models the accumulation of events and the involution models time reversal. With these definitions, topological completion derives the convolution algebra of measures, and the operations together generate stochastic calculus.

The operations of the $\ast$-algebra of observables are the unit, product and involution of step functions, acting downward in the string diagrams on the right of figure \ref{fig:classicalgroup} with the first two operations extended linearly and the last operation extended antilinearly. The unit models the unit constant, the product models multiplication and the involution models complex conjugation. With these definitions, topological completion derives the pointwise algebra of functions, and the operations together generate functional calculus.

Taken separately, neither product is reversible. They can, however, be combined in ways that are reversible, and these combinations relate stochastic integration and differentiation. Stochastic integration maps incremental events to their accumulations, and stochastic differentiation maps accumulated events to their increments:
\begin{align}
\mathsf{E}:\bigotimes_{i=1}^n[y_i]&\mapsto\bigotimes_{i=1}^n[y_1\ldots y_i] \\
\mathsf{D}:\bigotimes_{i=1}^n[x_i]&\mapsto\bigotimes_{i=1}^n[x_{i-1}^{-1}x_i^{\vphantom{-1}}] \notag
\end{align}
where the path is initialised with $x_0=1$. These operations, presented in figure \ref{fig:integrationanddifferentiation} as string diagrams, are reversible as they are mutually inverse:
\begin{center}
\includegraphics[width=0.9\linewidth]{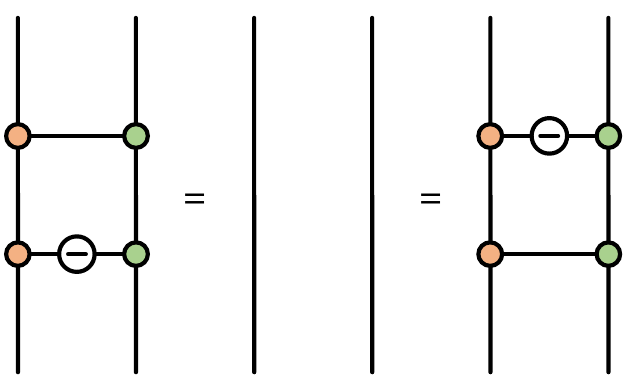}
\end{center}
Notably, both products and both involutions are needed in these definitions, and reversibility depends on all the axioms of the classical group. The model is then extended from discrete to continuous schedules in the refinement limit of the compounded operations.

The properties of classical groups that enable stochastic and functional calculus, and so provide a complete platform for the development of mathematical finance, can be abstracted as algebraic properties of the states and observables. In addition to their properties as $\ast$-algebras, the operations of states and observables must be complementary, together forming a $\ast$-bialgebra. The antipode -- the operation that enforces reversibility of integration and differentiation -- is formed as the composition of the two involutions and must satisfy the Hopf axiom. These are the defining properties of the quantum group.

Among quantum groups, the distinguishing feature of the classical group is the commutativity of its algebra of functions. Thanks to quantum group duality, there are two economic models that can be generated from the classical group: {\em classical economics}, where noncommutativity of measure is used as a resource in the stochastic calculus; and {\em coclassical economics}, where noncommutativity of measure is used as a resource in the functional calculus.

Beyond this, the quantum group utilises noncommutativity for both states and observables within the model of {\em quantum economics}. Noncommutativity of states means the accumulated impact of events depends on the order of their arrival. Noncommutativity of observables introduces a novel source of volatility for option pricing. By presenting the economic principles of pricing in purely algebraic terms, these effects are developed into new methodologies for the pricing of derivative securities.

\begin{figure*}[!t]
\centering
\begin{minipage}{0.38\textwidth}
\includegraphics[width=\linewidth]{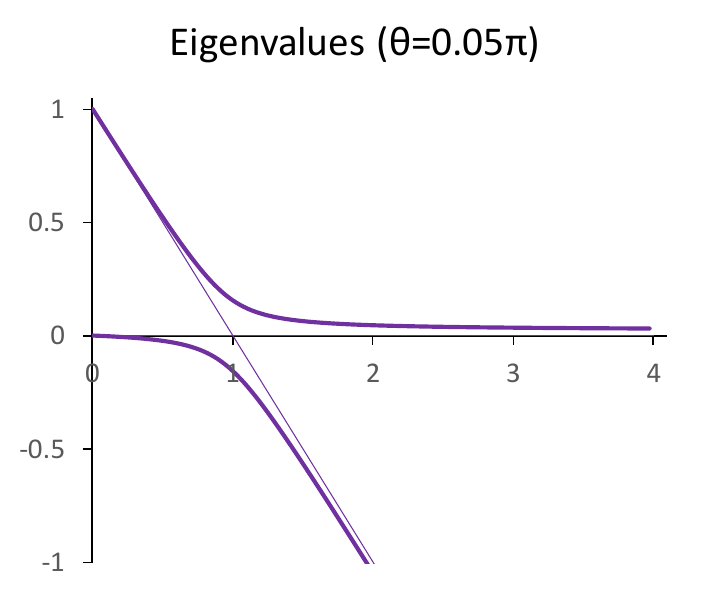}
\end{minipage}\qquad\qquad
\begin{minipage}{0.38\textwidth}
\includegraphics[width=\linewidth]{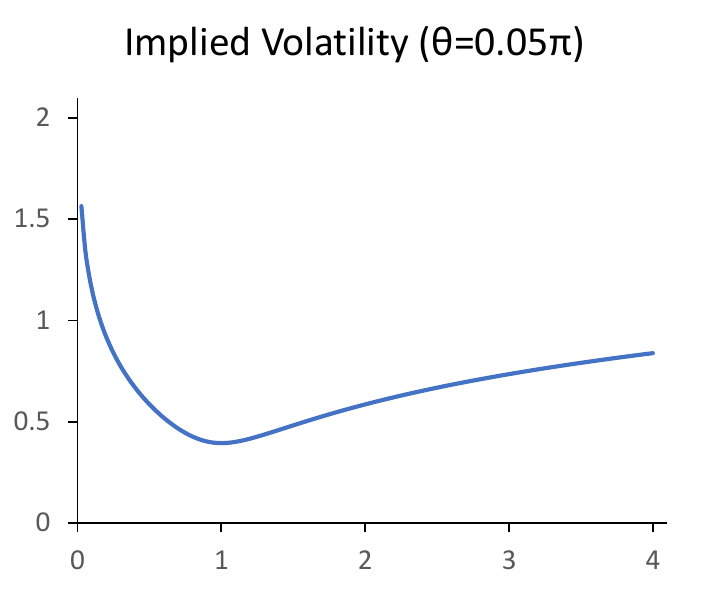}
\end{minipage}
\caption{Quantum relu activation with two internal dimensions matches classical squareplus activation, which is smooth and strictly greater than classical relu activation. Used as a model for options, this additional value generates a fat-tailed volatility smile as a consequence of the Heisenberg uncertainty of noncommuting matrices.}
\label{fig:twostatebinomial}
\end{figure*}

\begin{figure*}[!p]
\centering\includegraphics[width=0.85\linewidth]{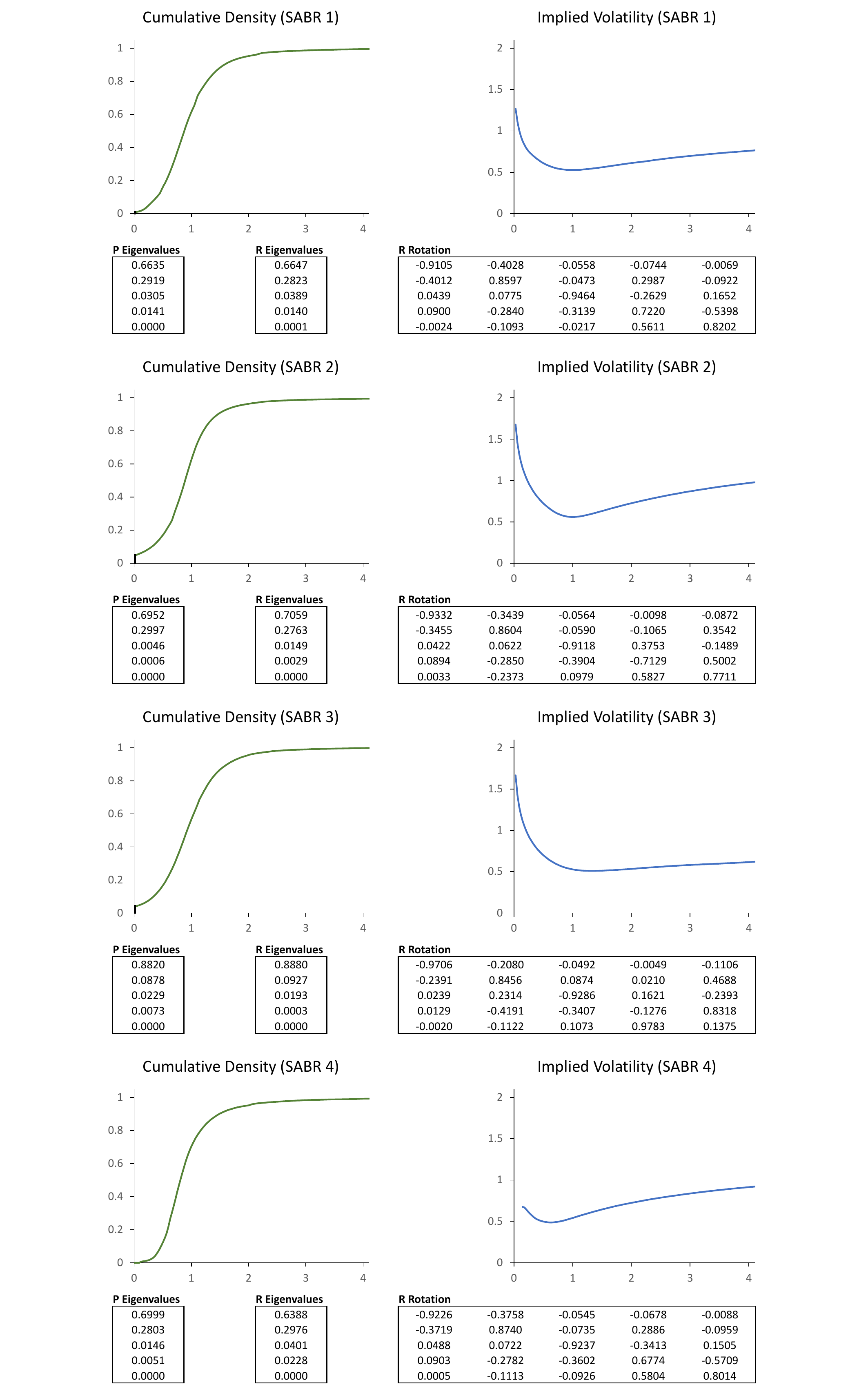}
\caption{The implied cumulative density and (unannualised) volatility smile in the quantum perceptron with five internal dimensions calibrated to four examples of the SABR model with parameters $(t,\sigma,\alpha,\beta,\rho)$ given by $(4,0.25,0.4,1,0)$, $(4,0.25,0.6,1,0)$, $(4,0.25,0.4,0.5,0)$, $(4,0.25,0.4,1,0.5)$ respectively.}
\label{fig:fivestatemultinomial}
\end{figure*}

Embedding noncommutativity in the perceptron model leads to quantum-inspired classical machine learning methods that are well adapted to applications in mathematical finance. Define the quantum perceptron with relu activation:
\begin{equation}
o[a_1,\ldots,a_n]:=\tr\!\left[\left(\sum\nolimits_{i=1}^na_iZ_i\right)^+\right]
\end{equation}
The model parameters $(Z_1,\ldots,Z_n)$ are symmetric matrices with matching dimensions, and the activation function by definition returns the sum of the positive eigenvalues of the matrix sum. When they commute with each other, the parameter matrices are simultaneously diagonalisable and quantum relu activation reduces to classical relu activation. More generally, the quantum perceptron accesses the nonlinear relationship between the inputs and the eigenvalues of the matrix sum they generate. This creates a larger and more interesting range of activation functions.

As an elementary example, consider the quantum perceptron:
\begin{equation}
o[k]:=\tr[(R-kP)^+]
\end{equation}
The simplest nontrivial case has two internal dimensions, with parameters standardised as:
\begin{equation}
P=
\begin{bmatrix}
1 & 0 \\ 
0 & 0
\end{bmatrix}
\qquad
R=U[\theta]
\begin{bmatrix}
1 & 0 \\ 
0 & 0
\end{bmatrix}
U[\theta]^\ast
\end{equation}
where the second matrix is rotated by angle $\theta$:
\begin{equation}
U[\theta]=
\begin{bmatrix}
\cos[\theta] & -\sin[\theta] \\ 
\sin[\theta] & \cos[\theta]
\end{bmatrix}
\end{equation}
This activation function is resolved as the positive eigenvalue:
\begin{equation}
o[k]=\frac{1}{2}(1-k)+\frac{1}{2}\sqrt{1-2k\cos[2\theta]+k^{2}}
\end{equation}
The eigenvectors of the matrices $P$ and $R$ are misaligned by the angle $\theta$ between $0$ and $\pi/2$. Rotation between the eigenvectors lifts and smooths relu activation to squareplus activation, demonstrated in figure \ref{fig:twostatebinomial} for the case $\theta=\pi/20$.

The quantum neural network formed from quantum perceptrons suffers in comparison to the classical neural network with equivalent architecture thanks to the additional compute required to resolve the eigenvalues. There are efficient classical algorithms for computing eigenvalues, however, and the richness of the activation functions created by the quantum perceptron means that fewer internal dimensions are needed for fitting.

As demonstrated in figure \ref{fig:twostatebinomial} and the calibrations to the SABR model in figure \ref{fig:fivestatemultinomial}, even in low internal dimensions a single quantum perceptron achieves a remarkable fit to real market volatility smiles. This is not a coincidence. In this essay, the quantum perceptron is derived as the most general arbitrage-free option price model on a finite set of states.

The complexities in the following exposition are a consequence of the ambition to encompass the entirety of classical pricing as a submodel. The worked examples are nonetheless drawn from the discrete model of the quantum perceptron whose technical overhead is much lighter, and the quantum neural network is here proposed as an effective tool for option pricing.

\section{Literature review}

This essay grew out of a series of presentations by the author at the Global Derivatives Trading and Risk Management Conference, and is an attempt to formalise and develop the ideas presented there. In the 2012 presentation ``Symmetry methods for the quadratic Gaussian Libor model'' \cite{McCloud2012}\cite{McCloud2012a}, the core operations of conditional expectation and measure change are identified as algebraic transformations, with the duality of these operations taking a central role. In the 2014 presentation ``In search of Schr\"{o}dinger’s cap: Pricing derivatives with quantum probability'' \cite{McCloud2014}\cite{McCloud2014a}\cite{McCloud2018}, noncommutativity is proposed as a resource for realistic and numerically efficient models of option pricing, using Heisenberg uncertainty as a source of volatility. The 2017 presentation ``From Quadratic Gaussian to Quantum Groups'' \cite{McCloud2017} connects these two strands, observing that both can be framed as quantum groups, and is the genesis for the current essay.

While recognisable option price models can be found as early as the thesis of Bachelier \cite{Bachelier1900}, modern derivative pricing theory is generally held to have started with the pioneering work of Black and Scholes \cite{Black1973} and Merton \cite{Merton1973} who complemented the stochastic approach to pricing with consideration of the risk-neutralisation benefits of dynamic hedging. This idea matured in the work of Harrison and Pliska \cite{Harrison1981}\cite{Harrison1983} and their contemporaries, and nowadays it is commonplace to assume the measure-theoretic equivalence of price with expectation as axiomatic. An entire ecosystem of volatility models has since been developed, including the popular SABR model \cite{Hagan2002} used here as the comparator for the quantum option price.

In many respects, this is the fusion of stochastic calculus with earlier work on portfolio optimisation by Markowitz \cite{Markowitz1952} and the classical measure implicit in the state prices of Arrow and Debreu \cite{Arrow1954}. Dynamic hedging does not eliminate market risk, but can mask it when combined with a stochastic model that limits the scope of future outcomes. Concerns on the limitations of classical probability are suggested in the oft-quoted observation of Keynes \cite{Keynes1937} that {\em ``\ldots the necessity for action and for decision compels us\ldots to behave exactly as we should if we had\ldots a series of prospective advantages and disadvantages, each multiplied by its appropriate probability, waiting to be summed.''} Shackle \cite{Shackle1969} is more explicit, arguing that classical probability stifles creativity and innovation in decision making, and numerous market calamities have since highlighted the negative consequences of groupthink.

At the same time that Keynes was gently casting shade on the role of classical probability in economic decision making, the pioneers of quantum mechanics were laying the foundations for a radically new approach to information and uncertainty. The early development of matrix models by Born, Jordan and Heisenberg \cite{Born1925}\cite{Born1926} evolved into the seminal works by von Neumann on the mathematical foundations of quantum mechanics \cite{Neumann1932} and the algebras that carry his name \cite{Neumann1949}, reviewed in the standard texts \cite{Conway2007}\cite{Kadison1997}\cite{Kadison1997a} along with more recent developments. The key results used here are the Gelfand-Naimark-Segal construction \cite{Gelfand1943}\cite{Segal1947} and the Radon-Nikodym theorem \cite{Nikodym1930}; this essay is heavily influenced by the approach of Gudder \cite{Gudder1979} in the generalisation of the results to noncommutative $\ast$-algebras.

Noncommutativity is the novel feature of quantum information and entanglement is the source of advantage in quantum algorithms, with famous examples including the Shor factorisation and Grover search algorithms described in the standard text \cite{Nielsen2010}. Modern interpretations position the theory in the context of monoidal categories, developed in the essential text by Mac Lane \cite{MacLane1971} with the diagrammatic language of Coecke and Kissinger \cite{Coecke2017} and Heunen and Vicary \cite{Heunen2019} demonstrating the elegance and power of the approach. In string diagrams, categorical proof is synonymous with algorithmic compilation, which is effectively deployed in the ZX calculus to improve the efficiency of quantum computation.

In this essay, the string diagrams of the ZX calculus are extended to quantum groups and applied in the stochastic and functional calculus. Quantum groups have been widely used in mathematics and physics, see \cite{Abe2004}\cite{Kassel1995}\cite{Timmermann2008} for introductions. Extending the self-duality observed in the quadratic Gaussian model, duality makes a vital contribution to the development here as it connects apparently unrelated valuation models through their common pairing, in the same way that holography connects distinct physical models through their shared partition function \cite{Ammon2015}\cite{Rangamani2017}. As with holography, this serves as a means to develop efficient numerical methods and inspires the invention of new models for pricing.

\section{The empirical category}

In proposing the category used throughout this essay as the locale for the model of information, the foremost consideration is its empirical interpretation. The system is represented as a pair of spaces, the states and observables, combined in experiments whose measurements are given by the pairing. Within this setup, an operation is equivalently defined by a semilinear map between the state spaces or the observable spaces of its domain and codomain systems, imposing consistency of these Schr\"{o}dinger and Heisenberg perspectives through the duality of their associated maps. Adhering to the empirical principle, two operations are then identified if they are indistinguishable by experiments. The empirical category is thus defined to be the set of all such bilinear pairings, with morphisms expressed as adjointable semilinear maps.

Created from the field $\mathbb{F}$ that provides its space of measurements, the automorphism group $\aut[\mathbb{F}]$ acts as signature group for the operations of the empirical category, in recognition of the internal symmetries in the measurement system presented by the automorphisms. Semilinearity as a concept is then defined for maps between spaces, and the operation is associated with semilinear maps that are compatible with the algebra and topology of its measurement field.

\subsection{Empirical algebra}

In the empirical perspective, the system $K$ is defined by its vector spaces of states $\StSpace[K]$ and observables $\ObSpace[K]$, with bilinear pairing:
\begin{equation}
\bullet:\StSpace[K]\times\ObSpace[K]\rightarrow\mathbb{F}
\end{equation}
that combines the state $\mathsf{z}\in\StSpace[K]$ and the observable $\mathsf{a}\in\ObSpace[K]$ to generate their joint valuation $\mathsf{z}\bullet\mathsf{a}\in\mathbb{F}$.

The operation $e\in\homspace{\alpha}{K}{L}$ with domain system $K$ and codomain system $L$ is defined by its dual semilinear actions on states and observables:
\begin{align}
\StSpace[e]&:\mathsf{z}\in\StSpace[K]\mapsto\mathsf{z}\circ e\in\StSpace[L] \\
\ObSpace[e]&:\mathsf{a}\in\ObSpace[L]\mapsto e\circ\mathsf{a}\in\ObSpace[K] \notag
\end{align}
The operation has signature $\alpha$ taken from the automorphism group $\aut[\mathbb{F}]$ of the measurement field; its state and observable maps are then $\alpha$-linear and $\alpha^{-1}$-linear respectively, so that they commute with scalar multiplication in the form:
\begin{align}
(\lambda\mathsf{z})\circ e&=\lambda^\alpha(\mathsf{z}\circ e) \\
e\circ(\lambda\mathsf{a})&=\lambda^{\alpha^{-1}}(e\circ\mathsf{a}) \notag
\end{align}
where the action of the signature is denoted by the superscript. The operation is empirically distinguished by its contractions with states and observables, captured in its valuation maps.
\begin{description}
\item[Schr\"{o}dinger picture:]The operation $e$ acts on the state $\mathsf{z}$ and is paired with the observable $\mathsf{a}$ in the valuation $\mathsf{z}\circ e\bullet\mathsf{a}$.
\item[Heisenberg picture:]The operation $e$ acts on the observable $\mathsf{a}$ and is paired with the state $\mathsf{z}$ in the valuation $\mathsf{z}\bullet e\circ\mathsf{a}$.
\end{description}
Motivated by these dual approaches to valuation, the final condition in the definition of the operation imposes consistency of the Schr\"{o}dinger and Heisenberg pictures:
\begin{equation}
\mathsf{z}\circ e\bullet\mathsf{a}=(\mathsf{z}\bullet e\circ\mathsf{a})^\alpha
\end{equation}
asserting the mutual adjointness of the state and observable maps.

Systems are the objects and operations are the arrows of the {\em empirical category}, which is equipped as a symmetric monoidal category. The empty system $1$ is defined by its states and observables:
\begin{align}
\StSpace[1]&:=\mathbb{C} \\
\ObSpace[1]&:=\mathbb{C} \notag
\end{align}
with pairing:
\begin{equation}
\lambda\bullet\mu:=\lambda\mu
\end{equation}
The concatenated system $K\otimes L$ that combines the systems $K$ and $L$ is defined by its states and observables:
\begin{align}
\StSpace[K\otimes L]&:=\StSpace[K]\otimes\StSpace[L] \\
\ObSpace[K\otimes L]&:=\ObSpace[K]\otimes\ObSpace[L] \notag
\end{align}
with pairing:
\begin{equation}
(\mathsf{z}\otimes\mathsf{y})\bullet(\mathsf{a}\otimes\mathsf{b}):=(\mathsf{z}\bullet\mathsf{a})(\mathsf{y}\bullet\mathsf{b})
\end{equation}
Completing the definitions, the identity and braid operations:
\begin{align}
\identity&\in\homspace{1}{K}{K} \\
\braid&\in\homspace{1}{K\otimes L}{L\otimes K} \notag
\end{align}
and the concatenation and composition of compatible operations:
\begin{align}
\otimes:\homspace{\alpha}{K}{L}\times{}&\homspace{\alpha}{M}{N}\to \\
&\homspace{\alpha}{K\otimes M}{L\otimes N} \notag \\
\circ:\homspace{\alpha}{K}{L}\times{}&\homspace{\beta}{L}{M}\to\homspace{\alpha\beta}{K}{M} \notag
\end{align}
are defined on states and observables:
\begin{alignat}{3}
&\text{\textbf{State:}} & &\qquad & &\text{\textbf{Observable:}} \\
&\mathsf{z}\circ\identity:=\mathsf{z} & &\qquad & &\identity\circ\mathsf{a}:=\mathsf{a} \notag \\
&(\mathsf{z}\otimes\mathsf{y})\circ\braid:=\mathsf{y}\otimes\mathsf{z} & &\qquad & &\braid\circ(\mathsf{a}\otimes\mathsf{b}):=\mathsf{b}\otimes\mathsf{a} \notag \\
&(\mathsf{z}\otimes\mathsf{y})\circ(e\otimes f):= & &\qquad & &(e\otimes f)\circ(\mathsf{a}\otimes\mathsf{b}):= \notag \\
&\qquad (\mathsf{z}\circ e)\otimes(\mathsf{y}\circ f) & &\qquad & &\qquad (e\circ\mathsf{a})\otimes(f\circ\mathsf{b}) \notag \\
&\mathsf{z}\circ(e\circ f):= & &\qquad & &(e\circ f)\circ\mathsf{a}:= \notag \\
&\qquad(\mathsf{z}\circ e)\circ f & &\qquad & &\qquad e\circ(f\circ\mathsf{a}) \notag
\end{alignat}
Note the compatibility conditions in these definitions: two operations are compatible for concatenation if and only if they have the same signature, and they are compatible for composition if and only if the codomain of the first operation matches the domain of the second operation.

The hom-space $\homspace{\alpha}{K}{L}$ of operations with signature $\alpha$ between the systems $K$ and $L$ has the structure of a vector space with definitions:
\begin{alignat}{2}
\mathsf{z}\circ(\lambda e)&:=\lambda^\alpha(\mathsf{z}\circ e)\quad & (\lambda e)\circ\mathsf{a}&:=\lambda(e\circ\mathsf{a}) \\
\mathsf{z}\circ(e\lambda)&:=\lambda(\mathsf{z}\circ e)\quad & (e\lambda)\circ\mathsf{a}&:=\lambda^{\alpha^{-1}}(e\circ\mathsf{a}) \notag
\end{alignat}
for scalar multiplication on both sides of the operation. The scalar transforms by the signature as it passes through the operation:
\begin{align}
\lambda^{\alpha^{-1}}(e\circ f)&=(\lambda^{\alpha^{-1}}e)\circ f=(e\lambda)\circ f= \\
&e\circ(\lambda f)=e\circ(f\lambda^\beta)=(e\circ f)\lambda^\beta \notag
\end{align}
for compatible operations $e$ and $f$ with signatures $\alpha$ and $\beta$ respectively.

States and observables can be located in the hom-spaces. Identify the signature $\alpha$ with the operation in $\homspace{\alpha}{1}{1}$:
\begin{equation}
\lambda\circ\alpha:=\lambda^\alpha\qquad\alpha\circ\lambda:=\lambda^{\alpha^{-1}}
\end{equation}
Composition with the signature then bijectively maps between terminal hom-spaces:
\begin{align}
\homspace{\alpha}{1}{K}&=\alpha\circ\homspace{1}{1}{K} \\
\homspace{\alpha}{K}{1}&=\homspace{1}{K}{1}\circ\alpha \notag
\end{align}
By definition, the operation $\mathsf{z}\in\homspace{1}{1}{K}$ is associated with dual maps:
\begin{align}
\lambda\in\mathbb{C}&\mapsto\lambda\circ\mathsf{z}\in\StSpace[K] \\
\mathsf{b}\in\ObSpace[K]&\mapsto\mathsf{z}\circ\mathsf{b}\in\mathbb{C} \notag
\end{align}
satisfying the duality condition:
\begin{equation}
(1\circ\mathsf{z})\bullet\mathsf{b}=\mathsf{z}\circ\mathsf{b}
\end{equation}
and the operation $\mathsf{a}\in\homspace{1}{K}{1}$ is associated with dual maps:
\begin{align}
\mathsf{y}\in\StSpace[K]&\mapsto\mathsf{y}\circ\mathsf{a}\in\mathbb{C} \\
\lambda\in\mathbb{C}&\mapsto\mathsf{a}\circ\lambda\in\ObSpace[K] \notag
\end{align}
satisfying the duality condition:
\begin{equation}
\mathsf{y}\circ\mathsf{a}=\mathsf{y}\bullet(\mathsf{a}\circ1)
\end{equation}
The associations $\mathsf{z}\leftrightarrow 1\circ\mathsf{z}$ and $\mathsf{a}\leftrightarrow\mathsf{a}\circ 1$ identify these operations in the state and observable spaces:
\begin{align}
\homspace{1}{1}{1}&=\mathbb{C} \\
\homspace{1}{1}{K}&=\StSpace[K] \notag \\
\homspace{1}{K}{1}&=\ObSpace[K] \notag
\end{align}
aligning with the convention for monoidal categories that the terminal hom-spaces $\homspace{1}{1}{K}$ and $\homspace{1}{K}{1}$ are adopted respectively as the state and observable spaces of the system $K$, with the hom-space $\homspace{1}{1}{1}$ as the space of scalars.

Duality is rigorously enforced in the definition of the empirical category. For the system $K$ there is a dual system $\Opposite{K}$ with states and observables:
\begin{align}
\StSpace[\Opposite{K}]&:=\ObSpace[K] \\
\ObSpace[\Opposite{K}]&:=\StSpace[K] \notag
\end{align}
and pairing defined by reversing the order, and for the operation $e\in\homspace{\alpha}{K}{L}$ there is a dual operation $\Opposite{e}\in\homspace{\alpha^{-1}}{\Opposite{L}}{\Opposite{K}}$ with actions:
\begin{align}
\mathsf{a}\circ\Opposite{e}&:=e\circ\mathsf{a} \\
\Opposite{e}\circ\mathsf{z}&:=\mathsf{z}\circ e \notag
\end{align}
Duality thus supports a holographic principle that generates two empirically distinct models from the same pairing. This discipline is continued later in the definition of the quantum group, which extends the definition of the classical group by requiring duality in all the properties of its operations.

\subsection{Empirical topology}

From hereon take the measurement field to be the complex field $\mathbb{C}$, and restrict the signature group to the group of continuous automorphisms $\{1,\ast\}$ comprising the identity and complex conjugation. With these restrictions, topology is introduced in the empirical category.

A distinction is made between the theoretical system $K$ assumed {\em a priori} as the bounds of investigability and the empirical system $\bar{K}$ discovered {\em a posteriori} from its valuations. Topology plays a key role in this transition, completing the system with convergent sequences of theoretical states and observables. Sequences that are empirically indistinguishable are then identified by quotienting out the null sequences whose valuations are zero.

In the first step, the state and observable spaces are expanded to include sequences. Introduce the convergence relation $\mathcal{C}$ and nullification relation $\mathcal{N}$ defined in figure \ref{fig:sequenceconvergence}. Locally convex topology is induced via seminorms:
\begin{equation}
p_\mathsf{a}[\mathsf{z}]:=|\mathsf{z}\bullet\mathsf{a}|=:p_\mathsf{z}[\mathsf{a}]
\end{equation}
whenever the state $\mathsf{z}$ and the observable $\mathsf{a}$ satisfy the convergence relation $\mathsf{z}\mathbin{\mathcal{C}}\mathsf{a}$. The families of seminorms generated by the subsets $\mathsf{A}\subset\ObSpace[K]^\mathbb{N}$ and $\mathsf{Z}\subset\StSpace[K]^\mathbb{N}$ induce topologies on the convergents $\mathcal{C}\mathsf{A}$ and $\mathsf{Z}\mathcal{C}$ respectively. The nature and domain of these topologies thus depend on the choice of seminorms.

A simple choice for the topology uses seminorms generated from the subsets $\ObSpace[K]$ and $\StSpace[K]$ of constant sequences. The pairing does not necessarily extend to the convergents $\mathcal{C}\ObSpace[K]$ and $\StSpace[K]\mathcal{C}$, however. This limitation is avoided with the topology induced on the smaller convergents $\mathcal{C}\StSpace[K]\mathcal{C}$ and $\mathcal{C}\ObSpace[K]\mathcal{C}$ by seminorms generated from the subsets $\StSpace[K]\mathcal{C}$ and $\mathcal{C}\ObSpace[K]$. The topology is more refined with these seminorms:
\begin{align}
\ObSpace[K]&\subset\StSpace[K]\mathcal{C} \\
\StSpace[K]&\subset\mathcal{C}\ObSpace[K] \notag
\end{align}
Following from the general properties of relations:
\begin{align}
\mathcal{C}\StSpace[K]\mathcal{C}&\subset\mathcal{C}\mathcal{C}\ObSpace[K]\mathcal{C} \\
\mathcal{C}\ObSpace[K]\mathcal{C}&\subset\mathcal{C}\StSpace[K]\mathcal{C}\mathcal{C} \notag
\end{align}
and:
\begin{align}
\mathcal{N}\StSpace[K]\mathcal{C}&\subset\mathcal{N}\mathcal{C}\ObSpace[K]\mathcal{C} \\
\mathcal{C}\ObSpace[K]\mathcal{N}&\subset\mathcal{C}\StSpace[K]\mathcal{C}\mathcal{N} \notag
\end{align}
so that states in $\mathcal{C}\StSpace[K]\mathcal{C}$ can be paired with observables in $\mathcal{C}\ObSpace[K]\mathcal{C}$, and the pairing is zero if either the state is in $\mathcal{N}\StSpace[K]\mathcal{C}$ or the observable is in $\mathcal{C}\ObSpace[K]\mathcal{N}$.

\begin{figure*}[!p]
\centering
\setlength{\tabcolsep}{0.025\linewidth}
\begin{tabular}{C{0.95\textwidth}}
Theoretical and Empirical Systems \\[0.5ex] \hline
\\[-1ex]
{\begin{minipage}[t]{0.95\textwidth}
Consider a relation $\mathcal{P}\subset\StSpace\times\ObSpace$ between two sets $\StSpace$ and $\ObSpace$, and use the notation $\mathsf{z}\mathbin{\mathcal{P}}\mathsf{a}$ to express the property $(\mathsf{z},\mathsf{a})\in\mathcal{P}$. The relation generates a concept of completion for subsets.
\end{minipage}}
\end{tabular}
\\[1ex]
\begin{tabular}{C{0.45\textwidth}|C{0.45\textwidth}}
{\begin{minipage}[t]{0.45\textwidth}
The relator $\mathsf{Z}\mathcal{P}\subset\ObSpace$ of the subset $\mathsf{Z}\subset\StSpace$ is defined by:
\begin{equation}
\mathsf{a}\in\mathsf{Z}\mathcal{P}\iff\forall\mathsf{z}\in\mathsf{Z}:\mathsf{z}\mathbin{\mathcal{P}}\mathsf{a}
\end{equation}
Relations satisfy the following properties.
\begin{description}
\item[Subset properties:]
\begin{gather}
\mathcal{P}\subset\mathcal{Q}\implies\mathsf{Z}\mathcal{P}\subset\mathsf{Z}\mathcal{Q} \\
\mathsf{Z}\subset\mathsf{Y}\implies\mathsf{Y}\mathcal{P}\subset\mathsf{Z}\mathcal{P}\iff\mathsf{Z}\subset\mathcal{P}\mathsf{Y}\mathcal{P} \notag
\end{gather}
\item[Completion properties:]
\begin{align}
\mathsf{Z}&\subset\mathcal{P}\mathsf{Z}\mathcal{P} \\
\mathsf{Z}\mathcal{P}&=\mathcal{P}\mathsf{Z}\mathcal{PP} \notag
\end{align}
\end{description}
\end{minipage}}
&
{\begin{minipage}[t]{0.45\textwidth}
The relator $\mathcal{P}\mathsf{A}\subset\StSpace$ of the subset $\mathsf{A}\subset\ObSpace$ is defined by:
\begin{equation}
\mathsf{z}\in\mathcal{P}\mathsf{A}\iff\forall\mathsf{a}\in\mathsf{A}:\mathsf{z}\mathbin{\mathcal{P}}\mathsf{a}
\end{equation}
Relations satisfy the following properties.
\begin{description}
\item[Subset properties:]
\begin{gather}
\mathcal{P}\subset\mathcal{Q}\implies\mathcal{P}\mathsf{A}\subset\mathcal{Q}\mathsf{A} \\
\mathsf{A}\subset\mathsf{B}\implies\mathcal{P}\mathsf{B}\subset\mathcal{P}\mathsf{A}\iff\mathsf{A}\subset\mathcal{P}\mathsf{B}\mathcal{P} \notag
\end{gather}
\item[Completion properties:]
\begin{align}
\mathsf{A}&\subset\mathcal{P}\mathsf{A}\mathcal{P} \\
\mathcal{P}\mathsf{A}&=\mathcal{PP}\mathsf{A}\mathcal{P} \notag
\end{align}
\end{description}
\end{minipage}}
\end{tabular}
\\[1ex]
\begin{tabular}{C{0.95\textwidth}}
{\begin{minipage}[t]{0.95\textwidth}
Specialise to the spaces of sequences of states and observables for a theoretical system $K$. The pairing of the state $\mathsf{z}\in\StSpace[K]^\mathbb{N}$ and the observable $\mathsf{a}\in\ObSpace[K]^\mathbb{N}$ is:
\begin{equation}
\mathsf{z}\bullet\mathsf{a}:=\lim_{m\to\infty}\lim_{n\to\infty}\mathsf{z}_m\bullet\mathsf{a}_n=\lim_{m,n\to\infty}\mathsf{z}_m\bullet\mathsf{a}_n=\lim_{n\to\infty}\lim_{m\to\infty}\mathsf{z}_m\bullet\mathsf{a}_n
\end{equation}
defined whenever the double sequence of valuations regularly converges, meaning that the limits on the right exist and are equal. The sequences satisfy the convergence relation $\mathsf{z}\mathbin{\mathcal{C}}\mathsf{a}$ when their pairing exists, and satisfy the nullification relation $\mathsf{z}\mathbin{\mathcal{N}}\mathsf{a}$ when their pairing exists and is zero. These relations between sequences satisfy $\mathcal{N}\subset\mathcal{C}$ and the following additional properties.
\end{minipage}}
\end{tabular}
\\[1ex]
\begin{tabular}{C{0.45\textwidth}|C{0.45\textwidth}}
{\begin{minipage}[t]{0.45\textwidth}
\begin{description}
\item[Linear property:]The convergent $\mathsf{Z}\mathcal{C}$ and nullifier $\mathsf{Z}\mathcal{N}$ are subspaces of $\ObSpace[K]^\mathbb{N}$.
\item[Constant property:]
\begin{equation}
\ObSpace[K]\subset\StSpace[K]\mathcal{C}
\end{equation}
\end{description}
The state space $\StSpace[\bar{K}]$ of the empirical system $\bar{K}$ is defined by the quotient:
\begin{equation}
\StSpace[\bar{K}]:=\mathcal{C}\StSpace[K]\mathcal{C}/\mathcal{N}\StSpace[K]\mathcal{C}
\end{equation}
expanding the space by topological completion and reducing it by factoring out the empirically indistinguishable states.
\end{minipage}}
&
{\begin{minipage}[t]{0.45\textwidth}
\begin{description}
\item[Linear property:]The convergent $\mathcal{C}\mathsf{A}$ and nullifier $\mathcal{N}\mathsf{A}$ are subspaces of $\StSpace[K]^\mathbb{N}$.
\item[Constant property:]
\begin{equation}
\StSpace[K]\subset\mathcal{C}\ObSpace[K]
\end{equation}
\end{description}
The observable space $\ObSpace[\bar{K}]$ of the empirical system $\bar{K}$ is defined by the quotient:
\begin{equation}
\ObSpace[\bar{K}]:=\mathcal{C}\ObSpace[K]\mathcal{C}/\mathcal{C}\ObSpace[K]\mathcal{N}
\end{equation}
expanding the space by topological completion and reducing it by factoring out the empirically indistinguishable observables.
\end{minipage}}
\end{tabular}
\\[1ex]
\begin{tabular}{C{0.95\textwidth}}
{\begin{minipage}[t]{0.95\textwidth}
The empirical system $\bar{K}$ is the topological completion of the theoretical system $K$ with pairing:
\begin{equation}
(\mathsf{z}+\mathcal{N}\StSpace[K]\mathcal{C})\bullet(\mathsf{a}+\mathcal{C}\ObSpace[K]\mathcal{N}):=\mathsf{z}\bullet\mathsf{a}
\end{equation}
\end{minipage}}
\end{tabular}
\caption{The empirical system is created from the theoretical system by expanding the state and observable spaces via topological completion and reducing them by eliminating indistinguishable valuations, leveraging the locally convex topology defined by the pairing.}
\label{fig:sequenceconvergence}
\end{figure*}

The second step proceeds by factoring out empirical equivalences. Define the empirical system $\bar{K}$ with state and observable spaces:
\begin{align}
\StSpace[\bar{K}]&:=\mathcal{C}\StSpace[K]\mathcal{C}/\mathcal{N}\StSpace[K]\mathcal{C} \\
\ObSpace[\bar{K}]&:=\mathcal{C}\ObSpace[K]\mathcal{C}/\mathcal{C}\ObSpace[K]\mathcal{N} \notag
\end{align}
The pairing:
\begin{equation}
(\mathsf{z}+\mathcal{N}\StSpace[K]\mathcal{C})\bullet(\mathsf{a}+\mathcal{C}\ObSpace[K]\mathcal{N}):=\mathsf{z}\bullet\mathsf{a}
\end{equation}
is well defined thanks to the preceding observations. This empirical system encapsulates all the experiments that can be constructed from the theoretical system it completes, and eliminates states and observables whose valuations are guaranteed to be zero.

\begin{figure*}[!t]
\centering
\setlength{\tabcolsep}{0.0\linewidth}
\begin{tabular}{C{0.23\textwidth}C{0.23\textwidth}C{0.23\textwidth}|C{0.31\textwidth}}
Unit & Product & Involution & Antipode \\ \hline
\begin{tabular}{c}\includegraphics[width=0.13\textwidth]{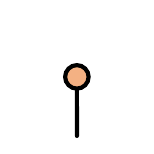}\\$\unita\in\Homspace{1}{K}{1}$\\[1ex]$\counita\in\Homspace{1}{1}{K}$\\\includegraphics[width=0.13\textwidth]{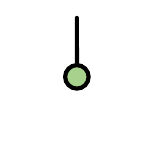}\end{tabular} &
\begin{tabular}{c}\includegraphics[width=0.13\textwidth]{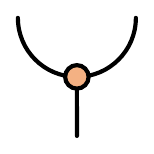}\\$\producta\in\Homspace{1}{K}{K\otimes K}$\\[1ex]$\coproducta\in\Homspace{1}{K\otimes K}{K}$\\\includegraphics[width=0.13\textwidth]{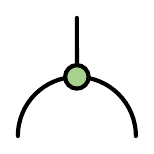}\end{tabular} &
\begin{tabular}{c}\includegraphics[width=0.13\textwidth]{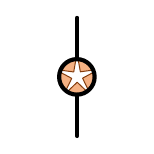}\\$\invol\in\Homspace{\ast}{K}{K}$\\[1ex]$\coinvol\in\Homspace{\ast}{K}{K}$\\\includegraphics[width=0.13\textwidth]{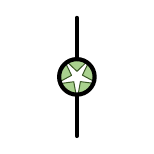}\end{tabular} &
\begin{tabular}{c}\includegraphics[width=0.15\textwidth]{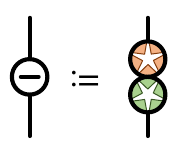}\\$\antipode:=\coinvol\circ\invol\in\Homspace{1}{K}{K}$\end{tabular}
\end{tabular}
\caption{The states and observables of the quantum group separately support unit, product and involution operations. Composing the two involutions generates the antipode of the quantum group.}
\label{fig:quantumgroupoperations}
\end{figure*}

Extending the empirical category to sequences, the operation $e\in\homspace{\alpha}{K}{L}^\mathbb{N}$ acts elementwise on states and observables:
\begin{align}
(\mathsf{z}\circ e)_n&:=\mathsf{z}_n\circ e_n \\
(e\circ\mathsf{a})_n&:=e_n\circ\mathsf{a}_n \notag
\end{align}
for $\mathsf{z}\in\StSpace[K]^\mathbb{N}$ and $\mathsf{a}\in\ObSpace[L]^\mathbb{N}$. By definition, the operation is {\em continuous}, $e\in\homspace{\alpha}{K}{L}_\mathcal{C}^\mathbb{N}$, if it preserves convergent sequences:
\begin{align}
\mathcal{C}\StSpace[K]\mathcal{C}\circ e&\subset\mathcal{C}\StSpace[L]\mathcal{C} \\
e\circ\mathcal{C}\ObSpace[L]\mathcal{C}&\subset\mathcal{C}\ObSpace[K]\mathcal{C} \notag
\end{align}
and it preserves null sequences:
\begin{align}
\mathcal{N}\StSpace[K]\mathcal{C}\circ e&\subset\mathcal{N}\StSpace[L]\mathcal{C} \\
e\circ\mathcal{C}\ObSpace[L]\mathcal{N}&\subset\mathcal{C}\ObSpace[K]\mathcal{N} \notag
\end{align}
and the operation is {\em null}, $e\in\homspace{\alpha}{K}{L}_\mathcal{N}^\mathbb{N}$, if it annihilates convergent sequences:
\begin{align}
\mathcal{C}\StSpace[K]\mathcal{C}\circ e&\subset\mathcal{N}\StSpace[L]\mathcal{C} \\
e\circ\mathcal{C}\ObSpace[L]\mathcal{C}&\subset\mathcal{C}\ObSpace[K]\mathcal{N} \notag
\end{align}
The continuous operation maps to an operation between the empirical systems $\bar{K}$ and $\bar{L}$ with:
\begin{align}
(\mathsf{z}+\mathcal{N}\StSpace[K]\mathcal{C})\circ e&:=\mathsf{z}\circ e+\mathcal{N}\StSpace[L]\mathcal{C}& \\
e\circ(\mathsf{a}+\mathcal{C}\ObSpace[L]\mathcal{N})&:=e\circ\mathsf{a}+\mathcal{C}\ObSpace[K]\mathcal{N}& \notag
\end{align}
and the continuous operation maps to the zero operation if and only if it is null. Taking the quotient of the space of continuous operations by the space of null operations identifies a hom-subspace:
\begin{align}
&\Homspace{\alpha}{K}{L}:= \\
&\qquad\homspace{\alpha}{K}{L}_\mathcal{C}^\mathbb{N}/\homspace{\alpha}{K}{L}_\mathcal{N}^\mathbb{N}\subset\homspace{\alpha}{\bar{K}}{\bar{L}} \notag
\end{align}
The theoretical hom-space $\homspace{\alpha}{K}{L}$ and the empirical hom-space $\Homspace{\alpha}{K}{L}$ are thus defined between the systems $K$ and $L$, where empirical operations are continuous operations between the completed systems that can be approximated to arbitrary precision by theoretical operations.

Concatenation and composition of compatible operations are also defined elementwise:
\begin{align}
(e\otimes f)_n&:=e_n\otimes f_n \\
(e\circ f)_n&:=e_n\circ f_n \notag
\end{align}
The composed operation is continuous if both operations are continuous, and the composed operation is null if either operation is null. Composition thus extends to the empirical hom-spaces. This is not necessarily the case for concatenation, however. The conditions for concatenation to be well defined on the empirical hom-spaces will not be considered further in this essay, except to note that the conditions are trivially satisfied for systems with finite-dimensional state and observable spaces. Worked examples in discrete systems are thus validated, and the validity of results more generally is conditional on the existence of concatenations for the empirical operations they involve.

\subsection{Quantum groups}

\begin{figure*}[!p]
\centering
\setlength{\tabcolsep}{0.0\linewidth}
\begin{tabular}{C{0.312\textwidth}C{0.332\textwidth}C{0.252\textwidth}C{0.104\textwidth}}
\multicolumn{4}{c}{$\ast$-Algebra} \\ \hline
\includegraphics[width=0.2808\textwidth]{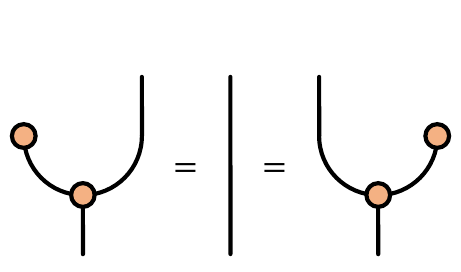} & \includegraphics[width=0.2988\textwidth]{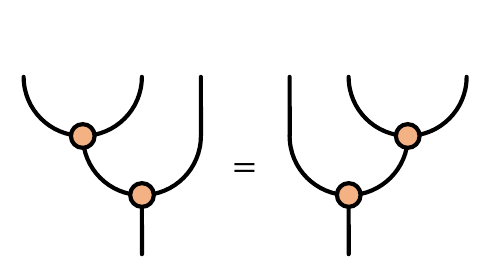} & \includegraphics[width=0.2268\textwidth]{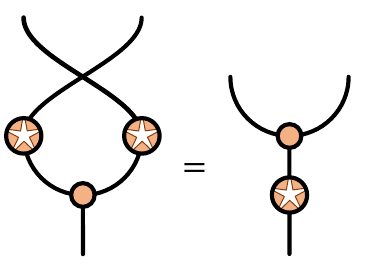} & \includegraphics[width=0.0936\textwidth]{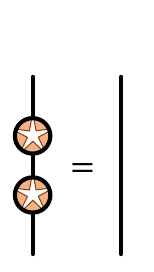} \\
\makebox[0pt]{$\producta\circ(\unita\otimes\identity)=\identity=\producta\circ(\identity\otimes\unita)$} & \makebox[0pt]{$\producta\circ(\producta\otimes\identity)=\producta\circ(\identity\otimes\producta)$} & \makebox[0pt]{$\producta\circ(\invol\otimes\invol)\circ\braid=\invol\circ\producta$} & \makebox[0pt]{$\invol\circ\invol=\identity$} \\[1ex]
\makebox[0pt]{$(\counita\otimes\identity)\circ\coproducta=\identity=(\identity\otimes\counita)\circ\coproducta$} & \makebox[0pt]{$(\coproducta\otimes\identity)\circ\coproducta=(\identity\otimes\coproducta)\circ\coproducta$} & \makebox[0pt]{$\braid\circ(\coinvol\otimes\coinvol)\circ\coproducta=\coproducta\circ\coinvol$} & \makebox[0pt]{$\coinvol\circ\coinvol=\identity$} \\
\includegraphics[width=0.2808\textwidth]{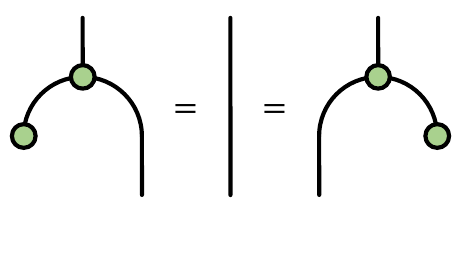} & \includegraphics[width=0.2988\textwidth]{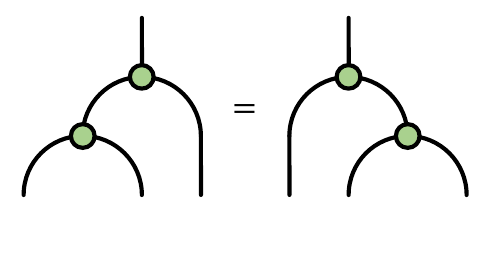} & \includegraphics[width=0.2268\textwidth]{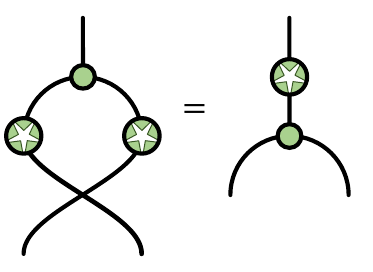} & \includegraphics[width=0.0936\textwidth]{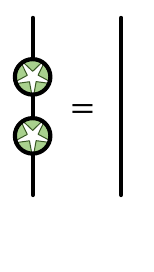}
\end{tabular}\vspace{-2ex}
\caption{The $\ast$-algebra axioms apply separately to the operations of states and observables, and require that they are unital, associative, antiautomorphic and involutive.}\vspace{2ex}
\label{fig:quantumgroupalgebraaxioms}
\begin{tabular}{C{0.33\textwidth}C{0.33\textwidth}C{0.33\textwidth}}
\multicolumn{3}{c}{$\ast$-Bialgebra} \\ \hline
\includegraphics[width=0.2448\textwidth]{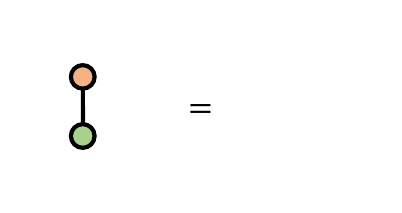} & \includegraphics[width=0.2448\textwidth]{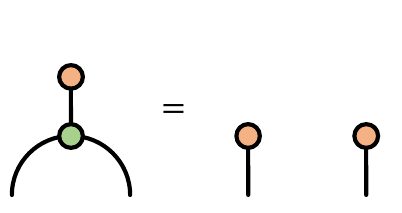} & \includegraphics[width=0.2448\textwidth]{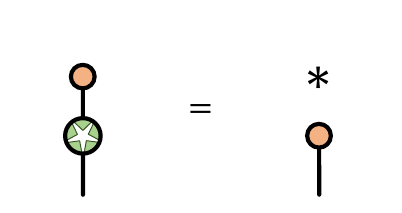} \\
\makebox[0pt]{$\counita\circ\unita=1$} & \makebox[0pt]{$\coproducta\circ\unita=\unita\otimes\unita$} & \makebox[0pt]{$\coinvol\circ\unita=\unita\circ{\large\ast}$} \\
\includegraphics[width=0.2448\textwidth]{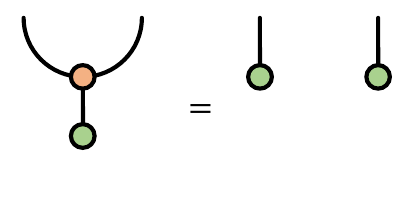} & \includegraphics[width=0.2448\textwidth]{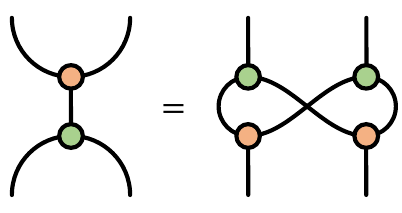} & \includegraphics[width=0.2448\textwidth]{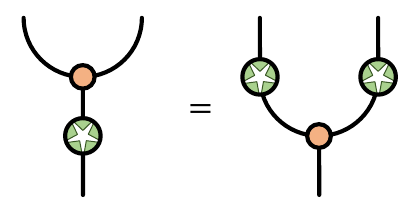} \\
\makebox[0pt]{$\counita\circ\producta=\counita\otimes\counita$} & \makebox[0pt]{$\coproducta\circ\producta=(\producta\otimes\producta)\circ(\identity\otimes\braid\otimes\identity)\circ(\coproducta\otimes\coproducta)$} & \makebox[0pt]{$\coinvol\circ\producta=\producta\circ(\coinvol\otimes\coinvol)$} \\
\includegraphics[width=0.2448\textwidth]{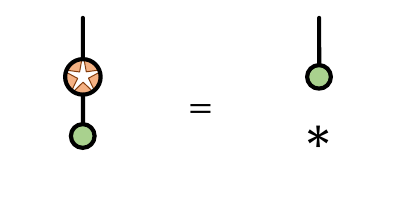} & \includegraphics[width=0.2448\textwidth]{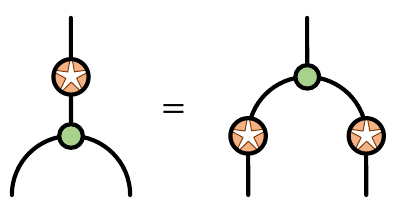} & \includegraphics[width=0.2448\textwidth]{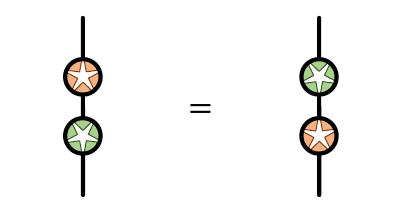} \\
\makebox[0pt]{$\counita\circ\invol={\large\ast}\circ\counita$} & \makebox[0pt]{$\coproducta\circ\invol=(\invol\otimes\invol)\circ\coproducta$} & \makebox[0pt]{$\coinvol\circ\invol=\invol\circ\coinvol$}
\end{tabular}
\caption{The $\ast$-bialgebra axioms require that the two $\ast$-algebras are complementary, expressed as commutation relations between the operations of states and observables.}\vspace{2ex}
\label{fig:quantumgroupbialgebraaxioms}
\begin{tabular}{C{\textwidth}}
Hopf $\ast$-Algebra \\ \hline
\\[-1ex]
\includegraphics[width=0.3888\textwidth]{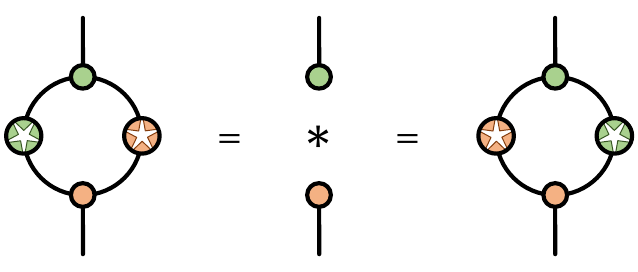} \\
\makebox[0pt]{$\producta\circ(\coinvol\otimes\invol)\circ\coproducta=\unita\circ{\large\ast}\circ\counita=\producta\circ(\invol\otimes\coinvol)\circ\coproducta$}
\end{tabular}
\caption{The Hopf $\ast$-algebra axiom relates all the operations of the quantum group, enabling the creation of the antipode as the composition of the two involutions.}
\label{fig:quantumgroupHopfaxiom}
\end{figure*}

\begin{figure*}[!t]
\centering
\setlength{\tabcolsep}{0.0\linewidth}
\begin{tabular}{C{0.25\textwidth}C{0.47\textwidth}C{0.28\textwidth}}
Unit & Product & Involution \\ \hline
\begin{tabular}{c}\includegraphics[width=0.18\textwidth]{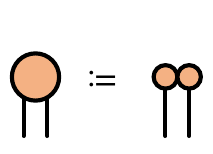}\\$\unita:=\unita\otimes\unita$\\[1ex]$\counita:=\counita\otimes\counita$\\\includegraphics[width=0.18\textwidth]{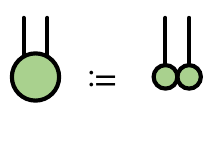}\end{tabular} &
\begin{tabular}{c}\includegraphics[width=0.34\textwidth]{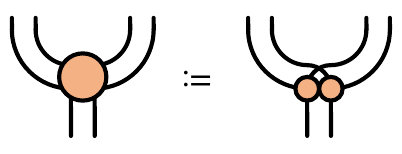}\\\makebox[0pt]{$\producta:=(\producta\otimes\producta)\circ(\identity\otimes\braid\otimes\identity)$}\\[1ex]\makebox[0pt]{$\coproducta:=(\identity\otimes\braid\otimes\identity)\circ(\coproducta\otimes\coproducta)$}\\\includegraphics[width=0.34\textwidth]{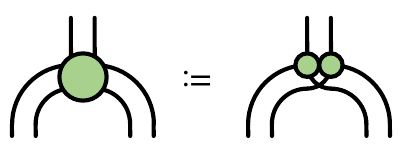}\end{tabular} &
\begin{tabular}{c}\includegraphics[width=0.2\textwidth]{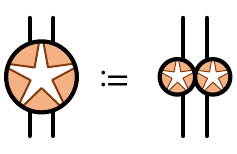}\\$\invol:=\invol\otimes\invol$\\[1ex]$\coinvol:=\coinvol\otimes\coinvol$\\\includegraphics[width=0.2\textwidth]{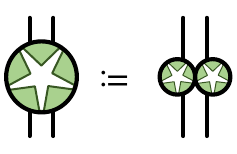}\end{tabular}
\end{tabular}
\caption{The concatenation of two quantum groups is also a quantum group. The structural operations of the concatenated quantum group are the concatenations of the structural operations of the component quantum groups.}
\label{fig:quantumgroupconcatenatedoperations}
\end{figure*}

With the apparatus of the empirical category in place, the quantum group is characterised by its extended lexicon of operations and the grammar they satisfy. The definition is motivated by the properties of classical groups, and is essentially equivalent to the definition of the classical group in the case when the product of observables is commutative. Removing this ugly asymmetry from the otherwise pristine duality between state and observable is the principal abstract achievement of the quantum group. The variance this generates has numerous applications, including as the origin of indeterminism in quantum mechanics and as a source of value for convexity in the pricing of derivative securities.

A system $K$ is here defined to be a quantum group if it has two families of operations acting as unit, product and involution on the states and observables, as shown in the string diagrams of figure \ref{fig:quantumgroupoperations}. The axioms that relate these structural operations, presented in figures \ref{fig:quantumgroupalgebraaxioms}, \ref{fig:quantumgroupbialgebraaxioms} and \ref{fig:quantumgroupHopfaxiom}, fall into three categories. The first set of axioms implies that the states and observables separately have the structure of $\ast $-algebras. By requiring them to commute, the second set of axioms combines the two $\ast $-algebras into a $\ast $-bialgebra. The final axiom then enables the creation of an antipode that generates the structure of dual Hopf $\ast $-algebras on the states and observables.

Quantum groups are closed under concatenation with the definitions in figure \ref{fig:quantumgroupconcatenatedoperations}, where the operation for the concatenated system, appearing on the left in these diagrams, is defined on the right in terms of the corresponding operations for its component systems. Concatenated quantum groups thus inherit the structural operations from their component systems. This ability to combine quantum groups allows systems to be built from component systems, a property that is used in the construction of dynamic systems on discrete schedules.

These operations, with the rules they satisfy, are sufficient for the application of quantum groups to mathematical finance. As will be demonstrated below, the state and observable algebras separately create stochastic and functional calculus, the bialgebra properties ensure they can be implemented consistently, and the Hopf axiom enforces the reversibility of integration and differentiation.

\section{Arbitrage and functional calculus}

Mathematical finance studies the relationship between the economic and pricing states, respectively quantifying the expected and present values of economic variables. Applied to undetermined observables, the economic state is inherently subjective, providing an assessment of economic conditions yet to be revealed. In contrast, the pricing state is marked to observed market prices, incorporating the market consensus and external factors such as liquidity and funding, and relegating subjective expectations to the unhedged convexities.

A sound platform for mathematical finance thus needs to establish the essential links between the economic and pricing states without being overly constraining. The constraints considered here prevent arbitrage and impose a loose relationship that allows both subjective and market expectations to influence price. Markets are not obliged to obey these principles, but they are satisfied to a reasonable approximation in normal conditions, and are a point of attraction even when markets are stressed.

In this section, the economy is modelled as a fixed system $K$ equipped with unit, product and involution operations:
\begin{align}
\unita&\in\Homspace{1}{K}{1} \\
\producta&\in\Homspace{1}{K}{K\otimes K} \notag \\
\invol&\in\Homspace{\ast}{K}{K} \notag
\end{align}
on the observables that satisfy the unital, associative, antiautomorphic and involutive properties of $\ast$-algebra. The standard notation convention is used to simplify expressions:
\begin{align}
1&:=\unita\circ 1 \\
\mathsf{ab}&:=\producta\circ(\mathsf{a}\otimes\mathsf{b}) \notag \\
\mathsf{a}^\ast&:=\invol\circ\mathsf{a}\circ\ast \notag
\end{align}
for the observables $\mathsf{a},\mathsf{b}\in\ObSpace[\bar{K}]$.

In the state $\mathsf{z}\in\StSpace[\bar{K}]$, observables are evaluated via the pairing. By definition, the state is {\em coreal} if it satisfies:
\begin{equation}
\mathsf{z}\bullet\mathsf{a}^\ast=(\mathsf{z}\bullet\mathsf{a})^\ast
\end{equation}
so that real observables have real valuations, and the state is {\em copositive} if it is coreal and satisfies:
\begin{equation}
\mathsf{z}\bullet\mathsf{aa}^\ast\geq 0
\end{equation}
so that positive observables have positive valuations. Copositivity immediately leads to the Cauchy-Schwarz inequality.
\begin{theorem}[Cauchy-Schwarz inequality]
The copositive state $\mathsf{z}\in\StSpace[\bar{K}]$ satisfies the inequality:
\begin{equation}
|\mathsf{z}\bullet\mathsf{ab}^\ast|^2\leq(\mathsf{z}\bullet\mathsf{aa}^\ast)(\mathsf{z}\bullet\mathsf{bb}^\ast)
\end{equation}
for the observables $\mathsf{a},\mathsf{b}\in\ObSpace[\bar{K}]$.
\end{theorem}
\noindent This initiates a chain of foundational results. With the economic principles proposed here, the valuation model has the following structure.
\begin{itemize}
\item The state and observable spaces are represented as mutually commutant von Neumann algebras acting on a Hilbert space, with a normalised vacuum vector $\bra{1}$ that is separating for states and generating for observables.
\item The representation $\mathsf{a}\mapsto[\mathsf{a}]$ of observables as operators is a $\ast$-algebra homomorphism. The representation $\mathsf{z}\mapsto[\mathsf{z}]$ of states as operators commutes with the representation of observables:
\begin{equation}
[\mathsf{z}][\mathsf{a}]=[\mathsf{a}][\mathsf{z}]
\end{equation}
\item The economic and pricing states are weight\-ed pure states with pairing:
\begin{equation}
\mathsf{z}\bullet\mathsf{a}=\braket{1|[\mathsf{z}][\mathsf{a}]|1}=\braket{1|[\mathsf{a}][\mathsf{z}]|1}
\end{equation}
The economic state is represented by the identity operator, and the pricing state is arbitrage-free if and only if its representing operator is positive.
\end{itemize}
Topological convergence conditions qualify these statements in infinite dimensions. The economic principles are the preconditions of the Gelfand-Naimark-Segal construction and the Radon-Nikodym theorem, outlined below with consideration of their technical limitations.

\subsection{Principle of equivalence}

In classical economics, the principle of equivalence states that the price of an Arrow-Debreu security can only be nonzero if the probability of the event it indicates is nonzero. The pricing measure is thus, in the measure-theoretic sense, dominated by the economic measure. This marriage of convenience between expectation and price creates a potentially toxic codependency that incentivises unrealistic economic assumptions. It is nonetheless an important theoretical device unlocking significant developments in mathematical finance, and will be assumed here.

The concept of equivalence is generalised using relations that empirically test whether an observable is zero. The elementary example is the nullification relation $\mathcal{N}$:
\begin{equation}
\mathsf{z}\mathbin{\mathcal{N}}\mathsf{a}\iff\mathsf{z}\bullet\mathsf{a}=0
\end{equation}
That the observable evaluates to zero in the state supports the hypothesis that the observable is zero, but does not present compelling evidence. Stronger evidence is provided by the one-sided and two-sided annihilation relations $\mathcal{R}$ and $\mathcal{A}$:
\begin{alignat}{3}
\mathsf{z}\mathbin{\mathcal{R}}\mathsf{a}&\iff{} & \forall\mathsf{c}:{}& & \mathsf{z}\bullet\mathsf{ac}=0& \\
\mathsf{z}\mathbin{\mathcal{A}}\mathsf{a}&\iff{} & \forall\mathsf{b},\mathsf{c}:{}& & \mathsf{z}\bullet\mathsf{bac}=0&=\mathsf{z}\bullet\mathsf{ca}^\ast\mathsf{b} \notag
\end{alignat}
adding the condition that proportional observables also evaluate to zero. The relations satisfy:
\begin{equation}
\mathcal{A}\subset\mathcal{R}\subset\mathcal{N}
\end{equation}
Applied to the subset $\mathsf{Y}\subset\StSpace[\bar{K}]$ of states, the one-sided annihilator $\mathsf{Y}\mathcal{R}$ is a right ideal:
\begin{equation}
(\mathsf{Y}\mathcal{R})\ObSpace[\bar{K}]=\mathsf{Y}\mathcal{R}
\end{equation}
and the two-sided annihilator $\mathsf{Y}\mathcal{A}$ is a $\ast$-ideal:
\begin{align}
\ObSpace[\bar{K}](\mathsf{Y}\mathcal{A})\ObSpace[\bar{K}]&=\mathsf{Y}\mathcal{A} \\
(\mathsf{Y}\mathcal{A})^\ast&=\mathsf{Y}\mathcal{A} \notag
\end{align}
Use the following notation for the cosets:
\begin{align}
\bra{\mathsf{a}}&:=\mathsf{a}+\mathsf{Y}\mathcal{R} \\
[\mathsf{a}]&:=\mathsf{a}+\mathsf{Y}\mathcal{A} \notag
\end{align}
These operations on the cosets are well defined:
\begin{align}
\bra{\mathsf{a}}\![\mathsf{b}]&:=\bra{\mathsf{ab}} \\
[\mathsf{a}][\mathsf{b}]&:=[\mathsf{ab}] \notag \\
[\mathsf{a}]^\ast&:=[\mathsf{a}^\ast] \notag
\end{align}
The distinguishable observables $\ObSpace[\bar{K}]/\mathsf{Y}\mathcal{A}$ obtained by factoring out the annihilated observables thus form a $\ast$-algebra faithfully represented as operators on $\ObSpace[\bar{K}]/\mathsf{Y}\mathcal{R}$ with $\bra{1}$ as generating vector. This observation is the basis for the Gelfand-Naimark-Segal construction.

Each economic agent is associated with an economic state that encapsulates the subjective expectations of the agent. The principle of equivalence dictates that an observable annihilated by all these economic states must also be annihilated by the pricing state.
\begin{description}
\item[Principle of equivalence:]The annihilator of the pricing state $\mathsf{z}$ contains the annihilator of the set of economic states $\mathsf{Y}$.
\end{description}
By the subset properties of relations, $\mathsf{Y}\mathcal{A}\subset\mathsf{z}\mathcal{A}$ if and only if $\mathsf{z}\in\mathcal{A}\mathsf{Y}\mathcal{A}$, and the pricing state $\mathsf{z}$ satisfying this principle is said to be dominated by the set $\mathsf{Y}$ of economic states. Thanks to the $\ast$-ideal property of $\mathsf{Y}\mathcal{A}$, the double annihilations for the one-sided and two-sided variants satisfy:
\begin{equation}
\mathsf{Y}\subset\mathcal{R}\mathsf{Y}\mathcal{R}\subset\mathcal{N}\mathsf{Y}\mathcal{A}=\mathcal{A}\mathsf{Y}\mathcal{A}
\end{equation}
Models that comply with the principle of equivalence thus reside entirely in the subsystem $K_\mathsf{Y}$ generated by the economic states, with state and observable spaces:
\begin{align}
\StSpace[K_\mathsf{Y}]&:=\mathcal{A}\mathsf{Y}\mathcal{A} \\
\ObSpace[K_\mathsf{Y}]&:=\ObSpace[\bar{K}]/\mathsf{Y}\mathcal{A} \notag
\end{align}
restricting to the dominated states and the distinguishable observables.

Within the system $K$, the community of agents is represented by their set $\mathsf{Y}$ of economic states. If the market satisfies the principle of equivalence, then the analysis of price can be conducted entirely within the subsystem $K_\mathsf{Y}$. This is a strong statement with significant implications for pricing: an event can only have nonzero price if at least one agent assigns the event nonzero probability. Dismissing events as impossible may be pragmatic for pricing but it comes at the cost of unmitigated model risk.

The economic states $\mathsf{Y}$ and their associated annihilated observables $\mathsf{A}$ and pricing states $\mathsf{Z}$ are related by:
\begin{align}
\mathcal{A}\mathsf{Y}\mathcal{A}&=:\mathsf{Z}=\mathcal{A}\mathsf{A}=\mathcal{A}\mathsf{Z}\mathcal{A} \\
\mathsf{Y}\mathcal{A}&=:\mathsf{A}=\mathsf{Z}\mathcal{A}=\mathcal{A}\mathsf{A}\mathcal{A} \notag
\end{align}
Two communities of agents whose possible pricing states align thus operate within the same subsystem, and any subspace of states or observables that is closed under double annihilation represents a unique equivalence class of communities with the same space of pricing states.

The pairing projects to the subsystem:
\begin{equation}
\mathsf{z}\bullet[\mathsf{a}]:=\mathsf{z}\bullet\mathsf{a}
\end{equation}
and the operation $\identity\in\Homspace{1}{K_\mathsf{Y}}{K}$ with:
\begin{align}
\mathsf{z}\circ\identity&:=\mathsf{z} \\
\identity\circ\mathsf{a}&:=[\mathsf{a}] \notag
\end{align}
represents the agents as an embedding of their subsystem. Extending this observation, the relationship between two communities of agents $\mathsf{X}$ and $\mathsf{Y}$ satisfying $\mathsf{Y}\mathcal{A}\subset\mathsf{X}\mathcal{A}$ is captured in the embedding operation $\identity\in\Homspace{1}{K_\mathsf{X}}{K_\mathsf{Y}}$ with:
\begin{align}
\mathsf{z}\circ\identity&:=\mathsf{z} \\
\identity\circ[\mathsf{a}]&:=[\mathsf{a}] \notag
\end{align}
In this case, the agents $\mathsf{Y}$ are said to dominate the agents $\mathsf{X}$ on the system $K$. The larger community of agents operates in a less constrained system with the potential to identify value where the smaller community assigns none.

Suppose the agents operate in two systems $K$ and $L$, with economic states $\mathsf{X}$ and $\mathsf{Y}$ respectively. When $e\in\Homspace{\alpha}{K}{L}$ is consistent with the corresponding subsystems, satisfying the consistency condition:
\begin{equation}
e\circ\mathsf{Y}\mathcal{A}\subset\mathsf{X}\mathcal{A}
\end{equation}
the operation projects to the subsystems as the operation $e\in\Homspace{\alpha}{K_\mathsf{X}}{L_\mathsf{Y}}$:
\begin{align}
\mathsf{z}\circ e&:=\mathsf{z}\circ e \\
e\circ[\mathsf{a}]&:=[e\circ\mathsf{a}] \notag
\end{align}
The consistency condition restricts the range of operations available to the agents. Reversing this observation, a model that assumes the existence of special operations must also impose consistency conditions on the annihilated observables if these operations are to be inherited by the agents. This is relevant in the development of stochastic and functional calculus later, as conditions are imposed to ensure the dynamic system inherits the operations of the quantum group.

\subsection{Principle of expectation}

The economic state $\mathsf{y}\in\StSpace[\bar{K}]$ generates the expected value of an observable via its pairing. As a model of expectation, the state is required to satisfy the following principle.
\begin{description}
\item[Principle of expectation:]The economic state $\mathsf{y}$ is copositive and normalised by $\mathsf{y}\bullet 1=1$.
\end{description}
For the real observable $\mathsf{a}$, satisfying $\mathsf{a}^\ast=\mathsf{a}$, define the mean and standard deviation:
\begin{align}
\mu[\mathsf{a}]&:=\mathsf{y}\bullet\mathsf{a} \\
\sigma[\mathsf{a}]&:=\sqrt{\mathsf{y}\bullet(\mathsf{a}-\mu[\mathsf{a}])^2} \notag
\end{align}
Coreality implies that the mean is real, and copositivity implies that the standard deviation is positive. These statistics estimate the observable and measure its uncertainty. When the standard deviation is nonzero, define the standardised observable:
\begin{equation}
\mathsf{a}^\perp:=\frac{\mathsf{a}-\mu[\mathsf{a}]}{\sigma[\mathsf{a}]}
\end{equation}
Standardisation centralises and normalises the observable to have zero mean and unit standard deviation.

For the pair of real observables $\mathsf{a}$ and $\mathsf{b}$, the Cauchy-Schwarz inequality strengthens the constraints on the standard deviations.
\begin{theorem}[Heisenberg uncertainty]
The standard deviations for the real observables $\mathsf{a}$ and $\mathsf{b}$ satisfy the inequality:
\begin{align}
\sigma[\mathsf{a}]^2\sigma[\mathsf{b}]^2\geq{}&(\mathsf{y}\bullet\frac{1}{2}(\mathsf{ab}+\mathsf{ba})-\mu[\mathsf{a}]\mu[\mathsf{b}])^2 \\
&+(\mathsf{y}\bullet\frac{1}{2i}(\mathsf{ab}-\mathsf{ba}))^2 \notag
\end{align}
\end{theorem}
\noindent This result follows from the decomposition:
\begin{equation}
\mathsf{ab}=\frac{1}{2}(\mathsf{ab}+\mathsf{ba})+i\,\frac{1}{2i}(\mathsf{ab}-\mathsf{ba})
\end{equation}
of the product into its real and imaginary components. The standard deviations of the two observables must therefore be strictly positive if either of the positive terms on the right side of the Heisenberg inequality is nonzero, meaning that neither observable is determined precisely.

Correlation is defined when the standard deviations of the real observables are nonzero:
\begin{equation}
\rho[\mathsf{a},\mathsf{b}]:=\mathsf{y}\bullet\mathsf{a}^\perp\mathsf{b}^\perp
\end{equation}
In general, the correlation is complex with real and imaginary components:
\begin{align}
\Re[\rho[\mathsf{a},\mathsf{b}]]&=\mathsf{y}\bullet\frac{1}{2}(\mathsf{a}^\perp\mathsf{b}^\perp+\mathsf{b}^\perp\mathsf{a}^\perp) \\
\Im[\rho[\mathsf{a},\mathsf{b}]]&=\mathsf{y}\bullet\frac{1}{2i}(\mathsf{a}^\perp\mathsf{b}^\perp-\mathsf{b}^\perp\mathsf{a}^\perp) \notag
\end{align}
The first term is the classical correlation. The second term is novel to the quantum model, and can only be nonzero if the observables do not commute. Heisenberg uncertainty then translates to the condition:
\begin{equation}
|\rho[\mathsf{a},\mathsf{b}]|\leq 1
\end{equation}
constraining the correlation to the unit disk of the complex plane.

\begin{figure}[!t]
\centering
\setlength{\tabcolsep}{0.0\linewidth}
\centering\includegraphics[width=0.4\linewidth]{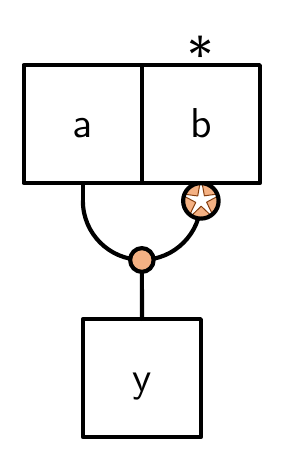}
\captionsetup{singlelinecheck=off}
\caption[.]{In this string diagram, the inner product of the observables $\mathsf{a}$ and $\mathsf{b}$ in the copositive state $\mathsf{y}$ is:\hfill\vspace{1ex}
{\begin{minipage}{\linewidth}
\setlength{\abovedisplayskip}{0pt}
\setlength{\belowdisplayskip}{0pt}
\begin{equation}
\braket{\mathsf{a}|\mathsf{b}}:=\mathsf{y}\bullet\mathsf{ab}^\ast=\mathsf{y}\bullet\producta\circ(\mathsf{a}\otimes(\invol\circ\mathsf{b}\circ\ast))
\end{equation}
\end{minipage}}\vspace{2ex}
using the operations of the observable $\ast$-algebra.\hfill}
\label{fig:innerproduct}
\end{figure}

\begin{figure*}[!t]
\centering
\setlength{\tabcolsep}{0.0\linewidth}
\begin{tabular}{C{0.5\textwidth}C{0.5\textwidth}}
\includegraphics[width=0.5\textwidth]{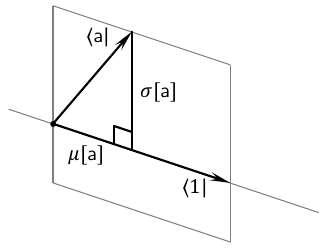} & \includegraphics[width=0.5\textwidth]{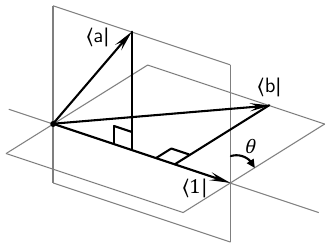}
\end{tabular}
\caption{The vacuum vector is the axis of determinism in the Gelfand-Naimark-Segal construction. In the left diagram, the vector $\bra{\mathsf{a}}$ decomposes into the parallel $\mu[\mathsf{a}]\bra{1}$ and a perpendicular of length $\sigma[\mathsf{a}]$, where $\mu[\mathsf{a}]$ and $\sigma[\mathsf{a}]$ are the mean and standard deviation of the real observable $\mathsf{a}$. In the right diagram, the cosine of the angle $\theta$ between the perpendiculars of the vectors $\bra{\mathsf{a}}$ and $\bra{\mathsf{b}}$ is the correlation norm $|\rho[\mathsf{a},\mathsf{b}]|$ between the real observables $\mathsf{a}$ and $\mathsf{b}$.}
\label{fig:GNSgeometry}
\end{figure*}

Copositivity of the economic state enables the Gelfand-Naimark-Segal construction for the observables, establishing a geometric interpretation of the economic model.
\begin{theorem}[Gelfand-Naimark-Segal]
The copositive state $\mathsf{y}\in\StSpace[\bar{K}]$ represents observables as operators on an inner product space.
\begin{itemize}
\item The space $\ObSpace[\bar{K}]/\mathsf{y}\mathcal{R}$ is an inner product space with the definition:
\begin{equation}
\bra{\mathsf{a}}\cdot\bra{\mathsf{b}}=\braket{\mathsf{a}|\mathsf{b}}:=\mathsf{y}\bullet\mathsf{ab}^\ast
\end{equation}
for the inner product.
\item The space $\ObSpace[\bar{K}]/\mathsf{y}\mathcal{A}$ is a $\ast$-subalgebra of the adjointable operators on $\ObSpace[\bar{K}]/\mathsf{y}\mathcal{R}$ and the map $\mathsf{a}\mapsto[\mathsf{a}]$ is a $\ast$-algebra homomorphism.
\item The copositive state $\mathsf{y}$ pairs with observables via the operator representation:
\begin{equation}
\mathsf{y}\bullet\mathsf{a}=\braket{1|[\mathsf{a}]|1}
\end{equation}
so that the state is a pure state with vacuum vector $\bra{1}\in\ObSpace[\bar{K}]/\mathsf{y}\mathcal{R}$.
\end{itemize}
\end{theorem}
\noindent The operations on the quotient spaces are defined for arbitrary states; copositivity contributes the inner product on the space $\ObSpace[\bar{K}]/\mathsf{y}\mathcal{R}$ and realises the space $\ObSpace[\bar{K}]/\mathsf{y}\mathcal{A}$ as a $\ast$-subalgebra of adjointable operators. Note that the conditions of annihilation simplify for the copositive state $\mathsf{y}$, with:
\begin{equation}
\mathsf{y}\mathbin{\mathcal{R}}\mathsf{a}\iff\mathsf{y}\bullet\mathsf{aa}^\ast=0
\end{equation}
and:
\begin{alignat}{2}
\mathsf{y}\mathbin{\mathcal{A}}\mathsf{a}&\iff\forall\mathsf{b}:{} & \mathsf{y}\bullet(\mathsf{ba})(\mathsf{ba})^\ast&=0 \\
&\iff\forall\mathsf{c}:{} & \mathsf{y}\bullet(\mathsf{ac})^\ast(\mathsf{ac})&=0 \notag
\end{alignat}
thanks to the Cauchy-Schwarz inequality.

This construction applies to any state that satisfies the conditions of copositivity. For the economic state, the construction translates statistical properties of observables into geometric properties of the corresponding vectors. The normalisation $\mathsf{y}\bullet 1=1$ implies the vacuum vector has unit norm:
\begin{equation}
\|\bra{1}\|=1
\end{equation}
The vector $\bra{\mathsf{a}}$ associated with the real observable $\mathsf{a}$ decomposes into vectors parallel and perpendicular to the vacuum vector $\bra{1}$:
\begin{equation}
\bra{\mathsf{a}}=\mu[\mathsf{a}]\bra{1}+\sigma[\mathsf{a}]\bra{\mathsf{a}^\perp}
\end{equation}
The vacuum vector is the axis of determinism: the real observable $\mathsf{a}$ has zero standard deviation if and only if the vector $\bra{\mathsf{a}}$ is parallel to the vacuum vector $\bra{1}$, in which case the vacuum vector is an eigenvector of the operator $[\mathsf{a}]$ with eigenvalue given by the mean. The correlation of the real observables $\mathsf{a}$ and $\mathsf{b}$ is the inner product of their perpendiculars:
\begin{equation}
\braket{\mathsf{a}^\perp|\mathsf{b}^\perp}=\rho[\mathsf{a},\mathsf{b}]
\end{equation}
whose norm is the cosine of the angular separation between the perpendiculars. The diagrams of figure \ref{fig:GNSgeometry} visualise these relationships.

\subsection{Principle of no-arbitrage}

The pricing state $\mathsf{z}\in\StSpace[\bar{K}]$ generates the present value of an observable via its pairing. As a model of price, the state is required to satisfy the following principle.
\begin{description}
\item[Principle of no-arbitrage:]The pricing state $\mathsf{z}$ is copositive.
\end{description}
Copositivity means that a positive observable has positive price, thereby preventing arbitrage. In this application, the observable prescribes the terminal value of a derivative security with settlement conditions specifying its delivery time and currency. The pricing state is then assumed to account for discounting and exchange into the numeraire currency.

The economic principles are combined to provide an operator representation of the price model on the inner product space of the economic model. Economic agents are associated with copositive and normalised states in a subset of economic states $\mathsf{Y}\subset\ObSpace[\bar{K}]$, and settlement conditions are associated with copositive states in a subset of pricing states $\mathsf{Z}\subset\mathcal{A}\mathsf{Y}\mathcal{A}\subset\StSpace[\bar{K}]$. Copositivity enables the Gelfand-Naimark-Segal construction on all these states, and the relationship between the inner product spaces they generate is explored in the Radon-Nikodym theorem.

To simplify the presentation, assume there is a copositive state $\mathsf{y}\in\mathcal{A}\mathsf{Y}\mathcal{A}$ that dominates all the economic states:
\begin{equation}
\mathsf{y}\mathcal{A}=\mathsf{Y}\mathcal{A}
\end{equation}
This state supports all the events anticipated by the economic agents. When the number of agents is finite, the dominant economic state is defined as any weighted average:
\begin{equation}
\mathsf{y}:=\sum\nolimits_a\omega_a\mathsf{y}_a
\end{equation}
of the economic states $\mathsf{y}_a$ associated with the agents $a$, with weights $\omega_a$ that are strictly positive and sum to one. The dominant economic state then averages across all the economic states, with weights that can be tuned to the prevalence and expertise of the agents.

Further assuming that the pricing state $\mathsf{z}$ is a copositive state in the subspace $\mathcal{R}\mathsf{y}\mathcal{R}\subset\mathcal{A}\mathsf{y}\mathcal{A}$ of states dominated by the economic state $\mathsf{y}$, the Radon-Nikodym theorem derives an operator representation of the pricing state in the inner product space of the economic state. The equivalence condition $\mathsf{y}\mathcal{R}\subset\mathsf{z}\mathcal{R}$ implies that the quotient map:
\begin{equation}
\mathsf{m}:\ObSpace[\bar{K}]/\mathsf{y}\mathcal{R}\to\ObSpace[\bar{K}]/\mathsf{z}\mathcal{R}
\end{equation}
on the cosets is a well-defined surjective operator that commutes with the operator representation of observables:
\begin{equation}
\mathsf{m}[\mathsf{a}]=[\mathsf{a}]\mathsf{m}
\end{equation}
Applying the Gelfand-Naimark-Segal construction, the operator $\mathsf{m}$ transforms from economic to pricing inner product, with both states expressed as pure states in their respective vacuums:
\begin{align}
\mathsf{y}\bullet\mathsf{a}&=\bra{1}\![\mathsf{a}]\cdot\bra{1} \\
\mathsf{z}\bullet\mathsf{a}&=\bra{1}\!\mathsf{m}[\mathsf{a}]\cdot\bra{1}\!\mathsf{m} \notag
\end{align}
If the transformation $\mathsf{m}$ is adjointable, define the representation:
\begin{equation}
[\mathsf{z}]:=\mathsf{mm}^\ast:\ObSpace[\bar{K}]/\mathsf{y}\mathcal{R}\to\ObSpace[\bar{K}]/\mathsf{y}\mathcal{R}
\end{equation}
of the pricing state as a positive operator on the economic inner product space. This operator also commutes with the operator representation of observables:
\begin{equation}
[\mathsf{z}][\mathsf{a}]=[\mathsf{a}][\mathsf{z}]
\end{equation}
and the economic and pricing states become:
\begin{align}
\mathsf{y}\bullet\mathsf{a}&=\braket{1|[\mathsf{a}]|1} \\
\mathsf{z}\bullet\mathsf{a}&=\braket{1|[\mathsf{z}][\mathsf{a}]|1}=\braket{1|[\mathsf{a}][\mathsf{z}]|1} \notag
\end{align}
expressed as weighted pure states in the economic vacuum.

The obvious shortcoming in this argument is the assumption of adjointability for the transformation between the inner product spaces. The Radon-Nikodym theorem addresses this shortcoming by imposing an additional boundedness condition on the relationship between the states.
\begin{theorem}[Radon-Nikodym]
The copositive state $\mathsf{y}\in\StSpace[\bar{K}]$ represents the dominated copositive state $\mathsf{z}\in\mathcal{R}\mathsf{y}\mathcal{R}$ as an operator on an inner product space. Let $\mathsf{m}:\ObSpace[\bar{K}]/\mathsf{y}\mathcal{R}\to\ObSpace[\bar{K}]/\mathsf{z}\mathcal{R}$ be the associated transformation of inner product spaces, with operator norm:
\begin{equation}
\left\Vert\mathsf{m}\right\Vert=\sqrt{\sup\{\mathsf{z}\bullet\mathsf{aa}^\ast:\mathsf{y}\bullet\mathsf{aa}^\ast=1\}}
\end{equation}
From the Gelfand-Naimark-Segal construction:
\begin{itemize}
\item The state $\mathsf{z}$ pairs with observables via the operator representation:
\begin{equation}
\mathsf{z}\bullet\mathsf{a}=\bra{1}\!\mathsf{m}[\mathsf{a}]\cdot\bra{1}\!\mathsf{m}
\end{equation}
so that the state is a pure state with vacuum vector $\bra{1}\!\mathsf{m}\in\ObSpace[\bar{K}]/\mathsf{z}\mathcal{R}$.
\item If $\mathsf{m}$ is bounded, $\left\Vert\mathsf{m}\right\Vert<\infty$, then it is adjointable on the Hilbert completions of the inner product spaces, and the state $\mathsf{z}$ pairs with observables via the operator representation:
\begin{equation}
\mathsf{z}\bullet\mathsf{a}=\braket{1|[\mathsf{z}][\mathsf{a}]|1}=\braket{1|[\mathsf{a}][\mathsf{z}]|1}
\end{equation}
so that the state is a weighted pure state with vacuum vector $\bra{1}\in\ObSpace[\bar{K}]/\mathsf{y}\mathcal{R}$ and positive weight operator $[\mathsf{z}]:=\mathsf{m}\mathsf{m}^\ast$ in the commutant of the representation of observables.
\end{itemize}
\end{theorem}
\noindent Acting on the Hilbert completion of $\ObSpace[\bar{K}]/\mathsf{y}\mathcal{R}$, the operator $[\mathsf{z}]$ is the {\em Radon-Nikodym weight} of the dominated copositive state $\mathsf{z}$ over the copositive state $\mathsf{y}$. In this construction, the extension to the Hilbert completion is needed to guarantee the existence of orthogonals in the definition of the operator adjoint. The operator $[\mathsf{a}]$ representing the observable $\mathsf{a}$ extends to the Hilbert completion of $\ObSpace[\bar{K}]/\mathsf{y}\mathcal{R}$ if and only if its operator norm:
\begin{equation}
\left\Vert[\mathsf{a}]\right\Vert=\sqrt{\sup\{\mathsf{y}\bullet(\mathsf{ba})(\mathsf{ba})^\ast:\mathsf{y}\bullet\mathsf{bb}^\ast=1\}}
\end{equation}
is finite, otherwise it is unbounded. Observables are thus represented as adjointable operators that may be unbounded but which are defined on a common dense subspace of a Hilbert space.

For a Hilbert space $\HilSpace$, define the commutation relation on the bounded operators $\mathsf{z}$ and $\mathsf{a}$:
\begin{equation}
\mathsf{z}\mathbin{\text{\textquotesingle}}\mathsf{a}\iff\mathsf{z}\mathsf{a}=\mathsf{a}\mathsf{z}
\end{equation}
The Gelfand-Naimark-Segal construction and the Radon-Nikodym theorem together imply that the price model is included in a system that has state space $\StOps$ and observable space $\ObOps$ represented as mutually commutant operator algebras, satisfying $\StOps\text{\textquotesingle}=\ObOps$ and $\text{\textquotesingle}\ObOps=\StOps$, so that the spaces are closed in the weak operator topology. This connection between pricing and von Neumann algebras is illuminating in both directions, as option exercise is critically dependent on the projections available in the observable algebra.

A derivative security is contractually defined by a finite collection of settlement observables across a range of settlement conditions, with present value generated by pairing with the corresponding pricing states. A fundamental requirement of the derivative is that its value is unambiguously defined at settlement. Since Heisenberg uncertainty imposes a baseline indeterminism on noncommuting observables, commutativity of the settlements is assumed in the definition.
\begin{definition}[Derivative security]
A derivative security comprises a finite collection of settlements $\{(c,\mathsf{a}_c)\}$, each described by its settlement conditions $c$ and the corresponding settlement observable $\mathsf{a}_c$. The settlement observables are real and mutually commutative.
\end{definition}
\noindent The present value of the settlement observable $\mathsf{a}_c$ is generated by pairing with the associated pricing state $\mathsf{z}_c$, and the derivative has present value:
\begin{equation}
\sum\nolimits_c\mathsf{z}_c\bullet\mathsf{a}_c=\sum\nolimits_c\braket{1|[\mathsf{z}_c][\mathsf{a}_c]|1}
\end{equation}
summing across all the outstanding settlements. This model for the price of the derivative satisfies the principles of equivalence and no-arbitrage, and novel features emerge from the operator representation of states and observables.

Consider for simplicity the case when the settlement observables are known, $\mathsf{a}_c=a_c1$. In this case, the derivative price is:
\begin{equation}
u:=\sum\nolimits_ca_c\braket{c|c}
\end{equation}
where the vector:
\begin{equation}
\bra{c}:=\bra{1}\![\mathsf{z}_c]^{1/2}
\end{equation}
generates the pricing state $\mathsf{z}_c$ as a pure state. Now suppose that the holder has the right but not the obligation to settle the derivative. Exercise is indicated by a projection $\mathsf{p}\in\ObOps$ with binary spectrum $\{0,1\}$, expressed as the properties:
\begin{equation}
\mathsf{p}^\ast=\mathsf{p}=\mathsf{p}^2
\end{equation}
and optimal exercise chooses the projection to maximise the option price:
\begin{equation}
o:=\sup\nolimits_\mathsf{p}\left\{\sum\nolimits_ca_c\braket{c|\mathsf{p}|c}\right\}
\end{equation}
ranging across the projections available in the von Neumann algebra of observables. A useful upper bound for the option price is easily discovered by optimising over all projections:
\begin{equation}
o\leq\tr\!\left[\left(\sum\nolimits_ca_c\hat{\mathsf{z}}_c\right)^+\right]
\end{equation}
with:
\begin{equation}
\hat{\mathsf{z}}_c:=\ket{c}\bra{c}
\end{equation}
where the trace is the sum of the positive eigenvalues of the self-adjoint operator in the brackets, evaluated from the roots of its characteristic polynomial. While the option price generally depends on the structure of the von Neumann algebra, the upper bound depends only on the finite-dimensional geometry expressed in the inner products of the state vectors:
\begin{equation}
\braket{c|d}=\braket{1|[\mathsf{z}_c]^{1/2}[\mathsf{z}_d]^{1/2}|1}
\end{equation}
It is thus an upper bound that applies to all models whose state vectors share these inner products.

This ability to simultaneously explore a constrained range of valuation models highlights a key benefit of the operator representation. An implicit criticism in the Keynesian understanding of uncertain knowledge, and further developed in the work of Shackle, is that classical probability expresses uncertainty against a backdrop of certitude on the state space topology. By embedding the observables as a subalgebra of operators, quantum probability provides the tools for pricing in a world of unknown unknowns. As an example, the upper bound for the option price depends only on the geometry of a finite set of known modes, and the result is valid for all models that include these modes even when they are insufficient to fully determine the topology. Such results are necessarily imprecise, reflecting the additional uncertainty that comes from the indeterminable state space.

\section{Discrete states and observables}

The most general valuation model with discrete states and observables is developed from its representation as operators on a finite-dimensional space. This model is easy to evaluate and avoids technical pitfalls, and can be embedded within common algorithms that discretise the states and observables. None of these statements carry merit, though, if the resulting model does not exhibit novel properties useful for real applications. Fortunately, the transition from classical to quantum significantly extends the phenomenology of the model, even in the finite-dimensional case. As an example of the approach, option pricing -- the most elementary challenge of mathematical finance -- is enriched by the quantum extension in ways that characterise the underlying algebra.

The mutually commutant von Neumann algebras of states and observables decompose as direct sums of von Neumann factors with trivial centres, each isomorphic to a matrix algebra. A convenient representation, used in the following, decomposes the Hilbert space:
\begin{equation}
\HilSpace=\bigoplus\nolimits_{i=1}^{n}\HilSpace_{i}
\end{equation}
where $\HilSpace_{i}=\mathbb{C}[\StDim_{i},\ObDim_{i}]$ is the space of complex matrices with $\StDim_{i}$ rows and $\ObDim_{i}$ columns equipped with the Hilbert-Schmidt inner product:
\begin{equation}
\braket{\phi_i|\psi_i}:=\tr[\phi_{i}^{\vphantom{\ast}}\psi_{i}^\ast]
\end{equation}
for the matrices $\phi_{i},\psi_{i}\in\HilSpace_{i}$. The von Neumann algebras $\StOps$ of states and $\ObOps$ of observables then decompose:
\begin{equation}
\StOps=\bigoplus\nolimits_{i=1}^{n}\StOps_{i}\qquad\ObOps=\bigoplus\nolimits_{i=1}^{n}\ObOps_{i}
\end{equation}
where $\StOps_{i}=\mathbb{C}[\StDim_{i}]$ is the space of complex matrices with $\StDim_{i}$ rows and columns and $\ObOps_{i}=\mathbb{C}[\ObDim_{i}]$ is the space of complex matrices with $\ObDim_{i}$ rows and columns. These spaces are represented on the Hilbert space via matrix multiplication:
\begin{equation}
\bra{\phi_i}\![\mathsf{z}_{i}]:=\bra{\mathsf{z}^t_i\phi_i^{\vphantom{t}}}\qquad\bra{\phi_i}\![\mathsf{a}_{i}]:=\bra{\phi^{\vphantom{\ast}}_i\mathsf{a}^{\vphantom{\ast}}_i}
\end{equation}
for the states $\mathsf{z}_i\in\StOps_i$ and the observables $\mathsf{a}_i\in\ObOps_i$. The representations commute:
\begin{equation}
\bra{\phi_i}\![\mathsf{z}_{i}][\mathsf{a}_{i}]=\bra{\mathsf{z}^t_i\phi_i^{\vphantom{t}}\mathsf{a}^{\vphantom{t}}_i}=\bra{\phi_i}\![\mathsf{a}_{i}][\mathsf{z}_{i}]
\end{equation}
and the states and observables are mutually commutant, $\StOps_i\hspace{-0.5pt}\text{\textquotesingle}=\ObOps_i$ and $\text{\textquotesingle}\ObOps_i=\StOps_i$.

With this representation, the pairing is expressed in terms of its vacuum matrices $\omega_i\in\HilSpace_i$:
\begin{align}
\mathsf{z}\bullet\mathsf{a}&=\sum\nolimits_{i=1}^{n}\braket{\omega_i|[\mathsf{z}_i][\mathsf{a}_i]|\omega_i} \\
&=\sum\nolimits_{i=1}^{n}\tr[\hat{\mathsf{z}}_i\mathsf{a}_i] \notag
\end{align}
where:
\begin{equation}
\hat{\mathsf{z}}_i:=\omega_i^\ast\mathsf{z}_i^t\omega_i^{\vphantom{\ast}}
\end{equation}
is the Radon-Nikodym weight of the state over the trace state.

The structural parameters of the valuation model are the pairs of strictly positive integers $((\StDim_{1},\ObDim_{1}),\ldots,(\StDim_{n},\ObDim_{n}))$ that dimension its von Neumann factors. Classical valuation sets each of these integers to one, reducing the model to evaluation against a discrete distribution with weights $(\hat{\mathsf{z}}_1,\ldots,\hat{\mathsf{z}}_n)$. If there is a novel contribution from quantum valuation, it must therefore emerge from the trace of noncommuting matrices in dimensions greater than one. From hereon, the focus is restricted to the case of a single von Neumann factor, $n=1$, and the subscripts indexing the factor are dropped from expressions.

\subsection{Option pricing}

In the application to mathematical finance, the vacuum matrix $\omega$ describes the economic state $\mathsf{y}$, and the pricing state $\mathsf{z}_c$ associated with settlement conditions $c$ scales this state:
\begin{align}
\mathsf{y}\bullet\mathsf{a}&=\tr[\hat{1}\mathsf{a}] \\
\mathsf{z}_c\bullet\mathsf{a}&=\tr[\hat{\mathsf{z}}_c\mathsf{a}] \notag
\end{align}
with weight matrices:
\begin{align}
\hat{1}&=\omega^\ast\omega \\
\hat{\mathsf{z}}_c&=\omega^\ast\mathsf{z}_c^t\omega \notag
\end{align}
The economic state is copositive by construction and is normalised by the condition:
\begin{equation}
\tr[\omega^\ast\omega]=1
\end{equation}
on the matrix $\omega$. The pricing state is then copositive when the matrix $\mathsf{z}_c$ is positive definite. This discrete model thus satisfies the principles of equivalence, expectation and no-arbitrage.

A derivative security is contractually defined by its settlements, which may occur at different times and in different currencies. The present value in the numeraire currency of each of these settlements is evaluated according to its settlement conditions. In this general case, the derivative has present value:
\begin{equation}
u:=\sum\nolimits_c\mathsf{z}_c\bullet\mathsf{a}_c=\tr\!\left[\sum\nolimits_c\hat{\mathsf{z}}_c\mathsf{a}_c\right]
\end{equation}
where the observable $\mathsf{a}_{c}$ is the payoff with settlement conditions $c$ and the state $\mathsf{z}_{c}$ is the corresponding pricing state, accounting for discounting and the exchange rate of the settlement. For the derivative to be well defined, the observables are furthermore assumed to be self-adjoint and mutually commuting.

Overlaying this underlying with optionality, if the holder has the right but not the obligation to settle, the exercise strategy needs to be incorporated into the option price. Exercise is indicated by a projection $\mathsf{p}$, a self-adjoint matrix with eigenvalues in $\{0,1\}$, which must commute with the underlying settlements for the option to be well defined. The option price is then:
\begin{equation}
o_\mathsf{p}:=\sum\nolimits_c\mathsf{z}_c\bullet\mathsf{a}_c\mathsf{p}=\tr\!\left[\left(\sum\nolimits_c\hat{\mathsf{z}}_c\mathsf{a}_c\right)\mathsf{p}\right]
\end{equation}
Optimal exercise maximises the option price:
\begin{equation}
o:=\sup\nolimits_\mathsf{p}\left\{\tr\!\left[\left(\sum\nolimits_c\hat{\mathsf{z}}_c\mathsf{a}_c\right)\mathsf{p}\right]\right\}
\end{equation}
taking the supremum over all projections that commute with the settlement observables. The projection that achieves this supremum represents the optimal exercise strategy for the option. Noncommutativity of the pricing weights introduces complexity in this calculation, preventing the simultaneous diagonalisation of the matrices in the sum, which leads to an option price that cannot be replicated in a classical discrete model.

Consider the case where the underlying settlements are known, $\mathsf{a}_{c}=a_{c}1$. In this case, the supremum is taken over all projections:
\begin{equation}
o=\tr\!\left[\left(\sum\nolimits_ca_c\hat{\mathsf{z}}_c\right)^+\right]
\end{equation}
thus demonstrating that the upper bound derived from the geometry of the Gelfand-Naimark-Segal construction is attained by the discrete von Neumann factor model. Optimal exercise corresponds to the projection onto the direct sum of the eigenspaces with positive eigenvalues of the bracketed matrix. The option price is the sum of the positive eigenvalues of the bracketed matrix, given by the positive roots of its characteristic polynomial.

The option price is evaluated as the sum of the positive eigenvalues for the difference of two positive-definite matrices:
\begin{equation}
o=\tr[(R-P)^{+}]
\end{equation}
paying the matrix $P$ and receiving the matrix $R$ defined by:
\begin{align}
P&=-\sum\limits_{a_c<0}a_c\hat{\mathsf{z}}_c \\
R&=\sum\limits_{a_c>0}a_c\hat{\mathsf{z}}_c \notag
\end{align}
The underlying has the structure of a swap, exchanging the settlements with $a_{c}<0$ for the settlements with $a_{c}>0$, and the option confers the right but not the obligation to enter the swap. This expression can be used to model the prices of common derivatives, such as foreign exchange options and interest rate swaptions.

Slightly generalising the presentation, consider the option to receive one unit of the settlements with $a_{c}>0$ in exchange for $k$ units of the settlements with $a_{c}<0$, where $k\geq 0$ is the strike of the option. For strikes near zero, the received settlements dominate and the option is always exercised. For high strikes, the paid settlements dominate and the option is never exercised. Inbetween, there is a regime where it may or may not be optimal to exercise.

By introducing the strike, the option price can be expressed in terms of an implied probability density $\pdf[s]$ for the swap rate $s$:
\begin{align}
\int_{s=0}^{\infty}(s-k)^{+}\pdf&[s]\,ds:= \\
&\tr[(R-kP)^{+}]/\tr[P] \notag
\end{align}
From this definition, the implied cumulative density $\cdf[s]$ is derived as:
\begin{align}
\cdf&[s]:= \\
&1+\left.\frac{d}{dk}\tr[(R-kP)^{+}]/\tr[P]\right\vert_{k=s} \notag
\end{align}
In the classical model, the probability density is discrete and the cumulative density is a step function. Thanks to the nonlinear relation between the strike and the eigenvalues of the option payoff, the probability density of the quantum model has components with both discrete and continuous support. This is evident even in the simplest case of the quantum binomial model.

\subsection{Quantum multinomial model}

In the binomial model with $\ObDim=2$, matrices have at most two eigenvalues. Without loss of generality, the two matrices $P$ and $R$ are expressed as:
\begin{alignat}{2}
P&={} &
&\begin{bmatrix}
p_{+} & 0 \\ 
0 & p_{-}
\end{bmatrix} \\
R&={} & U
&\begin{bmatrix}
r_{+} & 0 \\ 
0 & r_{-}
\end{bmatrix}
U^\ast \notag
\end{alignat}
where:
\begin{equation}
U=
\begin{bmatrix}
e^{i\psi}\cos[\theta] & -e^{-i\phi}\sin[\theta] \\ 
e^{i\phi}\sin[\theta] & e^{-i\psi}\cos[\theta]
\end{bmatrix}
\end{equation}
is a special unitary matrix in two dimensions. The eigenvalues of the matrices are assumed to satisfy $0\leq p_-\leq p_+$ and $0\leq r_-\leq r_+$. The coordinate frame is chosen to diagonalise $P$, and the angle $\theta$ and phase shifts $\phi$ and $\psi$ transform to the coordinate frame that diagonalises $R$. It is the rotation that creates the quantum features of the model.

The eigenvalues $\{u_{-}[k],u_{+}[k]\}$ of the matrix $R-kP$ are computed as the roots of its characteristic binomial:
\begin{align}
u_{\pm}&[k]= \\
&\frac{1}{2}\left((\bar{r}-k\bar{p})\pm\sqrt{\hat{r}^{2}-2k\hat{r}\hat{p}\cos[2\theta]+k^{2}\hat{p}^{2}}\right) \notag
\end{align}
where:
\begin{alignat}{3}
\bar{p}&=p_{+}+p_{-} & &\qquad & \hat{p}&=p_{+}-p_{-} \\
\bar{r}&=r_{+}+r_{-} & &\qquad & \hat{r}&=r_{+}-r_{-} \notag
\end{alignat}
The option price sums only the positive eigenvalues. Three regimes for the option price are delimited by the boundary strikes:
\begin{align}
k_{\pm}={}&\frac{\bar{r}\bar{p}-\hat{r}\hat{p}\cos[2\theta]}{\bar{p}^{2}-\hat{p}^{2}}\times \\
&\left(1\pm\sqrt{1-\frac{(\bar{r}^{2}-\hat{r}^{2})(\bar{p}^{2}-\hat{p}^{2})}{(\bar{r}\bar{p}-\hat{r}\hat{p}\cos[2\theta])^{2}}}\right) \notag
\end{align}
satisfying $0\leq k_{-}\leq k_{+}\leq\infty$. Degeneracies in this expression are resolved as limits.
\begin{itemize}
\item If $p_{-}=0$ and $r_-\ne 0$, then:
\begin{equation}
k_{-}=\frac{1}{2\bar{p}}\frac{\bar{r}^{2}-\hat{r}^{2}}{\bar{r}-\hat{r}\cos[2\theta]}\qquad k_{+}=\infty
\end{equation}
\item If $p_-\ne 0$ and $r_{-}=0$, then:
\begin{equation}
k_{-}=0\qquad k_{+}=2\bar{r}\frac{\bar{p}-\hat{p}\cos[2\theta]}{\bar{p}^{2}-\hat{p}^{2}}
\end{equation}
\item If $p_-=r_-=0$, then $k_-=0$ and $k_+=\infty$.
\end{itemize}
In the low-strike regime $k\leq k_{-}$, both eigenvalues are positive and the option is always exercised:
\begin{equation}
o[k]=\bar{r}-k\bar{p}
\end{equation}
In the high-strike regime $k_{+}\leq k$, both eigenvalues are negative and the option is never exercised:
\begin{equation}
o[k]=0
\end{equation}
In the mid-strike regime $k_{-}\leq k\leq k_{+}$, the eigenvalues satisfy $u_{-}[k]\leq 0\leq u_{+}[k]$ and the option price is the positive eigenvalue:
\begin{align}
o&[k]= \\
&\frac{1}{2}\left((\bar{r}-k\bar{p})+\sqrt{\hat{r}^{2}-2k\hat{r}\hat{p}\cos[2\theta]+k^{2}\hat{p}^{2}}\right) \notag
\end{align}
Differentiating the option price with respect to the strike generates the implied cumulative density:
\begin{align}
\cdf&[s]= \\
&\frac{1}{2}\left(1+\frac{\hat{p}}{\bar{p}}\frac{s\hat{p}-\hat{r}\cos[2\theta]}{\sqrt{\hat{r}^{2}-2s\hat{r}\hat{p}\cos[2\theta]+s^{2}\hat{p}^{2}}}\right) \notag
\end{align}
on the interval $k_{-}\leq s\leq k_{+}$, with cumulative density zero below this range and one above it. This is clearly not the cumulative density of a discrete distribution. The distribution has discrete probability density at the boundary points $s=k_{\pm}$, but also has continuous probability density inbetween. No classical discrete model is able to generate this distribution.

\begin{figure*}[!p]
\begin{minipage}{0.33\textwidth}
\includegraphics[width=\linewidth]{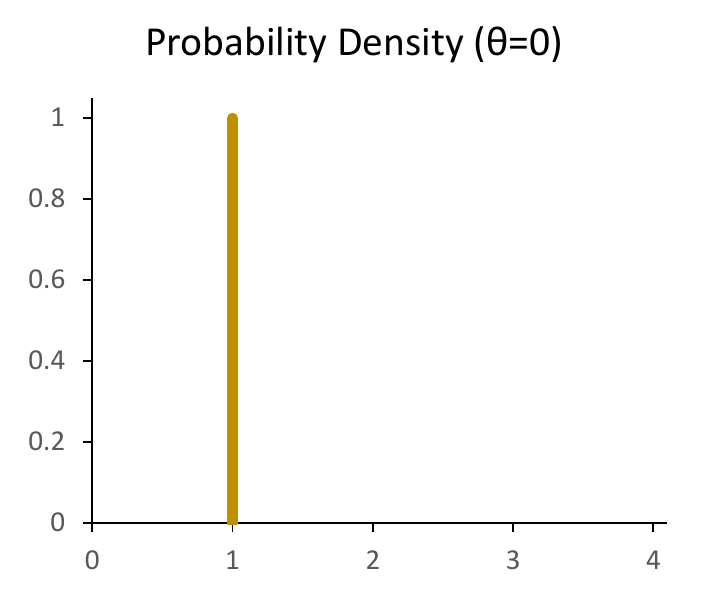}
\end{minipage}
\begin{minipage}{0.33\textwidth}
\includegraphics[width=\linewidth]{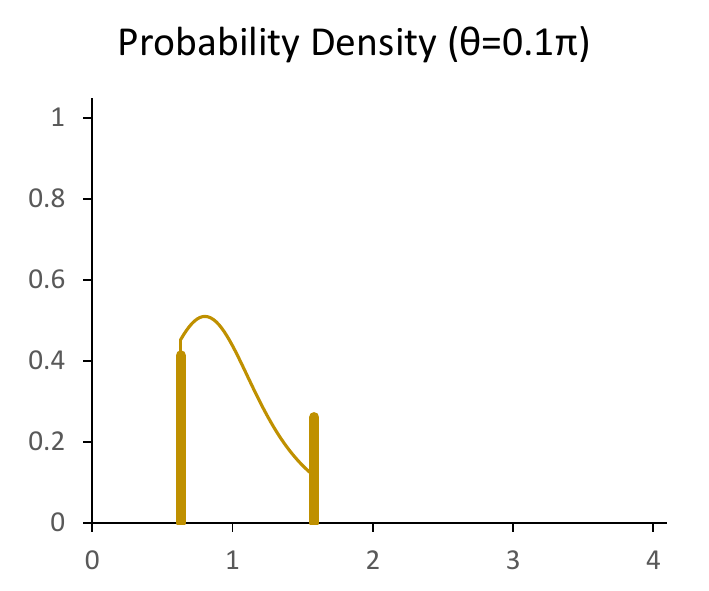}
\end{minipage}
\begin{minipage}{0.33\textwidth}
\includegraphics[width=\linewidth]{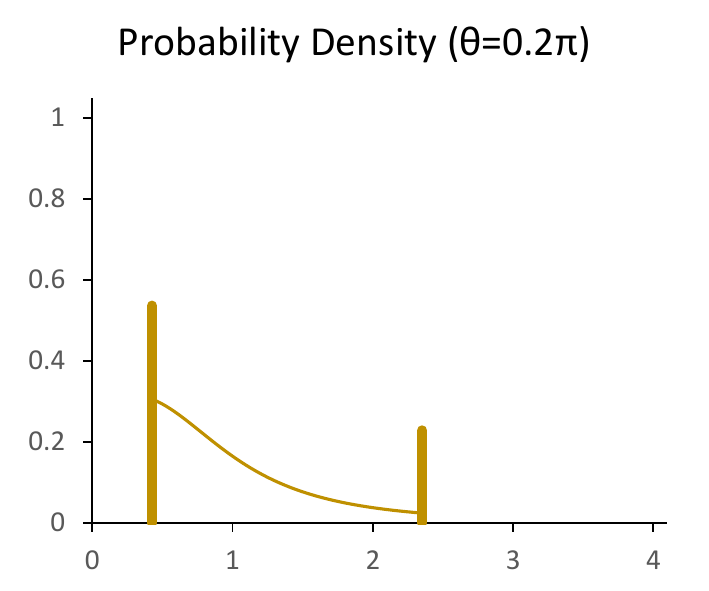}
\end{minipage}
\newline
\newline
\newline
\begin{minipage}{0.33\textwidth}
\includegraphics[width=\linewidth]{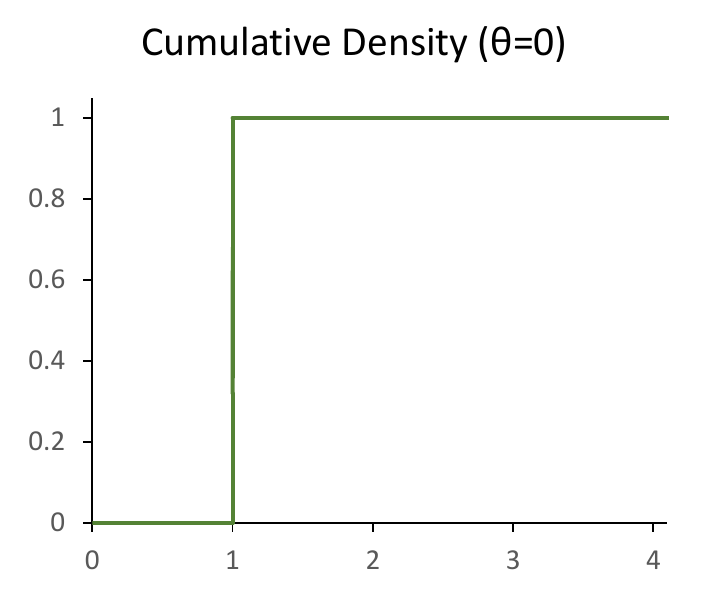}
\end{minipage}
\begin{minipage}{0.33\textwidth}
\includegraphics[width=\linewidth]{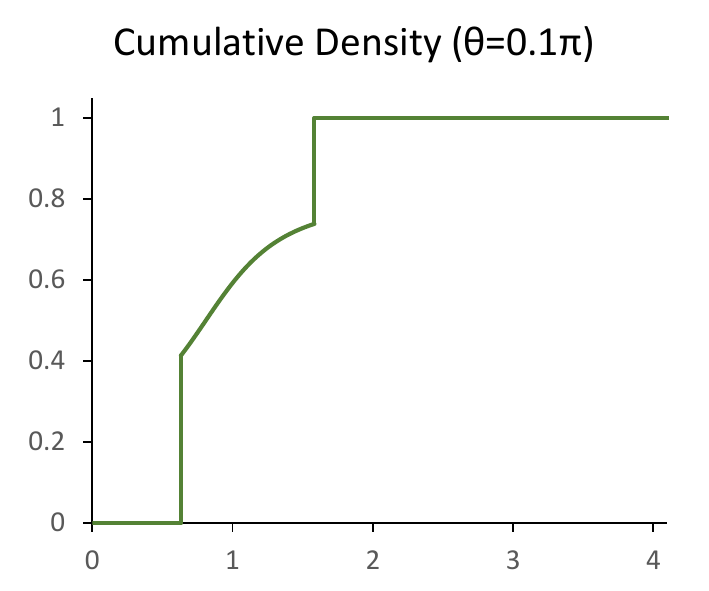}
\end{minipage}
\begin{minipage}{0.33\textwidth}
\includegraphics[width=\linewidth]{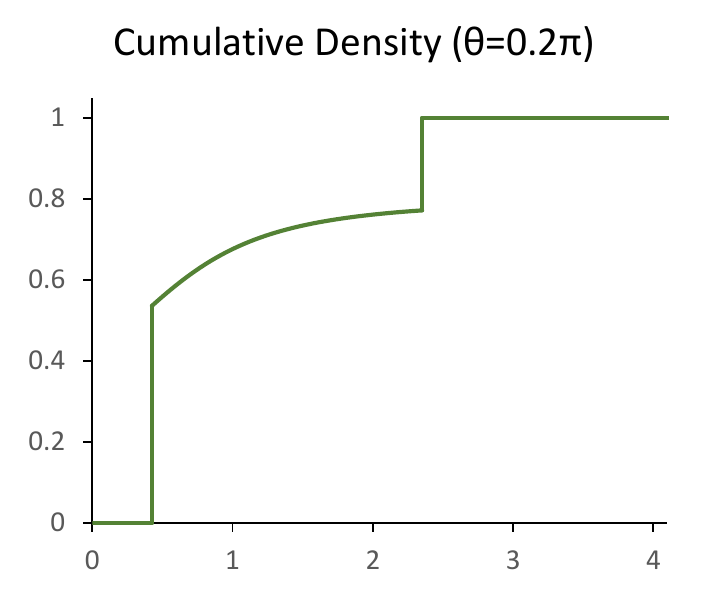}
\end{minipage}
\newline
\newline
\newline
\newline
\begin{minipage}{0.33\textwidth}
\includegraphics[width=\linewidth]{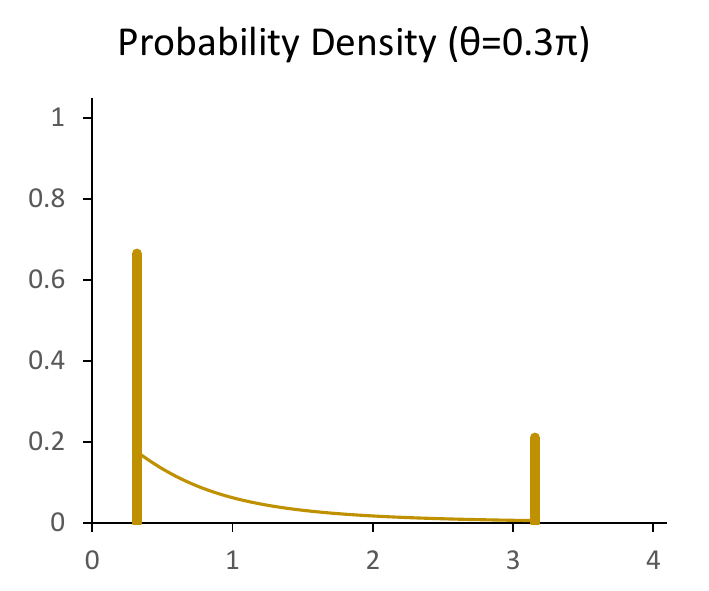}
\end{minipage}
\begin{minipage}{0.33\textwidth}
\includegraphics[width=\linewidth]{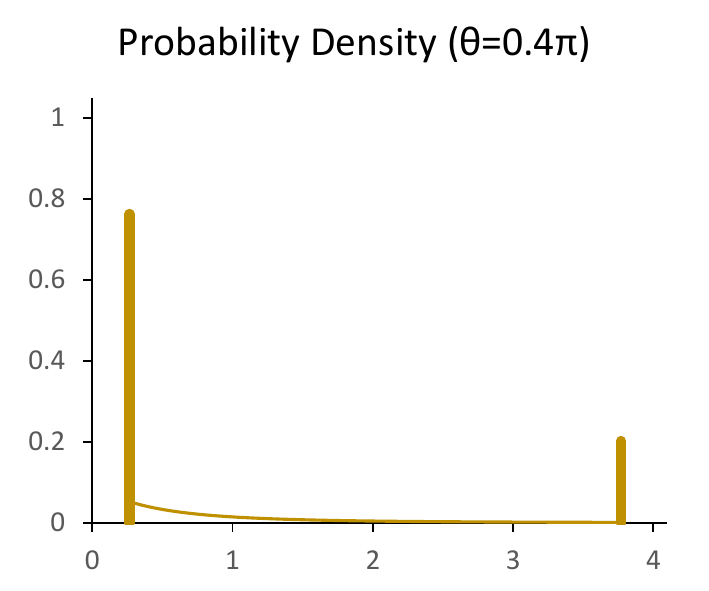}
\end{minipage}
\begin{minipage}{0.33\textwidth}
\includegraphics[width=\linewidth]{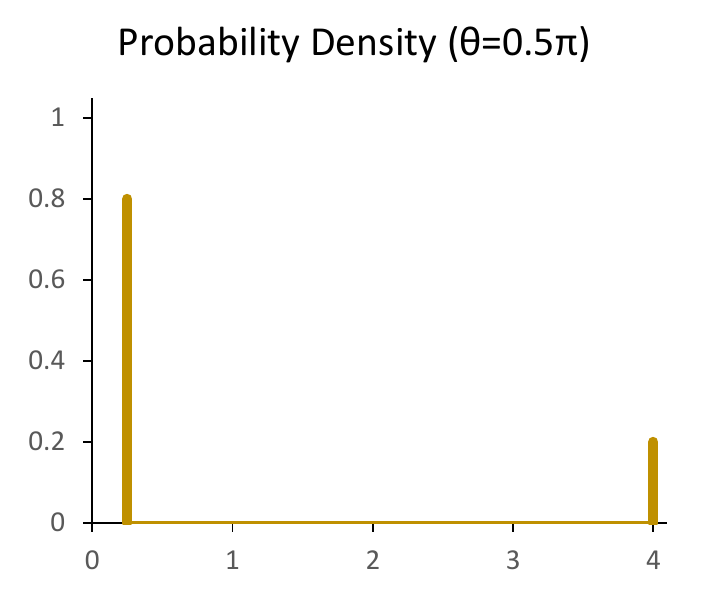}
\end{minipage}
\newline
\newline
\newline
\begin{minipage}{0.33\textwidth}
\includegraphics[width=\linewidth]{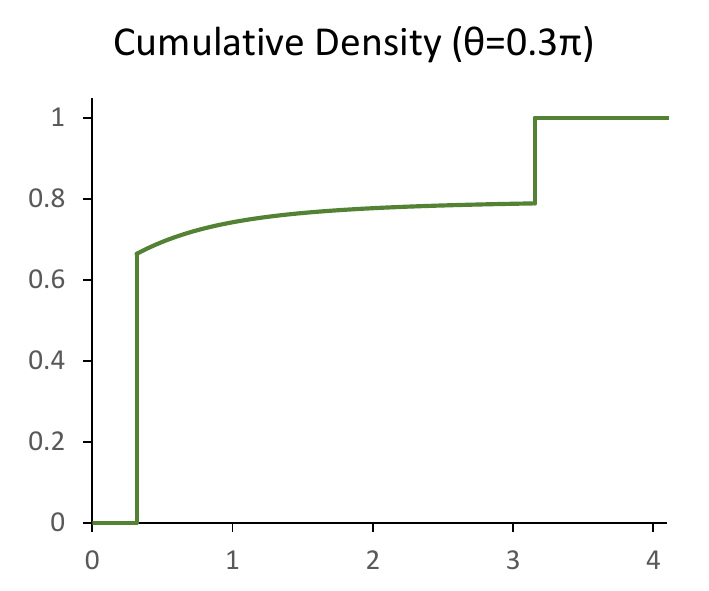}
\end{minipage}
\begin{minipage}{0.33\textwidth}
\includegraphics[width=\linewidth]{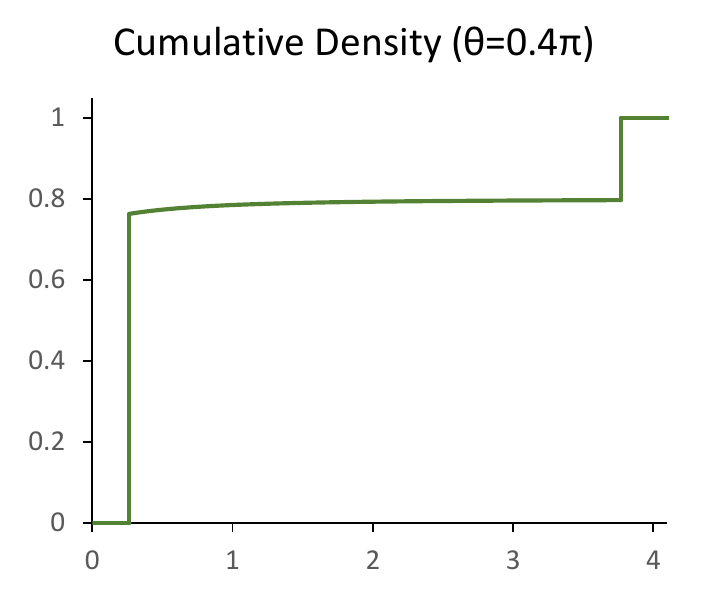}
\end{minipage}
\begin{minipage}{0.33\textwidth}
\includegraphics[width=\linewidth]{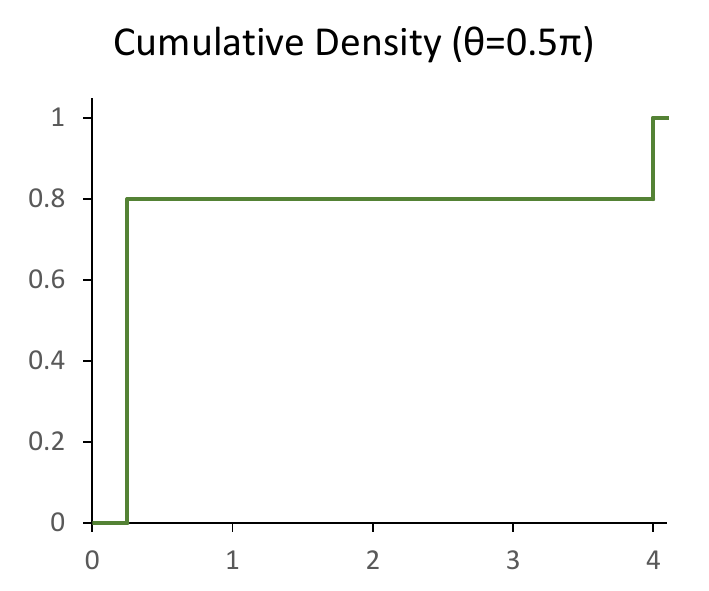}
\end{minipage}
\caption{The implied probability and cumulative densities for option pricing in the quantum binomial model with eigenvalues $0.2$ and $0.8$ for the pay and receive matrices. The distribution changes as the angle $\theta$ between the diagonalising bases for the two matrices varies. In the cases $\theta=0$ and $\theta=\pi/2$, the matrices are simultaneously diagonalisable and the result is a classical binomial model. As $\theta$ varies between these extremes, quantum tunnelling creates a continuous density between the two points of the discrete density.}
\label{fig:implieddensity1}
\end{figure*}

\begin{figure*}[!t]
\begin{minipage}{0.33\textwidth}
\includegraphics[width=\linewidth]{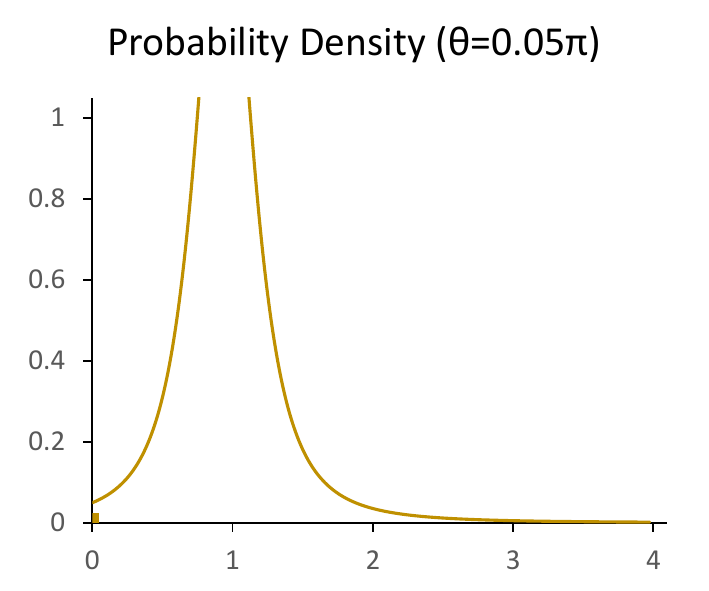}
\end{minipage}
\begin{minipage}{0.33\textwidth}
\includegraphics[width=\linewidth]{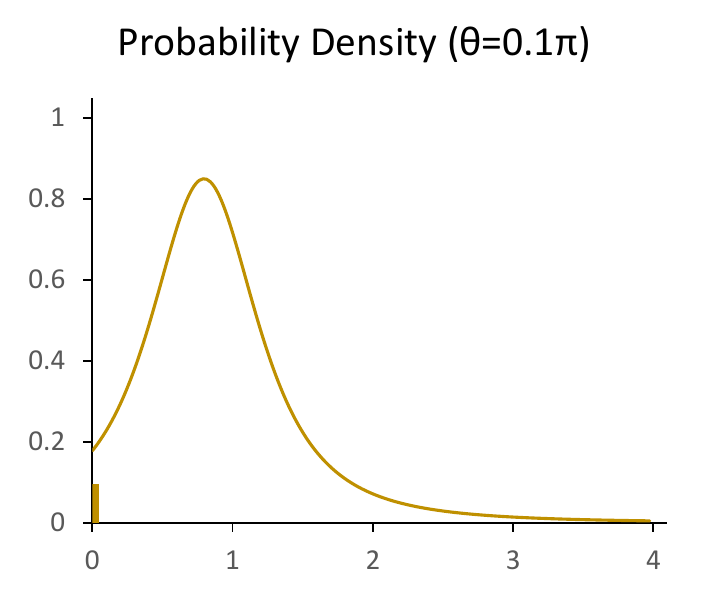}
\end{minipage}
\begin{minipage}{0.33\textwidth}
\includegraphics[width=\linewidth]{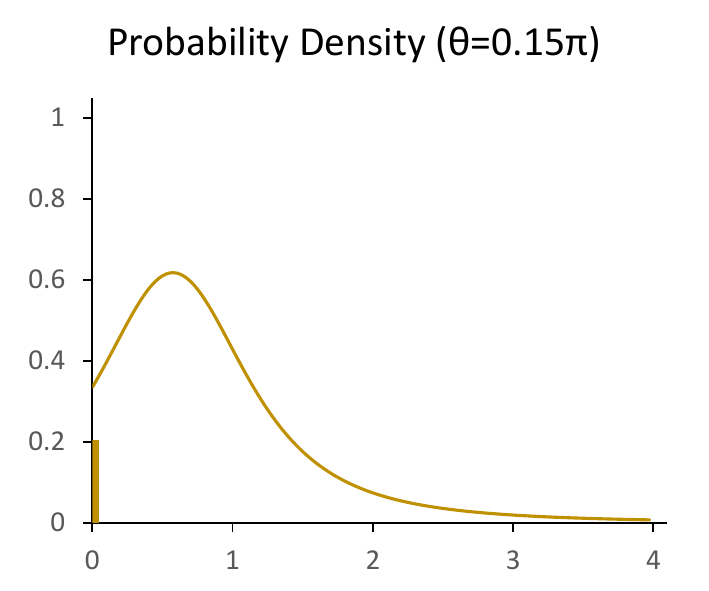}
\end{minipage}
\newline
\newline
\newline
\begin{minipage}{0.33\textwidth}
\includegraphics[width=\linewidth]{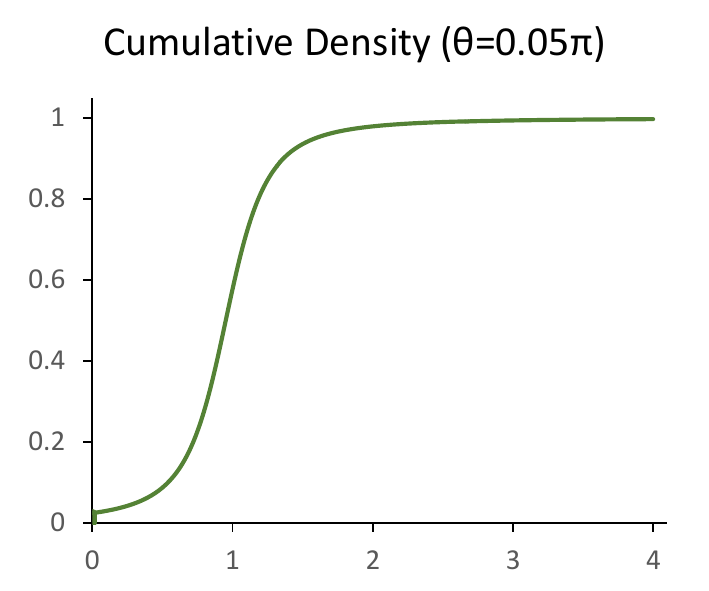}
\end{minipage}
\begin{minipage}{0.33\textwidth}
\includegraphics[width=\linewidth]{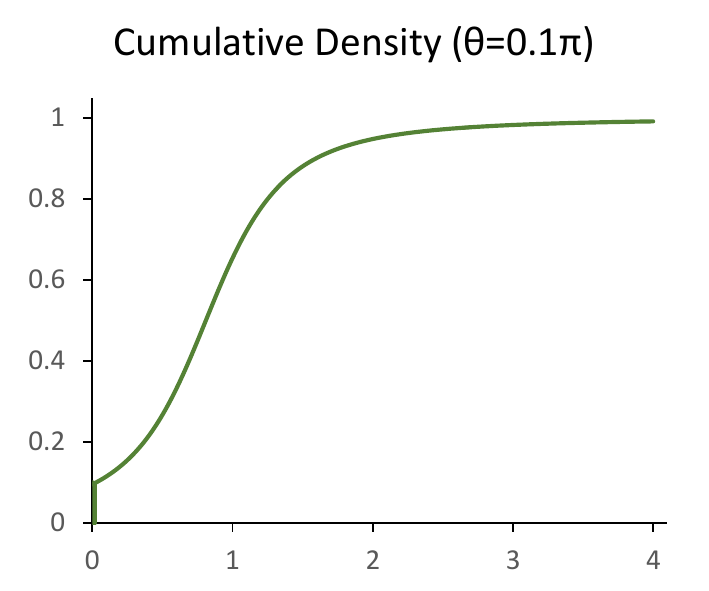}
\end{minipage}
\begin{minipage}{0.33\textwidth}
\includegraphics[width=\linewidth]{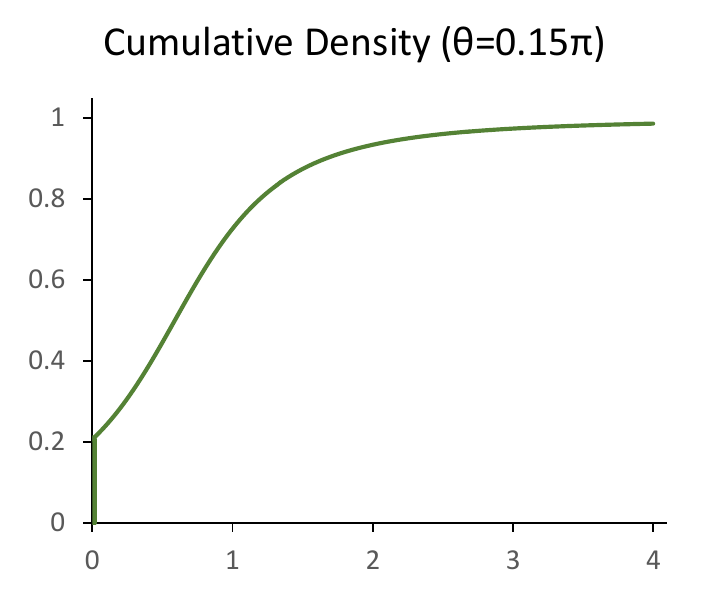}
\end{minipage}
\newline
\newline
\newline
\begin{minipage}{0.33\textwidth}
\includegraphics[width=\linewidth]{Graph-IVSFull1.pdf}
\end{minipage}
\begin{minipage}{0.33\textwidth}
\includegraphics[width=\linewidth]{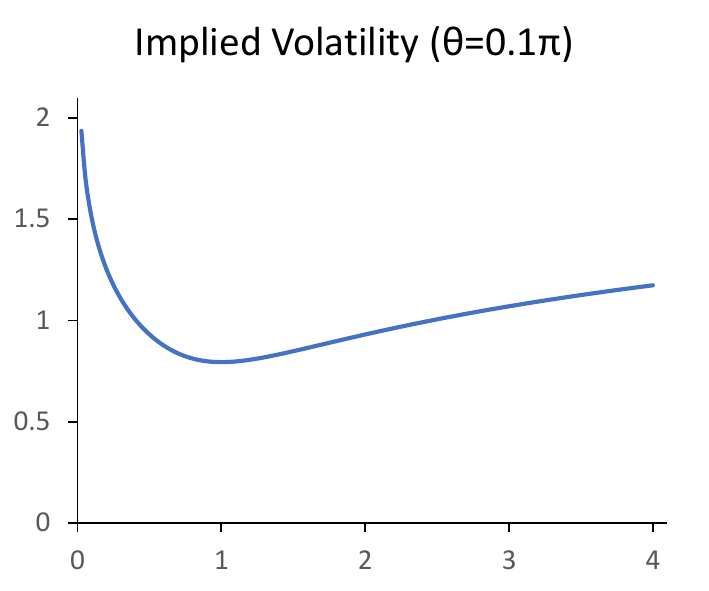}
\end{minipage}
\begin{minipage}{0.33\textwidth}
\includegraphics[width=\linewidth]{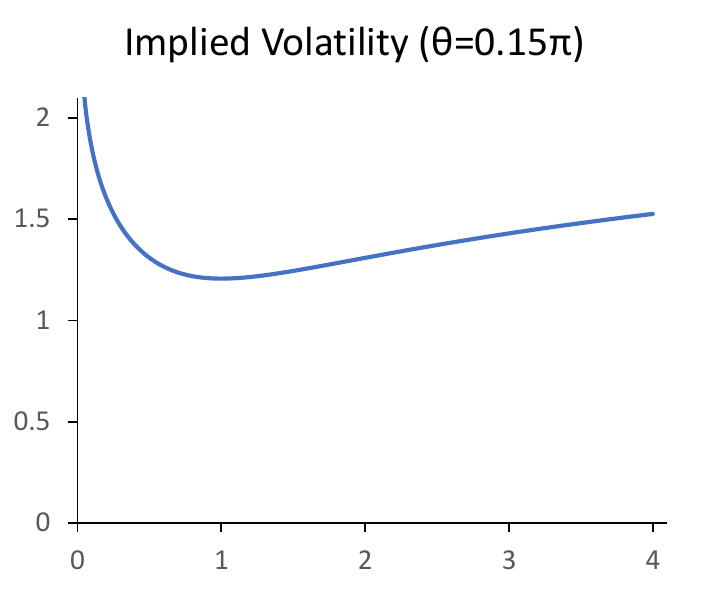}
\end{minipage}
\caption{The implied probability and cumulative densities and Black-Scholes implied volatility smiles for option pricing in the quantum binomial model with eigenvalues $0$ and $1$ for the pay and receive matrices. The distribution has a discrete density at zero and a continuous density on the upper half line. These distributions are similar to those generated by the SABR model.}\vspace{1ex}
\label{fig:implieddensity2}
\end{figure*}

\begin{figure*}[!t]
\centering
\begin{minipage}{0.38\textwidth}
\includegraphics[width=\linewidth]{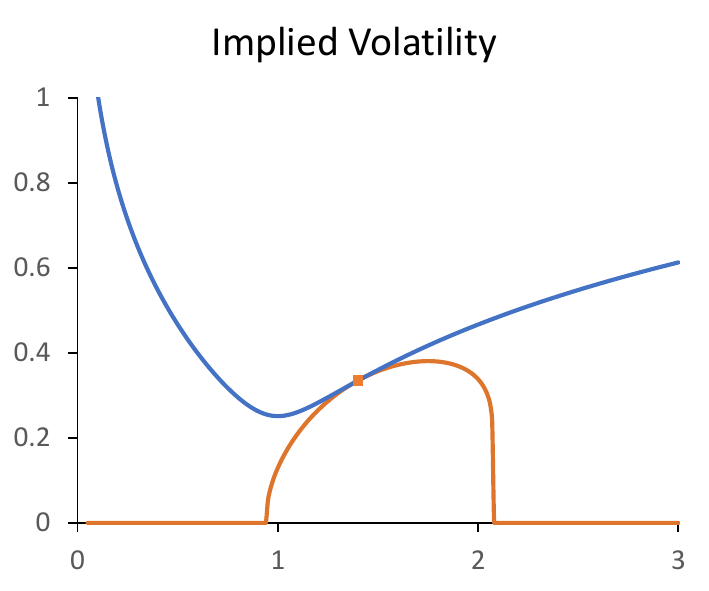}
\end{minipage}\qquad\qquad
\begin{minipage}{0.38\textwidth}
\includegraphics[width=\linewidth]{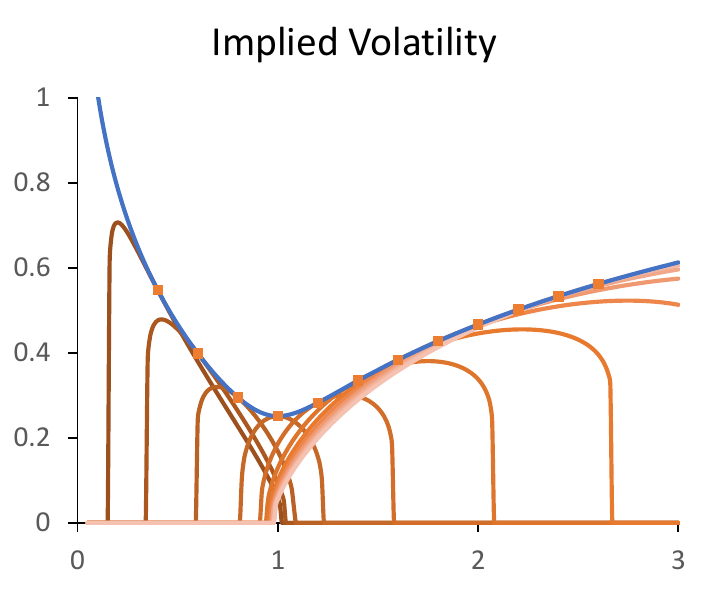}
\end{minipage}
\caption{Comparison of the implied volatility smiles for the classical and quantum binomial models. In the first graph, the classical distribution is supported on two points and the option has zero value beyond them, while the quantum distribution is supported on the upper half line and all options have positive value. In the second graph, a family of classical models is presented for binomial distributions that all have mean $1$ and root-variance $0.01$. The quantum implied volatility is the locus traced by the maximum classical implied volatility over this family. Quantum methods can thus be used to identify upper bounds for option prices when there are constraints on the moments.}
\label{fig:comparedensities}
\end{figure*}

In the binomial model that generates the graphs in figure \ref{fig:implieddensity1}, the matrices $P$ and $R$ both have eigenvalues $\{0.2,0.8\}$, and the graphs present the implied probability and cumulative densities for six different angles of separation $\theta$ between the diagonalising eigenbases of the matrices. The swap rate $s=R/P$ is well defined only when the two matrices are simultaneously diagonalisable, corresponding to the angles $\theta=0$ and $\theta=\pi/2$. In the case $\theta=0$, the matrices positively correlate and their simultaneous eigenvalues are $\{(0.2,0.2),(0.8,0.8)\}$, so the swap rate has a single eigenvalue $\{1\}$. In the case $\theta=\pi/2$, the matrices negatively correlate and their simultaneous eigenvalues are $\{(0.2,0.8),(0.8,0.2)\}$, so the swap rate has two eigenvalues $\{0.25,4\}$. These eigenvalues for the swap rate are the discrete support of the implied density.

For other values of $\theta$, the distribution has finite probabilities at two discrete points and a continuous density between these boundary points. The discrete points smoothly interpolate the bounding eigenvalues of the swap rate as the angle is varied. Quantum tunnelling effectively leaks probability into the interval spanned by the boundary points, implying a distribution with continuous support from a binomial model whose pay and receive matrices have at most two discrete eigenvalues. Dialling the angle from the positively correlated case $\theta=0$ to the negatively correlated case $\theta=\pi/2$ monotonically increases the variance of the resulting distribution.

The quantum binomial model has particularly interesting behaviour when the pay and receive matrices have zero eigenvalues, $p_-=r_-=0$, as the implied distribution is in this case supported on the upper half line. The resulting option price:
\begin{align}
o&[k]=\bar{p}\times \\
&\frac{1}{2}\left((f-k)+\sqrt{f^2-2fk\cos[2\theta]+k^2}\right) \notag
\end{align}
with $f:=\bar{r}/\bar{p}$ is similar to that produced by popular stochastic volatility models. In the examples of figure \ref{fig:implieddensity2}, both the matrices $P$ and $R$ are assumed to have eigenvalues $\{0,1\}$. The last set of graphs in this series represents the distribution by its implied volatility smile, which for each strike expresses the option price as the lognormal volatility used in the Black-Scholes formula to reproduce the price.

The first graph in figure \ref{fig:comparedensities} shows the difference between the implied volatilities of the classical and quantum binomial models. A classical model supported on two points necessarily has zero option value beyond these points, expressed as zero implied volatility outside a finite range of strikes. Thanks to quantum tunnelling, a more realistic implied volatility is generated by the quantum model. As demonstrated in the second graph, this quantum smile is the upper envelope of the family of classical smiles for binomial models of the swap rate with the same mean and root-variance, defined as the variance of the square-root of the swap rate.

In the quantum framework, the classical model is characterised by commutativity of the matrices $P$ and $R$, without loss of generality expressed as:
\begin{align}
P&=\bar{p}
\begin{bmatrix}
\cos[\alpha]^2 & 0 \\ 
0 & \sin[\alpha]^2
\end{bmatrix} \\
R&=\bar{r}
\begin{bmatrix}
\cos[\beta]^2 & 0 \\ 
0 & \sin[\beta]^2
\end{bmatrix}
\notag
\end{align}
in terms of angles $\alpha$ and $\beta$ satisfying $0<\alpha<\pi/2$ and $0\leq\beta\leq\pi/2$. The resulting option price is:
\begin{equation}
o[k]=\bar{p}(\omega_-(s_--k)^++\omega_+(s_+-k)^+)
\end{equation}
from the binomial distribution for the swap rate $s$ with support $(s_-,s_+)$ and weights $(\omega_-,\omega_+)$:
\begin{alignat}{3}
s_-&:=f\frac{\sin[\beta]^2}{\sin[\alpha]^2} & &\qquad & \omega_-&:=\sin[\alpha]^2 \\
s_+&:=f\frac{\cos[\beta]^2}{\cos[\alpha]^2} & &\qquad & \omega_+&:=\cos[\alpha]^2 \notag
\end{alignat}
The low-order moments of this distribution are:
\begin{align}
\mathbb{E}[1]&=1 \\
\mathbb{E}[s]&=f \notag \\
\mathbb{E}[s]-\mathbb{E}[\sqrt{s}]^2&=f\sin[\theta]^2 \notag
\end{align}
for the angle $\theta:=\beta-\alpha$. Fixing the mean and root-variance of $s$, the option price is a function of the free parameter $\alpha$:
\begin{align}
o[k]=\bar{p}&((f\sin[\alpha+\theta]^2-k\sin[\alpha]^2)^+ \\
&+(f\cos[\alpha+\theta]^2-k\cos[\alpha]^2)^+) \notag
\end{align}
Trigonometry simplifies this expression:
\begin{align}
f&\sin[\alpha+\theta]^2-k\sin[\alpha]^2=\frac{1}{2}(f-k) \\
&+\frac{1}{2}\sin[2\alpha+\gamma]\sqrt{f^2-2fk\cos[2\theta]+k^2} \notag \\
f&\cos[\alpha+\theta]^2-k\cos[\alpha]^2=\frac{1}{2}(f-k) \notag \\
&-\frac{1}{2}\sin[2\alpha+\gamma]\sqrt{f^2-2fk\cos[2\theta]+k^2} \notag
\end{align}
for an angle $\gamma$ that depends on the strike $k$ and the angle $\theta$ but not the angle $\alpha$, leading to:
\begin{align}
\sup&{}_\alpha\{o[k]\}=\bar{p}\times \\
&\frac{1}{2}\left((f-k)+\sqrt{f^2-2fk\cos[2\theta]+k^2}\right) \notag
\end{align}
for the maximum price of the option at strike $k$ over all classical binomial models with mean $f$ and root-variance $f\sin[\theta]^2$. The quantum option price is thus the maximum classical option price over a range of binomial models constrained to have matching low-order moments. This is an example of the more general result that identifies the quantum option price with the upper bound for the option price when the low-order moments of the distribution are known, derived from the Gelfand-Naimark-Segal geometry.

Higher dimensions in the quantum multinomial model generate more complex phenomenology from eigenvalues computed as the roots of the higher-order characteristic polynomial of the payoff matrix. As with the two-dimensional case, the price of the option is expressed as the sum of the positive roots of the polynomial, and complexity is generated in the valuation model thanks to the nontriviality of this exercise. The implied probability density is then a combination of a discrete distribution and a continuous distribution on its convex support, with structure characterised by the geometric relationship between the eigenbases of the pay and receive matrices. This can be used to closely match popular distributions such as the SABR model in as few as five discrete states, as demonstrated in the introductory examples.

Fundamental to the development of the option price model is the existence of a complete set of projections representing the exercise strategies for the option. Beyond finite dimensions, the model is enriched by the theory of von Neumann factors. Utilising the trace to extend the method to type I$_{\infty}$ and II factors, the results of this section remain valid subject to any necessary convergence conditions. For type III factors, the relationships between the state and observable algebras, encapsulated in the operators of Tomita-Takesaki theory, are leveraged to evaluate the option price. The distributions implied by the model characterise the von Neumann factor in the finite-dimensional case. The extent to which this statement holds more generally remains to be investigated.

\section{Time and stochastic calculus}

The progression from static to dynamic system introduces the notion of time as a partially-ordered set. A discrete schedule is defined to be a finite ordered sequence of times:
\begin{equation}
P=(p_{0},\ldots ,p_{n})
\end{equation}
with $p_{i-1}\leq p_{i}$, where $n$ is the number of intervals in the schedule. Introduce the following notation for the set, length, start and end of the schedule:
\begin{align}
\{P\}&:=\{p_{0},\ldots ,p_{n}\} \\
\left\vert P\right\vert &:=n \notag \\
P_{-}&:=p_{0} \notag \\
P_{+}&:=p_{n} \notag
\end{align}
Two consecutive schedules $P$ and $Q$ satisfying $P_{+}=Q_{-}$ are concatenated to create a new schedule $P\vee Q$:
\begin{equation}
P\vee Q:=(p_{0},\ldots ,p_{\left\vert P\right\vert }=q_{0},\ldots,q_{\left\vert Q\right\vert })
\end{equation}
with:
\begin{align}
\{P\vee Q\}&=\{P\}\cup \{Q\} \\
\left\vert P\vee Q\right\vert &=\left\vert P\right\vert +\left\vert Q\right\vert \notag \\
(P\vee Q)_{-}& =P_{-} \notag \\
(P\vee Q)_{+}& =Q_{+} \notag
\end{align}
Refinement is defined as the order relation $P\supset Q$ for schedules $P$ and $Q$ satisfying:
\begin{align}
\{P\}&\supset \{Q\} \\
P_{-}&=Q_{-}  \notag \\
P_{+}&=Q_{+}  \notag
\end{align}
For this relation to hold, the two schedules start and end at the same times, but the schedule $P$ refines the schedule $Q$ as all the intervals from $Q$ are mergers of intervals from $P$.

\begin{figure*}[!t]
\centering
\setlength{\tabcolsep}{0.0\linewidth}
\includegraphics[width=\linewidth]{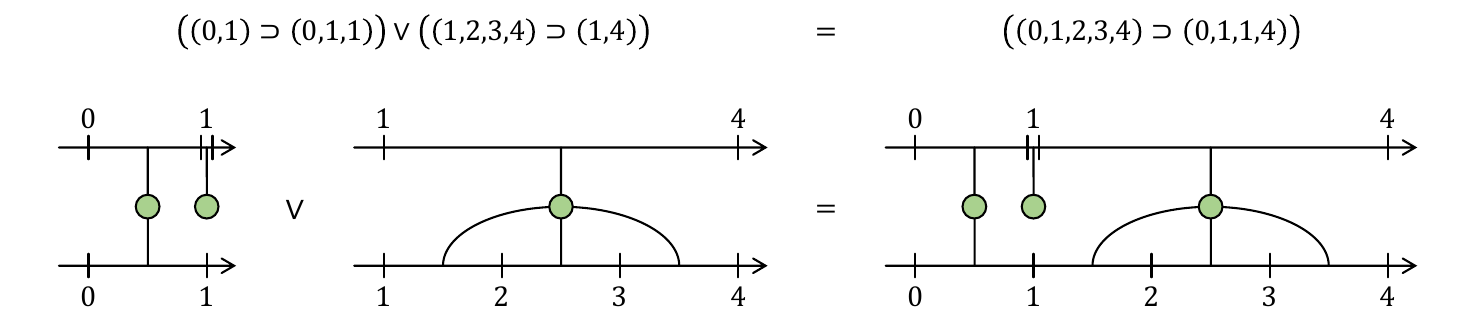}
\caption{The two refinements $(0,1)\supset(0,1,1)$ and $(1,2,3,4)\supset(1,4)$ are compatible for concatenation. String diagrams represent the concatenation by horizontal stacking.}\vspace{3ex}
\label{fig:timeconcatenation}
\includegraphics[width=\linewidth]{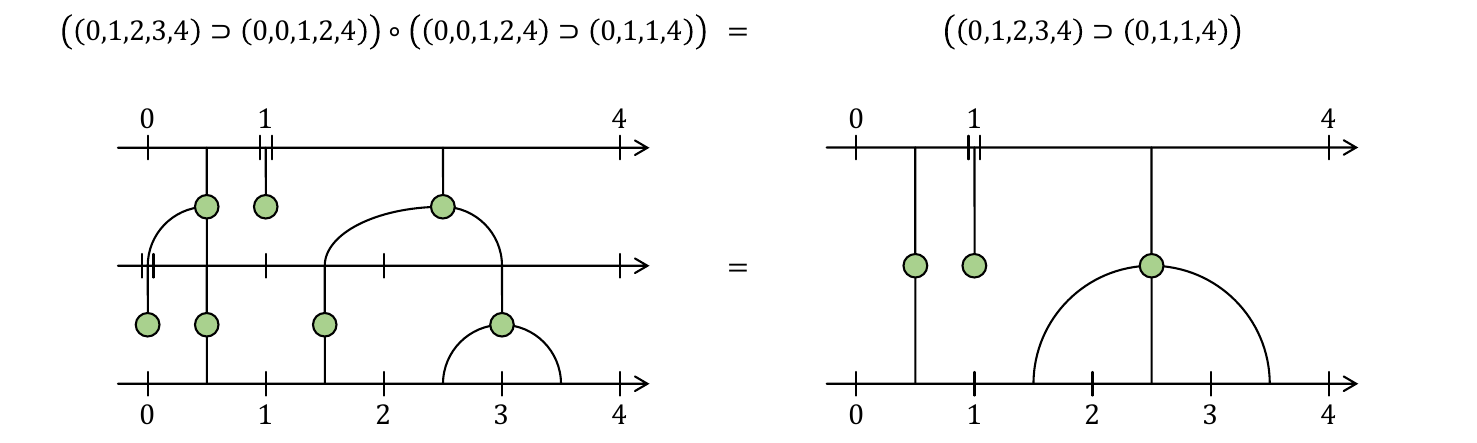}
\caption{The two refinements $(0,1,2,3,4)\supset(0,0,1,2,4)$ and $(0,0,1,2,4)\supset(0,1,1,4)$ are compatible for composition. String diagrams represent the composition by vertical stacking.}
\label{fig:timecomposition}
\end{figure*}

These definitions establish a bicategory structure on time, with $0$-cells given by the times, $1$-cells given by the schedules and $2$-cells given by the refinement relations. Concatenation and composition are implemented on compatible refinements with the definitions:
\begin{align}
(P\supset Q)\vee (R\supset S)&:=(P\vee R)\supset (Q\vee S) \notag \\
(P\supset Q)\circ (Q\supset R)&:=P\supset R
\end{align}
In the following, the time category is restricted to refinements with coarse schedules $Q$ satisfying $\left\vert Q\right\vert\geq 1$. Refinements are generated via concatenation from the total refinements $P\supset(P_{-},P_{+})$ for each schedule $P$, equivalently via concatenation and composition from the refinements:
\begin{align}
(p)&\supset (p,p) \\
(p,q,r)&\supset (p,r) \notag
\end{align}
and the trivial refinements $P\supset P$. These elementary refinements generalise the algebraic unit and product, and restricting to the subcategory they generate removes refinements such as $(p,p)\supset(p)$ that do not map to algebraic operations.

As presented in the examples of figures \ref{fig:timeconcatenation} and \ref{fig:timecomposition}, the string diagram for refinement links each interval in the coarse schedule with the intervals it merges in the refined schedule. The resemblance with string diagrams of the state algebra is intentional. The time category generalises algebra, and the dynamic system is implemented as a functor on refinements.

\subsection{Dynamic systems}

\begin{figure*}[!t]
\centering
\setlength{\tabcolsep}{0.0\linewidth}
\includegraphics[width=0.8\linewidth]{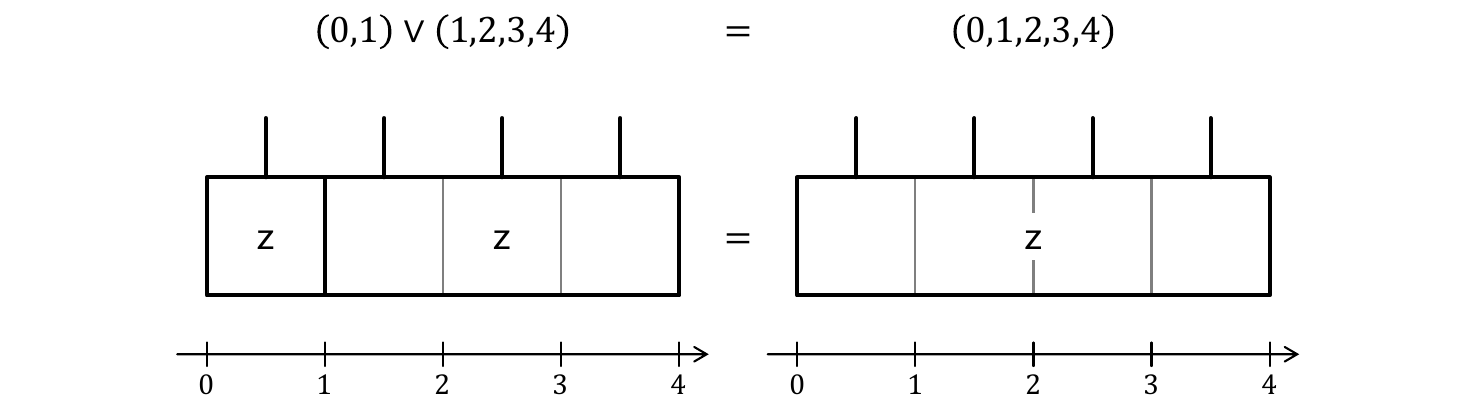}
\caption{The concatenation $(0,1)\vee(1,2,3,4)=(0,1,2,3,4)$ maps to this string diagram for the dynamic state, expressing compatibility of the state with the concatenation of refinements.}\vspace{3ex}
\label{fig:timestateconcatenation}
\includegraphics[width=0.8\linewidth]{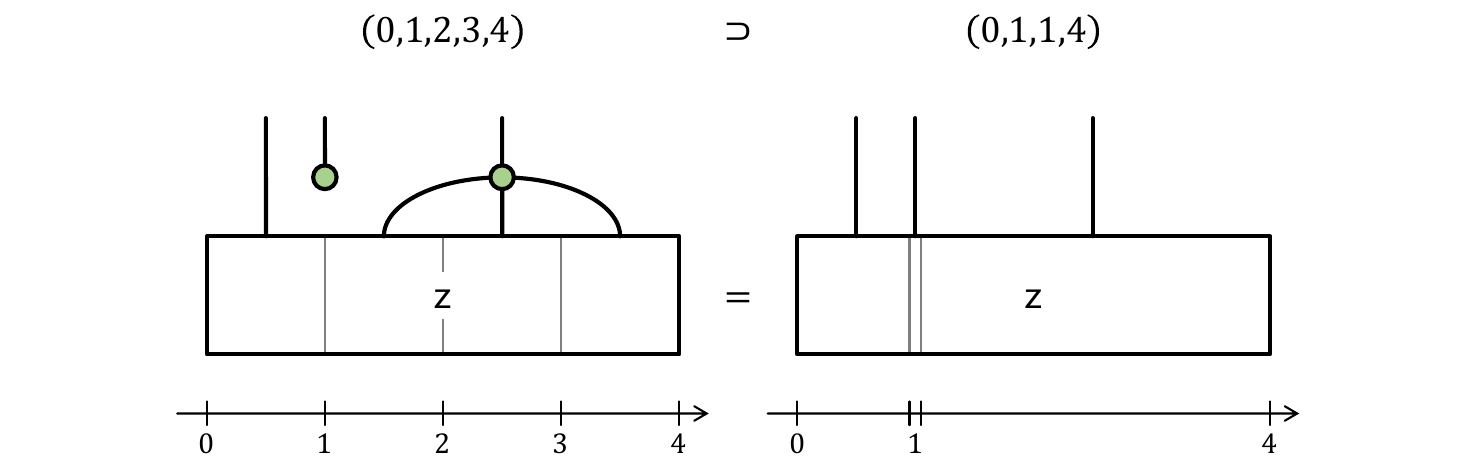}
\caption{The refinement $(0,1,2,3,4)\supset(0,1,1,4)$ maps to this string diagram for the dynamic state, expressing naturality of the state.}\vspace{3ex}
\label{fig:timestaterefinement}
\includegraphics[width=0.8\linewidth]{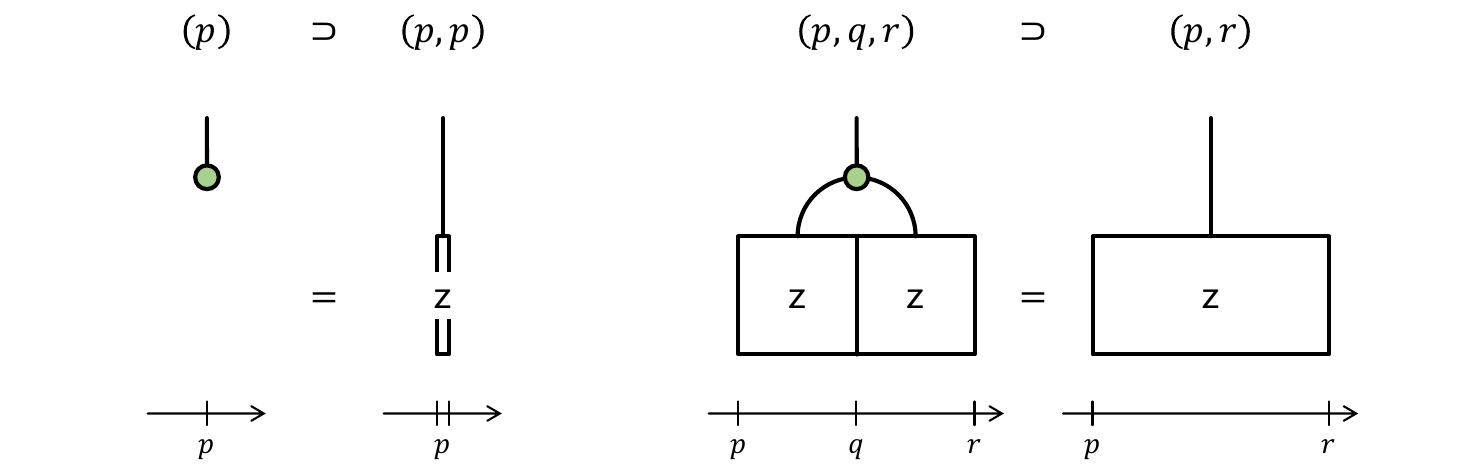}
\caption{The dynamic state forms a semigroup with natural properties mapping from the elementary refinements.}
\label{fig:statesemigroup}
\end{figure*}

In the dynamic system, the state on an interval expresses the total change over the interval. States on consecutive intervals are accumulated to generate the state on the merged interval, an operation that generalises the product. To facilitate this, a dynamic system is defined to be a monoidal functor from the time category to the empirical category. The functors are themselves objects of a monoidal category whose morphisms are the natural transformations between them. In this functor category, the monoidal unit $1$ is the trivial functor that maps all schedules to the empty system; a state of the dynamic system $K$ is then defined to be a natural transformation $\mathsf{z}\in \Homspace{1}{1}{K}$ from the unit functor.

Unpacking the definition, the dynamic system is implemented by an accumulation functor $K$ from the time category to the empirical category. The schedule $P$ maps to the system $K[P]$, consistent with concatenation in the form:
\begin{align}
K[(p)]&=1 \\
K[P\vee Q]&=K[P]\otimes K[Q] \notag
\end{align}
for compatible schedules $P$ and $Q$. Each refinement $P\supset Q$ maps to an operation:
\begin{equation}
K[P\supset Q]\in\Homspace{1}{K[P]}{K[Q]}
\end{equation}
that accumulates the state on all the merged intervals in the refinement. Functoriality becomes the properties:
\begin{align}
K[P\supset Q]\otimes K[R\supset S]&= \\
&\hspace{-1cm}K[(P\vee R)\supset (Q\vee S)] \notag \\
K[P\supset Q]\circ K[Q\supset R]&=K[P\supset R] \notag
\end{align}
for compatible refinements, so that string diagrams in the time category are replicated as string diagrams in the empirical category.

\begin{figure*}[!t]
\centering
\setlength{\tabcolsep}{0.0\linewidth}
\begin{tabular}{C{0.28\textwidth}C{0.44\textwidth}C{0.28\textwidth}}
Unit & Product & Involution \\ \hline
\begin{tabular}{c}\includegraphics[width=0.13\textwidth]{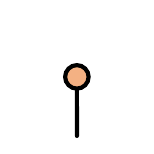}\\$\unita\in\Homspace{1}{K[(p,q)]}{1}$\\[1ex]$\counita\in\Homspace{1}{1}{K[(p,p)]}$\\\includegraphics[width=0.13\textwidth]{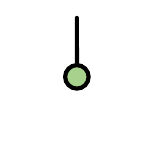}\end{tabular} &
\begin{tabular}{c}\includegraphics[width=0.13\textwidth]{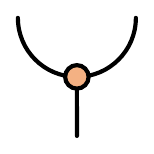}\\$\producta\in\Homspace{1}{K[(p,q)]}{K[(p,q)]\otimes K[(p,q)]}$\\[1ex]$\coproducta\in\Homspace{1}{K[(p,q)]\otimes K[(q,r)]}{K[(p,r)]}$\\\includegraphics[width=0.13\textwidth]{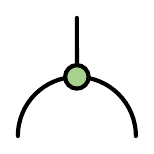}\end{tabular} &
\begin{tabular}{c}\includegraphics[width=0.13\textwidth]{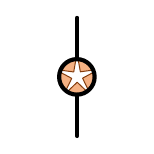}\\$\invol\in\Homspace{\ast}{K[(p,q)]}{K[(p,q)]}$\\[1ex]$\coinvol\in\Homspace{\ast}{K[(p,q)]}{K[(p,q)]}$\\\includegraphics[width=0.13\textwidth]{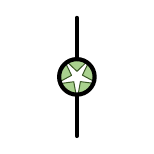}\end{tabular}
\end{tabular}\vspace{-2ex}
\caption{These operations on the dynamic system are needed to develop stochastic and functional calculus. The operations on the first row implement multiplication of observables on the interval $(p,q)$. The operations on the second row implement accumulation of states for the elementary refinements $(p)\supset(p,p)$ and $(p,q,r)\supset(p,r)$. The involutions are composed to create the antipode for reversibility of integration and differentiation.}
\label{fig:dynamicsystemoperations}
\end{figure*}

From these defining properties, the system $K[P]$ of the schedule $P=(p_0,\ldots,p_n)$ decomposes as:
\begin{equation}
K[P]=K[(p_0,p_1)]\otimes\cdots\otimes K[(p_{n-1},p_n)]
\end{equation}
and states are accumulated by the generalised unit and product operations:
\begin{align}
\counita&\in\Homspace{1}{1}{K[(p,p)]} \\
\coproducta&\in\Homspace{1}{K[(p,q)]\otimes K[(q,r)]}{K[(p,r)]} \notag
\end{align}
associated with the elementary refinements:
\begin{align}
\counita&:=K[(p)\supset(p,p)] \\
\coproducta&:=K[(p,q,r)\supset(p,r)] \notag
\end{align}
These elementary accumulators satisfy the unital and associative properties of algebra.

The dynamic state is implemented by a natural transformation $\mathsf{z}$ from the unit functor to the accumulation functor. The schedule $P$ maps to the state $\mathsf{z}[P]\in\Homspace{1}{1}{K[P]}$, consistent with concatenation in the form:
\begin{equation}
\mathsf{z}[P]\otimes\mathsf{z}[Q]=\mathsf{z}[P\vee Q]
\end{equation}
for compatible schedules $P$ and $Q$. Naturality becomes the property:
\begin{equation}
\mathsf{z}[P]\circ K[P\supset Q]=\mathsf{z}[Q]
\end{equation}
for each refinement $P\supset Q$. The operation accumulates the state over merged intervals, and naturality ensures consistency between these states.

Any string diagram in the time category maps to a string diagram in the empirical category. The dynamic state respects the concatenation and composition of refinements thanks to naturality and the functorial properties of the accumulators. With these consistency relations, the model of information supports the refinement of schedules, an essential ingredient in the stochastic calculus of states.

\subsection{Quantum group dynamics}

\begin{figure*}[!t]
\centering
\setlength{\tabcolsep}{0.0\linewidth}
\begin{tabular}{C{0.5\textwidth}|C{0.5\textwidth}}
Accessible States & Annihilated Observables \\[0.5ex] \hline
&\\[-0.5ex]
{\bf $\ast$-Subalgebra} & {\bf $\ast$-Ideal} \\[1.5ex]
\includegraphics[width=0.3\textwidth]{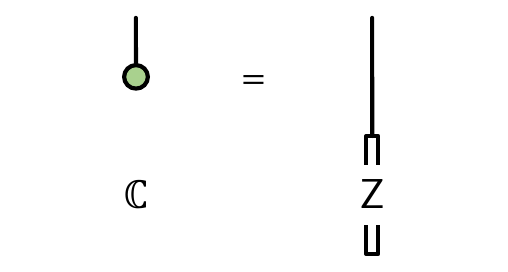} & \\
$\mathbb{C}\circ\counita=\mathsf{Z}[p,p]$ & \\[1.5ex]
\includegraphics[width=0.37\textwidth]{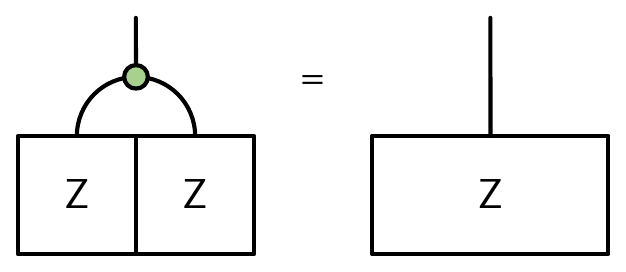} & \includegraphics[width=0.37\textwidth]{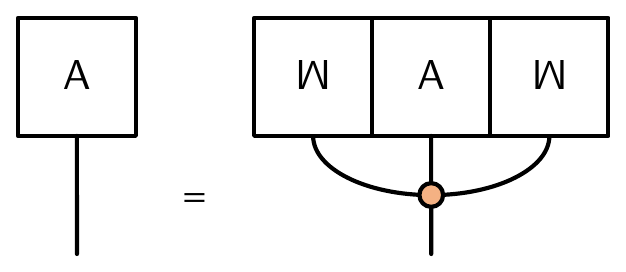} \\
$(\mathsf{Z}[p,q]\otimes\mathsf{Z}[q,r])\circ\coproducta=\mathsf{Z}[p,r]$ & $\mathsf{A}[p,q]=\producta\circ(\ObSpace[\bar{K}]\otimes\mathsf{A}[p,q]\otimes\ObSpace[\bar{K}])$ \\[1.5ex]
\includegraphics[width=0.3\textwidth]{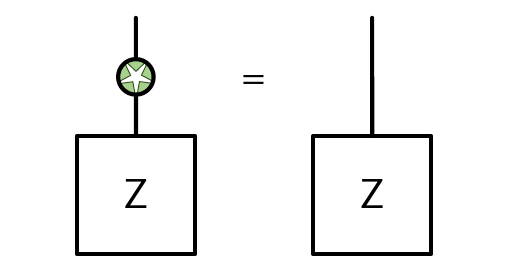} & \includegraphics[width=0.3\textwidth]{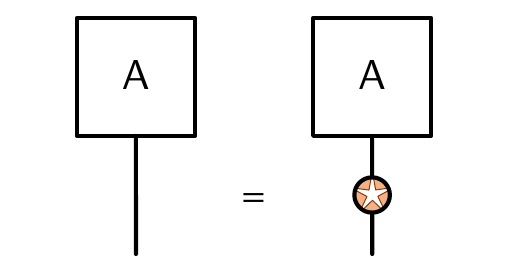} \\
$\mathsf{Z}[p,q]\circ\coinvol=\mathsf{Z}[p,q]$ & $\mathsf{A}[p,q]=\invol\circ\mathsf{A}[p,q]$ \\[2ex]
\end{tabular}
\caption{Stochastic calculus is enabled when the accessible states form a $\ast$-subalgebra. Functional calculus is enabled when the annihilated observables form a $\ast$-ideal.}
\label{fig:coidealcoalgebra}
\end{figure*}

Summarising earlier sections, the stochastic and functional calculus is constructed from the structural operations of figure \ref{fig:dynamicsystemoperations}, implementing the accumulation of states and the multiplication of observables. Observable involution engenders the concept of positivity and is used in the operator representation of states and observables. State involution has yet to be utilised, but this operation is essential to the relationship between integration and differentiation, and is composed with observable involution to create the antipode that implements reversibility.

In this section, the dynamic system is generated from a fixed quantum group $K$. Assigning the quantum group to each interval:
\begin{equation}
K[(p,q)]:=K
\end{equation}
defines a dynamic system that implements the state and observable algebras directly from the quantum group operations. With the quantum group as the {\em a priori} bounds of investigation, this definition gives the agent access to all states and observables on each interval.

More generally, if the {\em a posteriori} limitations of the agent are expressed by the subsets:
\begin{align}
\mathsf{Z}[p,q]&\subset\StSpace[\bar{K}] \\
\mathsf{A}[p,q]&\subset\ObSpace[\bar{K}] \notag
\end{align}
of accessible states and annihilated observables on each interval $(p,q)$, the dynamic system of the agent is modelled with the state and observable spaces:
\begin{align}
\StSpace[K[(p,q)]]&:=\mathsf{Z}[p,q] \\
\ObSpace[K[(p,q)]]&:=\ObSpace[\bar{K}]/\mathsf{A}[p,q] \notag
\end{align}
The pairing extends to this subsystem when the accessible states and annihilated observables are compatible:
\begin{equation}
\mathsf{Z}[p,q]=\mathcal{N}\mathsf{A}[p,q]
\end{equation}
Additional closure conditions enable stochastic and functional calculus.

\begin{figure*}[!t]
\setlength{\tabcolsep}{0.0\linewidth}
\centering\includegraphics[width=\linewidth]{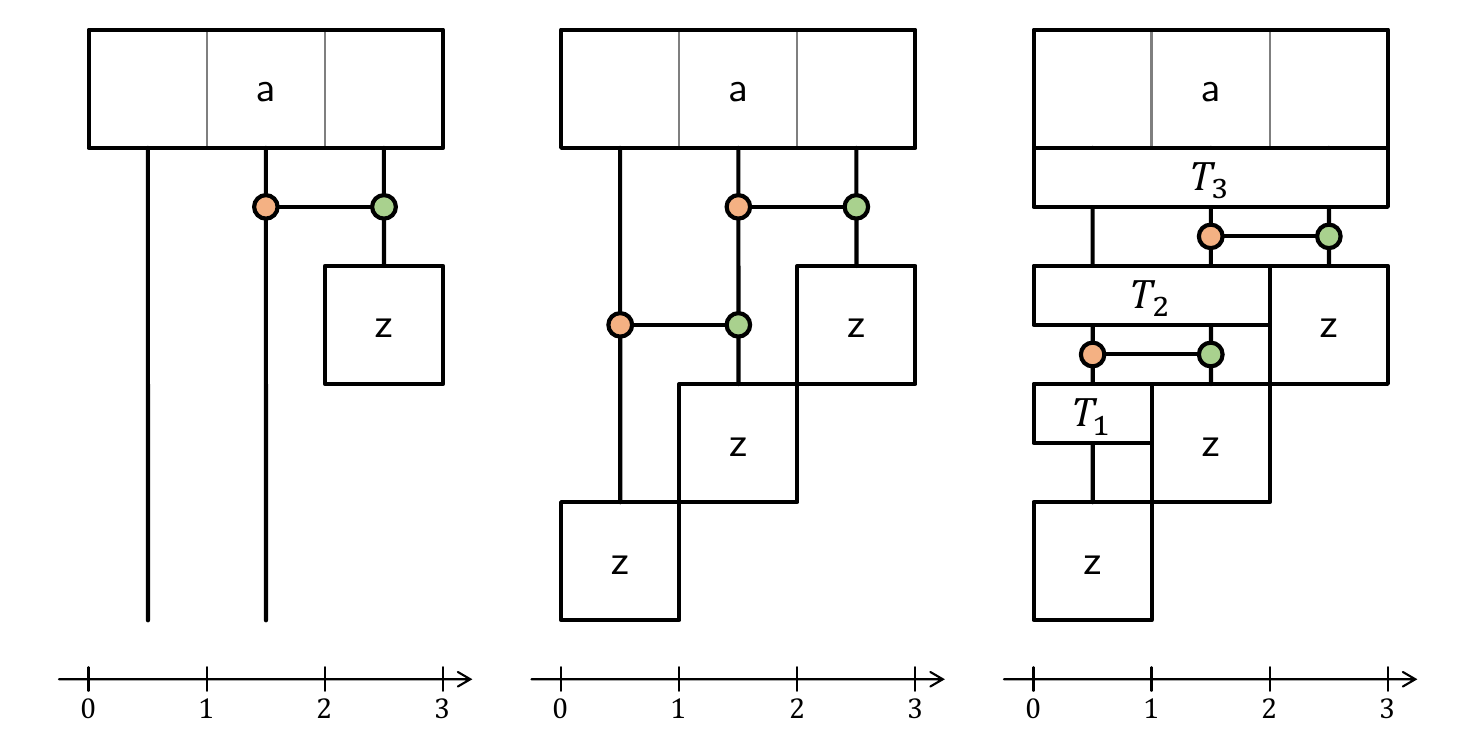}
\caption{These string diagrams demonstrate the use of integration in the evaluation of a path-dependent observable contingent on the state accumulated from time $0$ to times $1$, $2$ and $3$. The first example is conditional valuation, applying the state on the interval $(2,3)$ to generate an observable contingent on the state accumulated to times $1$ and $2$. The second example completes the valuation by iterating the action of the state on the third, second and first intervals. In the third example, this sequential application of the state is interlaced with functional transformations acting on the observable at each stage.}
\label{fig:valuationexample}
\end{figure*}

The dynamic system supports the accumulation of states when the accessible states form a $\ast$-subalgebra in the state algebra of the quantum group:
\begin{align}
\mathbb{C}\circ\counita&=\mathsf{Z}[p,p] \\
(\mathsf{Z}[p,q]\otimes\mathsf{Z}[q,r])\circ\coproducta&=\mathsf{Z}[p,r] \notag \\
\mathsf{Z}[p,q]\circ\coinvol&=\mathsf{Z}[p,q] \notag
\end{align}
The operations:
\begin{align}
\counita&\in\Homspace{1}{1}{K[(p,p)]} \\
\coproducta&\in\Homspace{1}{K[(p,q)]\otimes K[(q,r)]}{K[(p,r)]} \notag \\
\coinvol&\in\Homspace{\ast}{K[(p,q)]}{K[(p,q)]} \notag
\end{align}
are then inherited from the state algebra of the quantum group $K$, creating stochastic calculus on the dynamic system. The conditions on the accessible states are readily interpreted: the first condition imposes that the trivial interval accumulates no nontrivial states, and the second condition imposes that states accumulated over merged intervals are the products of states accumulated over the subintervals.

The dynamic system supports the multiplication of observables when the annihilated observables form a $\ast$-ideal in the observable algebra of the quantum group:
\begin{align}
\mathsf{A}[p,q]&=\producta\circ(\ObSpace[\bar{K}]\otimes\mathsf{A}[p,q]\otimes\ObSpace[\bar{K}]) \\
\mathsf{A}[p,q]&=\invol\circ\mathsf{A}[p,q] \notag
\end{align}
The operations:
\begin{align}
\unita&\in\Homspace{1}{K[(p,q)]}{1} \\
\producta&\in\Homspace{1}{K[(p,q)]}{K[(p,q)]\otimes K[(p,q)]} \notag \\
\invol&\in\Homspace{\ast}{K[(p,q)]}{K[(p,q)]} \notag
\end{align}
are then inherited from the observable algebra of the quantum group $K$, creating functional calculus on the dynamic system. The annihilated observables are the observables that are indistinguishable from zero, and the conditions of this result require that they are closed under multiplication with arbitrary observables and under complex conjugation.

In mathematical finance, the accessible states and annihilated observables are created from the set of economic states associated with the community of agents:
\begin{align}
\mathsf{Z}[p,q]&=\mathcal{A}\,\mathsf{Y}[p,q]\,\mathcal{A} \\
\mathsf{A}[p,q]&=\mathsf{Y}[p,q]\,\mathcal{A} \notag
\end{align}
The annihilated observables form a $\ast$-ideal by construction, thanks to the properties of the annihilation relation. The accessible states form a $\ast$-subalgebra when the economic states on merged intervals match the accumulated economic states on their subintervals.

\subsection{Integration and differentiation}

\begin{figure*}[!p]
\centering
\setlength{\tabcolsep}{0.0\linewidth}
\begin{tabular}{C{0.5\textwidth}C{0.5\textwidth}}
\includegraphics[width=0.4\textwidth]{Operation-Integration.pdf} & \includegraphics[width=0.4\textwidth]{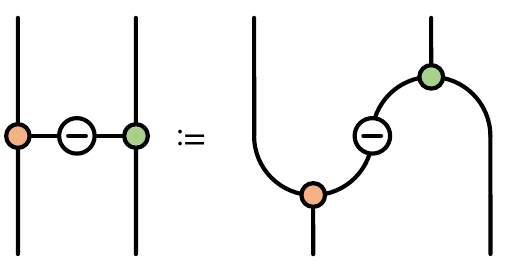} \\
$\mathsf{E}:=(\producta\otimes\identity)\circ(\identity\otimes\coproducta)$ & $\mathsf{D}:=(\producta\otimes\identity)\circ(\identity\otimes\antipode\otimes\identity)\circ(\identity\otimes\coproducta)$
\end{tabular}
\caption{Forward-starting and spot-starting deconstructions of the two-interval schedule are interchanged by integration and differentiation, implemented as the entanglement and disentanglement operations of the quantum group.}\vspace{1ex}
\label{fig:definitionintegrationanddifferentiation}
\begin{tabular}{C{\textwidth}}
\includegraphics[width=\textwidth]{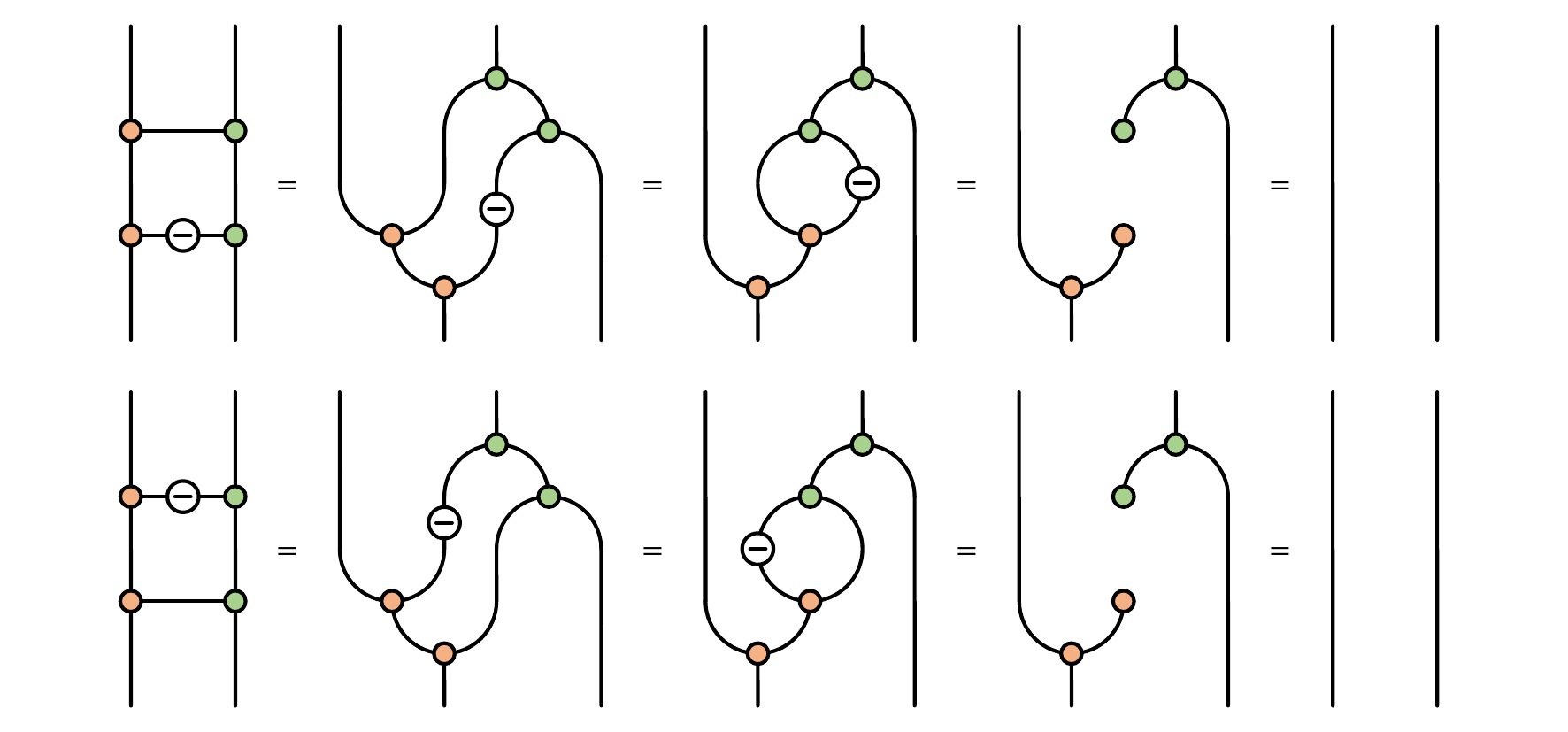} \\
$\mathsf{D}\circ\mathsf{E}=\identity\otimes\identity=\mathsf{E}\circ\mathsf{D}$
\end{tabular}
\caption{Integration and differentiation are reversible operations. In these string diagrams, the first and last steps respectively use associativity and unitality of the state and observable algebras. The middle step then employs the Hopf axiom of the quantum group.}\vspace{1ex}
\label{fig:invertibilityintegrationanddifferentiation}
\begin{tabular}{C{0.5\textwidth}C{0.5\textwidth}}
\includegraphics[width=0.4\textwidth]{Operation-IntegrationMultiple.pdf} & \includegraphics[width=0.4\textwidth]{Operation-DifferentiationMultiple.pdf} \\
$\mathsf{E}\in\Homspace{1}{\bigotimes\nolimits_{i=1}^n K[(p_{i-1},p_i)]}{\bigotimes\nolimits_{i=1}^n K[(p_0,p_i)]}$ & $\mathsf{D}\in\Homspace{1}{\bigotimes\nolimits_{i=1}^n K[(p_0,p_i)]}{\bigotimes\nolimits_{i=1}^n K[(p_{i-1},p_i)]}$
\end{tabular}
\caption{Integration and differentiation extend to discrete schedules with multiple intervals by concatenation and composition. These operations are also reversible, and extend to continuous schedules in the limit of refinement.}
\label{fig:scheduleintegrationanddifferentiation}
\end{figure*}

The empirical constraints of the agent are expressed in the subspaces of accessible states and annihilated observables, assumed from hereon to satisfy the conditions of the previous section. In the transition from discrete to continuous dynamics, the continuous schedule is obtained in the refinement limit of the discrete schedule. The technical challenge in this construction lies in the transition between integral and differential perspectives, whose reversibility depends nontrivially on the state and observable algebras and on the antipode generated as the composition of their involutions.

For the schedule $P=(p_0,\ldots,p_n)$, the perspectives differ in the way they deconstruct the schedule into intervals:
\begin{align}
(p_0,p_1),(p_1,p_2),&\ldots,(p_{n-1},p_n) \\
(p_0,p_1),(p_0,p_2),&\ldots,(p_0,p_n) \notag
\end{align}
interchanged using integration and differentiation operations
\begin{equation}
\bigotimes\nolimits_{i=1}^n K[(p_{i-1},p_i)]\xrightleftharpoons[\mathsf{D}]{\mathsf{E}}\bigotimes\nolimits_{i=1}^n K[(p_0,p_i)]
\end{equation}
Starting with the two-interval schedule, integration and differentiation are the operations:
\begin{align}
\mathsf{E}&:=(\producta\otimes\identity)\circ(\identity\otimes\coproducta) \\
\mathsf{D}&:=(\producta\otimes\identity)\circ(\identity\otimes\antipode\otimes\identity)\circ(\identity\otimes\coproducta) \notag
\end{align}
as presented in the string diagrams of figure \ref{fig:definitionintegrationanddifferentiation}. From the derivation in figure \ref{fig:invertibilityintegrationanddifferentiation}, they are mutually inverse:
\begin{equation}
\mathsf{D}\circ\mathsf{E}=\identity\otimes\identity=\mathsf{E}\circ\mathsf{D}
\end{equation}
The string diagrams of figure \ref{fig:scheduleintegrationanddifferentiation} then define integration and differentiation on the discrete schedule as the repeated application of the two-interval operations. Categorical coherence of these constructions means that discrete approximations to the continuous schedule can be refined to arbitrary degree, and the continuous integral is well defined if the refinement limit converges.

Consider the path-dependent observable:
\begin{equation}
\mathsf{a}_n\in\Homspace{1}{\bigotimes\nolimits_{i=1}^n K[(p_0,p_i)]}{1}
\end{equation}
contingent on the state accumulated from time $p_0$ to times $p_1,\ldots,p_n$. Terminating with this observable, define an iterative valuation scheme:
\begin{equation}
\mathsf{a}_{i-1}:=((\otimes^{i-2}\identity)\otimes\mathsf{C}_i)\circ T_i[\mathsf{a_i}]
\end{equation}
where $T_i$ is a functional transformation applied to the observable and $\mathsf{C}_i$ is the operation of conditional valuation:
\begin{equation}
\mathsf{C}_i:=(\identity\otimes\mathsf{z}[(p_{i-1},p_i)])\circ\mathsf{E}
\end{equation}
From the viewpoint of the observable, conditional valuation presents the final two intervals in the differential perspective and then applies the dynamic state $\mathsf{z}$ to the second interval, as demonstrated in the first string diagram of figure \ref{fig:valuationexample}. The scheme is completed with the valuation:
\begin{equation}
a:=\mathsf{z}[(p_0,p_1)]\bullet T_1[\mathsf{a}_1]
\end{equation}
as demonstrated in the third string diagram of figure \ref{fig:valuationexample}. This is a typical sequence in the pricing of derivative securities, where the termsheet prescribes both the terminal payoff and additional payments or termination clauses at earlier times. The structural operations of the quantum group are essential for this construction, consistently capturing the accumulation of states and the convexities between observables, and equipping the valuation model with a notion of integration and differentiation. The quantum group thus provisions the valuation model with all the operations necessary to express and evaluate the derivative.

\begin{figure}[!t]
\centering
\setlength{\tabcolsep}{0.0\linewidth}
\centering\includegraphics[width=0.5\linewidth]{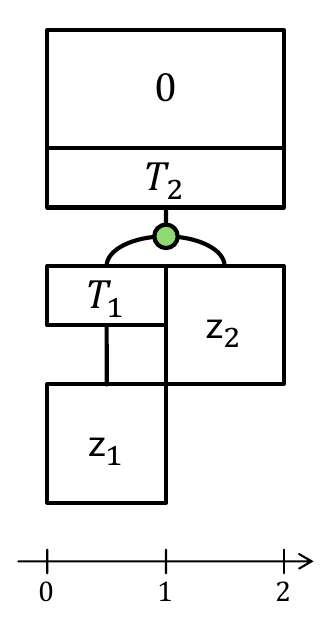}
\captionsetup{singlelinecheck=off}
\caption{In this two-interval example, the Bermudan option terminates with zero payoff at time $2$. At times $1$ and $2$ the holder of the option also has the right to exercise into an underlying security, implemented on the payoff by the transformations $T_1$ and $T_2$ respectively.}
\label{fig:bermudanoption}
\end{figure}

\begin{figure*}[!t]
\centering
\setlength{\tabcolsep}{0.0\linewidth}
\begin{tabular}{C{0.166\textwidth}C{0.166\textwidth}C{0.166\textwidth}|C{0.166\textwidth}C{0.166\textwidth}C{0.166\textwidth}}
$[X]$ & $[X]$ & $[X]$ & $1$ & $[X]\hspace{0.35cm}\otimes\hspace{0.35cm}[Y]$ & $[X]$ \\
\includegraphics[width=0.13\textwidth]{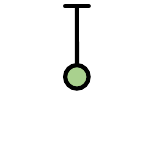} & \includegraphics[width=0.13\textwidth]{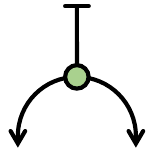} & \includegraphics[width=0.13\textwidth]{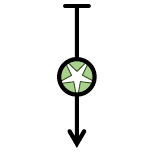} & \includegraphics[width=0.13\textwidth]{Operation-ObservableUnitMap.pdf} & \includegraphics[width=0.13\textwidth]{Operation-ObservableProductMap.pdf} & \includegraphics[width=0.13\textwidth]{Operation-ObservableInvolutionMap.pdf} \\[-0.5ex]
$(1\in X)$ & $[\coproducta^{-1}[X]]$ & $[X^{-1}]$ & $[K]$ & $[X\cap Y]$ & $[X]$ \\[1ex] \hline
&&&&&\\[-1.5ex]
$[1]$ & $[xy]$ & $[x^{-1}]$ & $1$ & $[x]\hspace{0.4cm}\otimes\hspace{0.4cm}[x]$ & $[x]$ \\
\includegraphics[width=0.13\textwidth]{Operation-StateUnitMap.pdf} & \includegraphics[width=0.13\textwidth]{Operation-StateProductMap.pdf} & \includegraphics[width=0.13\textwidth]{Operation-StateInvolutionMap.pdf} & \includegraphics[width=0.13\textwidth]{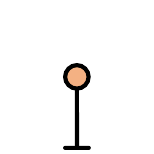} & \includegraphics[width=0.13\textwidth]{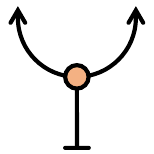} & \includegraphics[width=0.13\textwidth]{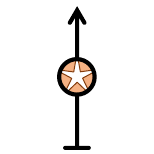} \\[-0.5ex]
$1$ & $[x]\hspace{0.4cm}\otimes\hspace{0.4cm}[y]$ & $[x]$ & $[x]$ & $[x]$ & $[x]$
\end{tabular}
\caption{The operations of $\ast$-algebra are defined on the point measures and digital functions of a classical group, linearly and topologically completed as the convolution algebra of measures and the pointwise algebra of functions. On the top row, the operations are defined by their actions on digital functions. On the bottom row, the operations are defined by their actions on point measures.}
\label{fig:classicalquantumgroup}
\end{figure*}

In the two-interval example of figure \ref{fig:bermudanoption}, the Bermudan option references an underlying security with price $\mathsf{u}_1$ at time $1$ and price $\mathsf{u}_2$ at time $2$, each assumed to be contingent only on the state accumulated to the observation time. Terminating with zero, the derivative security acquires value through its optionality:
\begin{equation}
b=\mathsf{z}_1\bullet T_1[(\identity\otimes\mathsf{z}_2)\circ\coproducta\circ T_2[0]]
\end{equation}
where the options are implemented by the transformations:
\begin{align}
T_2[\mathsf{b}_2]&:=(1-\mathsf{p}_2)\mathsf{b}_2+\mathsf{p}_2(\mathsf{u}_2-k_2) \\
T_1[\mathsf{b}_1]&:=(1-\mathsf{p}_1)\mathsf{b}_1+\mathsf{p}_1(\mathsf{u}_1-k_1) \notag
\end{align}
and the exercise decisions are expressed in the projections $\mathsf{p}_1$ and $\mathsf{p}_2$. Optimal exercise then maximises the price of the derivative security over all projections that commute with the price of the underlying security.

More complex derivative securities are similarly constructed, with path dependence incorporated through the structure of the dynamic system, and optionality embedded with the use of projections to express exercise decisions.

\section{Classical groups}

The valuation model of classical economics is set membership. For the measurable group $K$, theoretical measures $\StSpace[K]$ are spanned by the point measures $[x]$ associated with elements $x\in K$ and theoretical functions $\ObSpace[K]$ are spanned by the digital functions $[X]$ associated with measurable subsets $X\subset K$, with pairing:
\begin{equation}
[x]\bullet [X]:=(x\in X)
\end{equation}
Topological completion generates spaces $\StSpace[\bar{K}]$ and $\ObSpace[\bar{K}]$ of empirical measures and functions that are approximated by discrete measures and step functions:
\begin{align}
\mathsf{z}&=\sum_{n=1}^\infty\zeta_n[x_n] \\
\mathsf{a}&=\sum_{n=1}^\infty\alpha_n[X_n] \notag
\end{align}
convergent in the sense that the limits:
\begin{align}
\mathsf{z}\bullet [X]&=\sum_{n=1}^\infty\zeta_n(x_n\in X) \\
[x]\bullet\mathsf{a}&=\sum_{n=1}^\infty\alpha_n(x\in X_n) \notag
\end{align}
exist for all measurable subsets $X$ and elements $x$ of the group.

Algebra is defined on both sides of the pairing. On the left, it is the convolution algebra of discrete measures with unit, product and involution:
\begin{align}
1&:=[1] \\
[x][y]&:=[xy] \notag \\
[x]^\ast&:=[x^{-1}] \notag
\end{align}
On the right, it is the pointwise algebra of step functions with unit, product and involution:
\begin{align}
1&:=[K] \\
[X][Y]&:=[X\cap Y] \notag \\
[X]^\ast&:=[X] \notag
\end{align}
This recasting into the language of quantum groups has novelty in the dual interpretation of the information model. Exploiting the duality between states and observables, the holographic principle is applied by interchanging the roles of measure and function, generating two inequivalent valuation models -- {\em classical} and {\em coclassical} economics -- from the same pairing.

Classical economics is established on the system $K$ whose states are measures and whose observables are functions on the group. In contrast, coclassical economics is established on the system $\Opposite{K}$ whose states are functions and whose observables are measures on the group. The existence of two valuation models based on the same classical group is a direct consequence of quantum group duality.

As an example from classical economics, generate the space of pricing states $\mathsf{Z}$ from a fixed $\sigma$-ideal $\mathcal{N}$ of null sets. In this model, the pricing state $\mathsf{z}\in\mathsf{Z}$ evaluates null sets to zero:
\begin{equation}
\mathsf{z}\bullet[X]=0
\end{equation}
for all null sets $X\in\mathcal{N}$. The common ideal of null sets identifies the negligible events in the model, and price resolves to integral. This is the foundation for classical pricing. The null sets constrain the range of possibilities for the economy, and the pricing state respects this through the principle of equivalence. Extending to the dynamic system, time maps into the ideal space of the measurability structure, and the model has a natural concept of filtration that is compatible with the refinement of schedules.

As an example from coclassical economics, generate the space of pricing states $\mathsf{Z}$ from a fixed finite-dimensional measurable representation:
\begin{equation}
\rho:K\to\mathbb{C}[\ObDim]
\end{equation}
of the group. In this model, the pricing state $\mathsf{z}\in\mathsf{Z}$ is a representative function:
\begin{equation}
\mathsf{z}\bullet[x]=\tr[\hat{\mathsf{z}}\rho[x]]
\end{equation}
associated with a matrix $\hat{\mathsf{z}}\in\mathbb{C}[\ObDim]$. This model is discrete and numerically efficient, evaluated from the eigenvalues for the payoff matrices. As demonstrated earlier, the model accesses a wide range of credible price distributions, comparable to classical models with continuous support, thanks to noncommutativity. Extending to the dynamic system, time maps into the set of representations for the group, and the model leverages the algebraic and topological structure of this set. Groups with a rich and well-understood representation theory, such as discrete or compact groups, can then be used as the basis for numerical methods in pricing.

These two models present radically different approaches to pricing. They are the endpoints of the spectrum of price models that can be generated from quantum groups, and demonstrate the potential of quantum group duality to create novel numerical methods. Extrapolating this philosophy, known quantum groups with well-understood representations generate practical solutions for pricing, utilising noncommutativity as a resource for both states and observables.

\section{Conclusion}

As algebraic and topological constructions, the functional calculus and the stochastic calculus are founded upon dual Hopf $\ast$-algebras that are perfectly symmetric in the axioms of the quantum group. The dynamic system is then generated from dual pairs of state $\ast$-subalgebras and observable $\ast$-ideals. While the rules of the quantum group are symmetric, its subalgebra-ideal structures may not be. The two dual interpretations of the pairing thus potentially lead to inequivalent models of information, and even to distinct geometries for time.

This is evident in the classical group, where the choice resolves the states as either measures with common null sets or as representative functions with fixed representation. The holographic principle, which observes that both information models are implemented by the same pairing, is a feature of the axiomatic duality of state and observable, and extends to any dynamic system constructed from a quantum group.

All the operations necessary to create a feature-complete model of mathematical finance are found within the quantum group, and noncommutativity is a practical resource in this model. The approach significantly extends the efficacy of discrete numerical methods, using the eigenvalue map to explore a far wider range of realistic price distributions. By providing the tools to express the elementary economic principles, the quantum group is ideally suited to the pricing application.

In this essay, potential applications of quantum groups in mathematical finance are introduced with the intention to motivate further investigations. On the foundational side, future development will root the approach robustly within mathematical analysis and explore in more depth the connection with von Neumann algebras. On the practical side, a complete demonstration of the approach for pricing derivative securities is supported by further development of the utilities for discrete models, with links to the representation theory of discrete or compact groups.

The objective is to plant the tools of quantum information theory firmly within the library of numerical methods available for real applications in mathematical finance, extending the phenomenology of established methods such as finite difference, Monte Carlo or machine learning schemes.

\end{document}